\documentclass[twocolumn]{aastex61}
\usepackage{amssymb}
\usepackage{multirow}
\usepackage{morefloats}
\usepackage{amsmath}
\usepackage{bigstrut}
\usepackage{booktabs}
\usepackage{natbib}
\usepackage{float}
\usepackage{graphicx,epsfig,fancyhdr,epsf,txfonts,epstopdf}
\usepackage{latexsym,bbm}
\usepackage{lineno}
\usepackage{color}
\usepackage{ulem}
\usepackage{stfloats}

\received{XXX}
\revised{YYY}
\accepted{ZZZ}

\shortauthors{Zhang et al.}

\begin{document}

\shorttitle{X-ray IDV of PKS 2155--304}
%\shortauthors{Zhang et al.}

\title{X-ray Intraday Variability of the TeV Blazar PKS 2155$-$304 with {\it Suzaku} during 2005 $-$ 2014}

\author{Zhongli Zhang}\thanks{Email: zzl@shao.ac.cn}
\affiliation{Shanghai Astronomical Observatory, Chinese Academy of Sciences, Shanghai 200030, People's Republic of China}
\affil{Key Laboratory of Radio Astronomy, Chinese Academy of Sciences, 210008 Nanjing, People's Republic of China}

\author{Alok C.\ Gupta}\thanks{Email: acgupta30@gmail.com}
\affiliation{Key Laboratory for Research in Galaxies and Cosmology, Shanghai Astronomical Observatory, Chinese Academy of Sciences, Shanghai 200030, People's Republic of China}
\affiliation{Aryabhatta Research Institute of Observational Sciences (ARIES), Manora Peak, Nainital -- 263001, India}

\author{Haritma Gaur}\thanks{Email: harry.gaur31@gmail.com}
\affiliation{Aryabhatta Research Institute of Observational Sciences (ARIES), Manora Peak, Nainital -- 263001, India}

\author{Paul J. Wiita}
\affiliation{Department of Physics, The College of New Jersey, 2000 Pennington Rd., Ewing, NJ 08628-0718, USA}

\author{Tao An}
\affiliation{Shanghai Astronomical Observatory, Chinese Academy of Sciences, Shanghai 200030, People's Republic of China}
\affil{Key Laboratory of Radio Astronomy, Chinese Academy of Sciences, 210008 Nanjing, People's Republic of China}

\author{Yang Lu}
\affiliation{Shanghai Astronomical Observatory, Chinese Academy of Sciences, Shanghai 200030, People's Republic of China}

\author{Shida Fan}
\affiliation{School of Physics and Astronomy, Shanghai Jiao Tong University, 800 Dongchuan Road, Minhang, Shanghai 200240, 
People's Republic of China}

\author{Haiguang Xu}
\affiliation{School of Physics and Astronomy, Shanghai Jiao Tong University, 800 Dongchuan Road, Minhang, Shanghai 200240, 
People's Republic of China}
\affiliation{IFSA Collaborative Innovation Center, Shanghai Jiao Tong University, 800 Dongchuan Road, Minhang, Shanghai 200240, 
People's Republic of China}

\begin{abstract}
\noindent
 We have examined 13 pointed observations of the TeV emitting high synchrotron peak blazar PKS 2155$-$304, taken by the {\it Suzaku} satellite 
throughout its operational period. We found that the blazar showed large-amplitude intraday variabilities in the soft (0.8 -- 1.5 keV) and the hard (1.5 -- 8.0 keV) bands in the light curves.  Spectral variability on intraday timescales is estimated using the hardness ratio.  The blazar usually becomes harder when brighter and vice versa, following the typical behavior of high synchrotron peak blazars. The power spectral density (PSD) analyses of 11 out of 13 light curves in the total energy (0.8 -- 8.0  keV) are found to be red-noise dominated, with power-law spectral indices that span a large range, from $-$2.81 to $-$0.88. Discrete correlation function analyses of all the 13 light curves between the soft and the hard bands show that they are well correlated and peak at, or very close to, zero lag. This indicates that the emissions in soft and hard bands are probably cospatial and emitted from the same population of leptons. Considering fluxes versus variability timescales, we found no correlation on intraday timescales, implying that X-ray emission from  PKS 2155$-$304 is not dominated by simple changes in the Doppler factor. 
We briefly discuss the most likely emission mechanisms responsible for the observed flux and spectral variabilities and place constraints on magnetic field strength and Lorentz factors of the electrons emitting the X-rays in the most likely scenario.        
\end{abstract}


\section{Introduction}
\label{sec:introduction}
\noindent
Blazars constitute a small subset of the class of radio-loud active galactic nuclei (AGNs) consisting of BL Lacertae objects (BL Lacs) and flat spectrum radio quasars (FSRQs). Blazars exhibit strong flux and spectral variability throughout the electromagnetic (EM) spectrum, typically have core dominated radio structures, and their radiation at all wavelengths is predominantly non-thermal and is also strongly polarized ($>$ 3\%) in both optical and (less remarkably) radio  bands. According to a unified model of radio-loud AGNs based on the angle between the line of sight and the approaching relativistic jet, blazar jet makes an angle of $\lesssim$ 10$^{\circ}$ from the observer's line of sight \citep[e.g.,][]{Urry1995}. The spectral energy distributions (SEDs) of blazars' spectra are well represented by double-humped structures. The first hump (lower energy) of the SED peaks in infrared through X-ray bands and is clearly dominated by synchrotron emission from the relativistic jet, while the second, higher energy hump peaks in $\gamma$-rays (GeV to TeV energies) and arises from leptonic and/or hadronic-based processes \citep[e.g.,][]{Kirk1998,2003APh....18..593M,2004NewAR..48..367K,Gaur2010}. According to the leptonic scenarios, the high-energy hump is from inverse Compton (IC) scattering of the surrounding photon fields \citep[e.g.,][]{2009MNRAS.397..985G}: for the synchrotron photons themselves this is called the synchrotron-self-Compton (SSC) mechanism, whereas for fields external to the jet, as from the accretion disc or broad line region, it is called the external Comptonization (EC) mechanism. The hadronic model invokes proton-synchrotron and/or proton-photon cascade processes \citep[e.g.,][]{2003APh....18..593M}. \\ 
\\
Across the EM spectrum blazars have shown detectable flux variations on diverse timescales ranging from a few minutes to years. Flux variations on timescales from a few minutes to less than a day are variously called micro-variability \citep{1989Natur.337..627M}, intraday variability (IDV) \citep[e.g.,][]{Wagner1995}, or intra-night variability \citep[e.g.,][]{2009MNRAS.399.1622G}. Changes in flux on the timescales from days to a few months are often called short term variability (STV) while flux variations over timescales of months to years are commonly referred to as long term variability (LTV) \citep[e.g.,][]{Gupta2004}. \\
\\
The blazar PKS 2155$-$304 ($\alpha_{2000.0} =$ 21h 58m 52.07s, $\delta_{2000.0} =-$30$^{\circ}$13$^{'}$32.1$^{''}$) was listed in an early catalogue of BL Lac objects \citep{1980ApJS...43...57H}, has a redshift $z =$ 0.116$\pm$0.002 \citep{1993ApJ...411L..63F} and is the brightest object in the UV and X-ray bands in the southern hemisphere. It was classified as TeV blazar by detection of very high energy (VHE) $\gamma-$ray emission by the Durham {\it MK6} telescopes \citep{1999ApJ...513..161C} and later confirmed as a VHE $\gamma-$ray emitter by {\it H.E.S.S.} (High Energy Stereoscopic System) at a 45$\sigma$ detection \citep{2005A&A...442..895A}. 
Thanks to the new space and ground based $\gamma-$ray observing facilities, e.g., {\it Fermi}, {\it H.E.S.S.}, {\it MAGIC} (Major Atmospheric Gamma-ray Imaging Cerenkov), {\it VERITAS} (Very Energetic Radiation Imaging Telescope Array System), etc., there has been a revolution in TeV $\gamma-$ray astronomy and now a substantial number of TeV $\gamma-$ray emitting objects have been detected, including at least 73 blazars\footnote{http://tevcat.uchicago.edu/}. \\ 
\\
The TeV emitting blazars are among the prime targets for observations at X-ray and lower $\gamma-$ray energies. 
Extensive observations in these energy bands have been made of PKS 2155$-$304  \citep[e.g.,][]{1979ApJ...229L..53S,1980ApJ...237L..11S,1984ApJ...278L..99C,1986ApJ...306L..71M,1993ApJ...404..112S,1995ApJ...454L..93V,1998A&A...333L...5G,1999ApJ...513..161C,1999ApJ...521..552C,2000ApJ...528..243K,2001ApJ...554..274E,2002ApJ...572..762Z,2006ApJ...637..699Z,2005A&A...430..865A,2005A&A...442..895A,2007ApJ...664L..71A,2008PhRvL.101q0402A,2009ApJ...696L.150A,Gaur2010,2017ApJ...850..209G,2013PhRvD..88j2003A,2014MNRAS.444.1077K,2014MNRAS.444.3647B,2017AJ....153....2M,2010A&A...520A..83H,2017A&A...598A..39H,2017A&A...600A..89H}. 
It is one of the most extensively studied blazars due to its strong flux, as well as frequent polarization and spectral 
variations. Hence it has been a prime target for several simultaneous multi-wavelength observational 
campaigns for extended periods of time \citep[e.g.,][]{1989ApJ...341..733T,1993ApJ...411..614U,1997ApJ...486..799U,1994A&A...288..433B,1995ApJ...438..108C,1995ApJ...438..120E,1997ApJ...486..770P,1997ApJ...486..784P,2005A&A...442..895A,2009ApJ...696L.150A,2009A&A...502..749A,2007ApJ...671...97O,2012A&A...539A.149H,2019MNRAS.484..749C}. \\ 
\\
Since 1979, PKS 2155$-$304 has been observed on many occasions by various X-ray missions. The first X-ray observation of PKS 2155--304, from the High Energy Astronomy Observatory {\it (HEAO-1)} was reported by \citet{1979ApJ...229L..53S}, and variation by a factor of 2 in the energy range 0.5--20 keV was found on a timescale of 6 hours \citep{1980ApJ...237L..11S}. From the observations by {\it EXOSAT LE} in October -- November 1983 in the energy range 1--6 keV, an overall variation of a factor of 10 was reported \citep{1986ApJ...306L..71M}.  Additional {\it EXOSAT} observations of this blazar showed that the variability is more pronounced in a harder X-ray band (1--6 keV) compared to a softer X-ray band (0.1--2 keV), and the power spectrum of these variations shows red noise-type variability, with the power increasing steeply towards lower frequencies \citep{1991ApJ...380...78T}. In the observations made with the Large Area Counter (LAC) onboard the {\it Ginga} satellite in 1988 and 1989, the source exhibited variability by a factor of 7 in the 2 $-$ 6 keV band, and showed a break in the spectrum at $\sim$ 4 keV \citep{1993ApJ...404..112S}. Extensive {\it ROSAT} PSPC observations of  PKS 2155--304 carried out between 12--15 November 1991, reported the source in the bright state, and discovered a rapid variation in X-ray flux up to 30\% on an IDV timescale \citep{1994A&A...288..433B}. A continuous $\sim$ 100 ks observation in 0.1--100 keV by {\it BeppoSAX} showed that the blazar was in an intermediate intensity level compared to earlier observations, and between 0.1--10 keV the spectrum is convex with energy index gradually steepening from 1.1 to 1.6 keV \citep{1998A&A...333L...5G}.  In a detailed cross-correlation and power-density spectrum analysis of the X-ray light curves (LCs) observed with {\it BeppoSAX} in 1996 and 1997, no large-amplitude variability on hour-like timescale was noticed, and LCs in different X-ray bands were strongly correlated \citep{1999ApJ...527..719Z}. The blazar was monitored with the {\it ASCA} satellite in 1994 May when the 2--10 keV flux changed by a factor of 2 on a timescale of 30 ks, and the spectral evolution was tracked by a clockwise loop in the flux versus the photon-index plot \citep{2000ApJ...528..243K}. Extensive X-ray flux and spectral variability studies of PKS 2155--304 with {\it XMM-Newton} reported large-amplitude flux, cross-correlation, and spectral variability on diverse timescales \citep[e.g.,][]{2006ApJ...637..699Z,2008ApJ...682..789Z,Gaur2010,2017ApJ...850..209G,2014MNRAS.444.3647B,2016NewA...44...21B}. Recently, IDV flux and spectral variability of PKS 2155$-$304 was studied using {\it NuStar} data in the energy range of 3--79 keV \citep{Pandey2017}. Multi-wavelength SEDs of PKS 2155$-$304 at various epochs of observations have been fit with with  standard models, such as the synchrotron/Compton models, with possible contributions from accretion disks model and gravitational lensing etc. \citep[for a review see][]{2020Galax...8...64G}.  Because the synchrotron peak of this HBL is in the UV-EUV region, its X-ray emission, at least out to 10 keV, is dominated by the falling part of the synchrotron hump \citep[e.g.][]{2009ApJ...696L.150A,2016ApJ...831..142M}.   {\it XMM-Newton} data taken between 2009 and 2014 indicate that on occasions there is some flattening of the SED above 6 keV which may be explained by IC of synchrotron photons from the slower outer sheath of the jet by the faster moving electrons in the spine \citep{2017ApJ...850..209G}; the
 first evidence of  IC X-ray emission below 10 keV from this source seems to be in one of the {\it XMM-Newton} observations taken in 2006 \citep{2008ApJ...682..789Z}. \\
\\
Occasionally, detections of quasi-periodic oscillations (QPOs) were claimed in several blazars \citep[e.g.,][]{2008ApJ...679..182E,2009ApJ...690..216G,2019MNRAS.484.5785G,2009A&A...506L..17L,2014JApA...35..307G,2018Galax...6....1G,2016ApJ...832...47B,2019MNRAS.487.3990B} and several other AGNs \citep[e.g.,][]{2008Natur.455..369G,2014JApA...35..307G,2018Galax...6....1G,2018A&A...616L...6G} on diverse timescales and  in different EM bands. PKS 2155$-$304 is one of the AGNs, more specifically blazars, which has possibly the maximum number of claimed QPO detections on diverse timescales on different times and in different EM bands.  Early, but weak, evidence for UV and optical QPOs of a period of $\sim$ 0.7 days was found in \citet{1993ApJ...411..614U}. In an inhomogeneous data set in UBVRI bands collected for $\sim$ 17 years, some evidence of QPOs with periods of 4 and 7 years was reported \citep{2000A&A...355..880F}. More recently, \citet{2014ApJ...793L...1S} found a QPO with a period of $\sim$ 315 days by using VRIJHK photometric data of PKS 2155$-$304 taken from 2005--2012, which supported the findings of \citet{2014RAA....14..933Z} who used inhomogeneous data over $\sim$ 35 years collected from 25 different astronomical groups. There is some evidence of a QPO in optical polarization with a period of 13 minutes in PKS 2155$-$304, which is the only such claim in any AGN so far \citep{2016MNRAS.462L..80P}. In the LCs extracted in the 0.3--10 keV energy range using {\it XMM-Newton} during 24 pointed observations, one LC shows a reasonably strong QPO of a period of 4.6 hours \citep{2009A&A...506L..17L} and there was a hint of QPO with a period of 5.5$\pm$1.3 ks in another LC \citep{Gaur2010}. Recently, using extensive {\it Fermi-LAT} $\gamma-$ray data taken during 2008--2016, a QPO with a period of 1.74$\pm$0.13 years was claimed \citep{2017ApJ...835..260Z}. \\    
\\
Flux variations on IDV timescales have been reported in many blazars in different EM bands based on continuous monitoring for a few hours \citep[e.g.,][]{1989Natur.337..627M,2004MNRAS.348..176S,2006A&A...451..435M,2008AJ....135.1384G,Gaur2010,2015MNRAS.451.1356K,2018MNRAS.480.4873A,2019ApJ...884..125Z}. Most of these observations were not evenly sampled. There were only a few continuous and evenly sampled observations of blazars in X-ray and optical EM bands which lasted for more than a few days \citep[e.g.,][]{2001ApJ...563..569T,2013ApJ...766...16E,2019ApJ...884..125Z}. The most puzzling variations in blazars are those happening on the IDV timescales and they may be directly related to  activity in the close vicinity of the central supermassive black hole (SMBH). The level of variability strongly depends on the SMBH mass \citep[e.g,][]{2004ApJ...617..939M} and it can also help to constrain the size of the emitting region. 
Also, the occasional apparent detections of QPOs in blazars on IDV timescales are among the least understood phenomena in blazar research \citep[e.g.,][and references therein]{2014JApA...35..307G,2018Galax...6....1G}. \\
\\ 
With the above motivation, we have used blazar data taken from  X-ray missions {\it XMM-Newton, NuStar, Chandra} to study variability on a variety of timescales  \citep[e.g.,][]{Gaur2010,2014MNRAS.444.3647B,2015MNRAS.451.1356K,2016MNRAS.462.1508G,2018ApJ...859...49P,2018MNRAS.480.4873A}.  We have already analyzed the three lengthy {\it Suzaku} observations of the brightest northern hemisphere TeV blazar, Mrk 421 \citep{2019ApJ...884..125Z}. In the present study, we have taken all the archived {\it Suzaku} satellite observations of the brightest TeV blazar in the southern sky, PKS 2155$-$304, which were taken throughout nearly its entire mission lifetime (2005 -- 2015). These observations are very useful to study flux and spectral variations on IDV timescales, to search for QPOs on IDV timescales and to better characterize LTV flux and spectral variations. So, the data analyzed in this paper provides us with an excellent opportunity to increase our understanding of the X-ray multi-band flux and spectral behavior of one of the most interesting and peculiar blazars, PKS 2155$-$304. \\      
\\
The paper is structured as follows. In Section 2, we discuss the {\it Suzaku} public archival data of the blazar PKS 2155$-$304 and its reduction. Section 3 provides a brief summary of the analysis techniques used in the paper, which follow the approach of \citet{2019ApJ...884..125Z}. In Sections 4 and 5, we present our results and then discuss them. Our conclusions are summarized in Section 6.           

\begin{table*}
\centering
\caption{The {\it Suzaku} observations of PKS 2155--304.}
\label{tab:observation}
%\vspace{0.3cm}
\begin{tabular}{ccccccccc}\hline\hline
 ObsID     & Date       & MJD  &Elapse\tablenotemark{a} & GTI\tablenotemark{b} & Win.\tablenotemark{c}   & XI0 Rate \tablenotemark{d} &  Src Rate \tablenotemark{e}  &  Bkg Rate \tablenotemark{f}\\ 
           &            &      & (ks)                   & (ks)                  &                         & (count s$^{-1}$)           &  (count s$^{-1}$)            & (count s$^{-1}$)           \\\hline
700012010 & 2005-11-30 & 53704 &  136.8  &  63.9 & 1/8  &  6.93    &   ~8.21   &  $1.52\times10^{-2}$   \\
101006010 & 2006-05-01 & 53856 &  ~81.0  &  38.6 & 1/4  &  3.20    &   ~4.45   &  $2.34\times10^{-2}$   \\
102020010 & 2007-04-22 & 54212 &  ~24.0  &  12.0 & 1/4  &  6.51    &   ~9.23   &  $3.55\times10^{-2}$   \\
103011010 & 2008-05-12 & 54598 &  ~43.7  &  23.1 & 1/4  &  7.39    &   11.31   &  $4.27\times10^{-2}$   \\
104004010 & 2009-05-27 & 54978 &  155.6  &  62.4 & 1/4  &  4.78    &   ~7.28   &  $3.01\times10^{-2}$   \\
105001010 & 2010-04-27 & 55313 &  157.5  &  63.5 & 1/4  &  1.88    &   ~2.88   &  $1.96\times10^{-2}$   \\
106011010 & 2011-04-26 & 55677 &  127.9  &  60.6 & 1/4  &  3.82    &   ~6.15   &  $2.61\times10^{-2}$  \\
107010010 & 2012-04-27 & 56044 &  141.0  &  61.9 & 1/4  &  0.86    &   ~1.87   &  $2.11\times10^{-2}$  \\
107009010 & 2012-10-30 & 56230 &  ~53.3  &  21.4 & 1/4  &  2.58    &   ~3.38   &  $2.07\times10^{-2}$  \\
108010010 & 2013-04-23 & 56405 &  129.4  &  53.4 & 1/4  &  1.85    &   ~2.75   &  $1.95\times10^{-2}$  \\
108009010 & 2013-10-30 & 56595 &  ~42.7  &  23.8 & 1/4  &  1.01    &   ~1.21   &  $1.35\times10^{-2}$  \\
109011010 & 2014-04-24 & 56771 &  123.0  &  64.0 & 1/4  &  2.31    &   ~3.16   &  $1.82\times10^{-2}$  \\
109010010 & 2014-10-30 & 56960 &  ~59.1  &  20.4 & 1/4  &  1.69    &   ~2.28   &  $1.55\times10^{-2}$  \\\hline
\end{tabular}\\
\noindent
\tablenotemark{a}{Total elapsed time of the observation.}\\  
\tablenotemark{b}{Total clean GTI of the observation.}\\
\tablenotemark{c}{XIS window mode (1/4 or 1/8).}\\
\tablenotemark{d}{Total XIS 0 CCD count rate across the whole energy band.}\\
\tablenotemark{e}{Source count rate in 0.8$-$8 keV energy range of CCDs XIS 0+XIS 3.}\\
\tablenotemark{f}{Background count rate in 0.8$-$8 keV energy range of CCDs XIS 0+XIS 3.}\\
\end{table*}

\section{{\it Suzaku} Archival Data Reduction}
\label{sec:observation}

\noindent
PKS 2155$-$304 was observed by the Japanese X-ray observatory {\it Suzaku} 13 times during the course of nine years from 2005 to 2014 (Table \ref{tab:observation}). This is by far the greatest number of observations it made of any blazar: no other blazar had more than five. {\it Suzaku} was a near-earth satellite with its orbit apogee of 568 km and orbital period of 5752 seconds. It carried a soft X-ray Imaging Spectrometer \citep[XIS;][]{Koyama2007} with energy resolution of $\sim$ 100 eV, and hard X-ray detectors \citep[HXD;][]{Takahashi2007} which achieved an energy coverage up to $>$100 keV. The XIS contained four CCDs named XIS 0 to 3. Among them, XIS 0, 2, and 3 are front-illuminated (FI) CCDs, which are more accurately calibrated than the back-illuminated (BI) CCD XIS 1. Only two observations (ID 700012010 and 101006010) have data from XIS 2 because it stopped working on 9 November 2006. Since PKS 2155$-$304 is always at least moderately bright, for uniformity, we only utilized data from XISs 0 and 3 for all the observations  in this study. There were no promising detections by the HXD because the signals are comparable to the instrument background. \\

\begin{figure}
\begin{center}
\epsscale{1.0}
\includegraphics[width=8.65cm,angle=0]{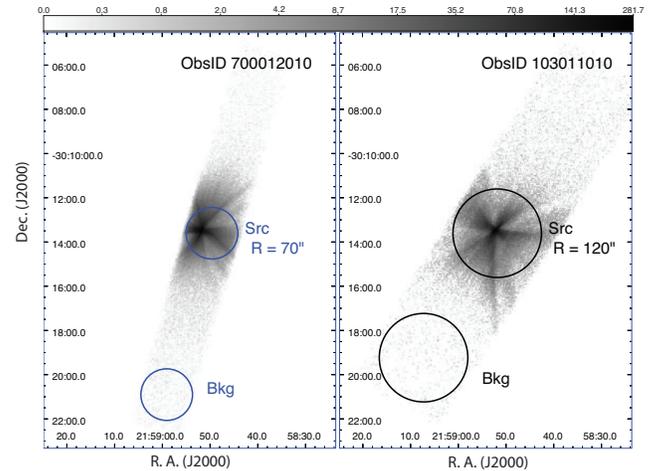}
\figcaption{{\it Suzaku} observation images of XIS 0 with 1/8 (ObsID 700012010; left panel) and 1/4 window modes (ObsID 103011010 taken as example; right panel). The figures are in logarithmic scales with reversed gray colors. The source region of ObsID 700012010 is a circle with radius of $70^{\prime\prime}$ centered at R.A.= $-21^{h}58^{m}49.59^{s}$ and Dec. = $-30^{\circ}13^{\prime}36.48^{\prime\prime}$, and the source regions of ObsID 103011010 and the other 11 observations are circles with radii of $120^{\prime\prime}$ centered at R.A.= $-21^{h}58^{m}51.84^{s}$ and Dec.= $-30^{\circ}13^{\prime}36.48^{\prime\prime}$. The background regions are taken from the edge of the CCD with the same sizes of their source regions.}
\label{fig:obs}
\end{center}
\end{figure}

\noindent
We utilized the cleaned XIS events in the archive, which were processed by the HEADAS \citep[v6.18][]{Blackburn1995} software. For the six observations before 2011, the processing pipeline version is PROCVER 3.0.22.43 using the XIS calibration database of version 20151005, while for the seven observations from 2011 the PROCVER version is 3.0.22.44 using the XIS calibration version 20160607.  The GRADE of the events is 0, 2, 3, 4, or 6. Because {\it Suzaku} is near the Earth, the good time intervals (GTIs) were first extracted from when the telescope's sight was not blocked by the Earth. Moreover, the initiation of time intervals after exiting from the South Atlantic Anomaly (SAA) were set to be longer than 436 s, and so the associated high background intervals also were excluded. As a result, the GTI of each observation is usually approximately half of the total elapsed time (Table \ref{tab:observation}).  \\
\\
The observations are mostly in 1/4 window mode with a time resolution of 2 seconds, while the brightest one, ObsID 700012010 is in 1/8 window mode with time resolution of 1 second. With such settings the pile-up effects were negligible \citep{Yamada2012}. We used circular regions to include most of the source counts for the study, and circles with the same radii near the edge of the CCDs to extract the background (see details in Figure 1). Count rates in source and background regions were then calculated.  As shown in Table \ref{tab:observation} the XIS background is $\lesssim$ 1\% of the source count, and was subtracted in the following analyses.\\
\\
We plot the background-subtracted light curves and hardness ratios through the elapsed times of the 13 observations in Figure \ref{fig:curve}. The total energy range is 0.8--8.0 keV, and is divided into soft (0.8--1.5 keV) and hard (1.5--8.0 keV) sub-bands to have comparable counts in each band. To show the GTIs, we plot the full LCs from 0.8 $-$ 8 keV with time binning of 128 seconds in the top panels of Figure \ref{fig:curve}. For other studies through the paper, the time binning was set to be 5752 seconds, which is exactly the orbital period of {\it Suzaku}, to most evenly sample the source in time and make the most homogenous GTI fraction in each bin. The range of the GTI fraction between bins is mainly caused by the interruption of the SAA, which obviously varies on timescales of an hour because of the spinning of the Earth. However, we consider any discrepancy arising from this to be negligible, because the source did not commonly show large intrinsic variation within one orbit of {\it Suzaku}. 
\begin{figure*}
    \centering
    \begin{tabular}{lcr}
\includegraphics[width=5.7cm,angle=0]{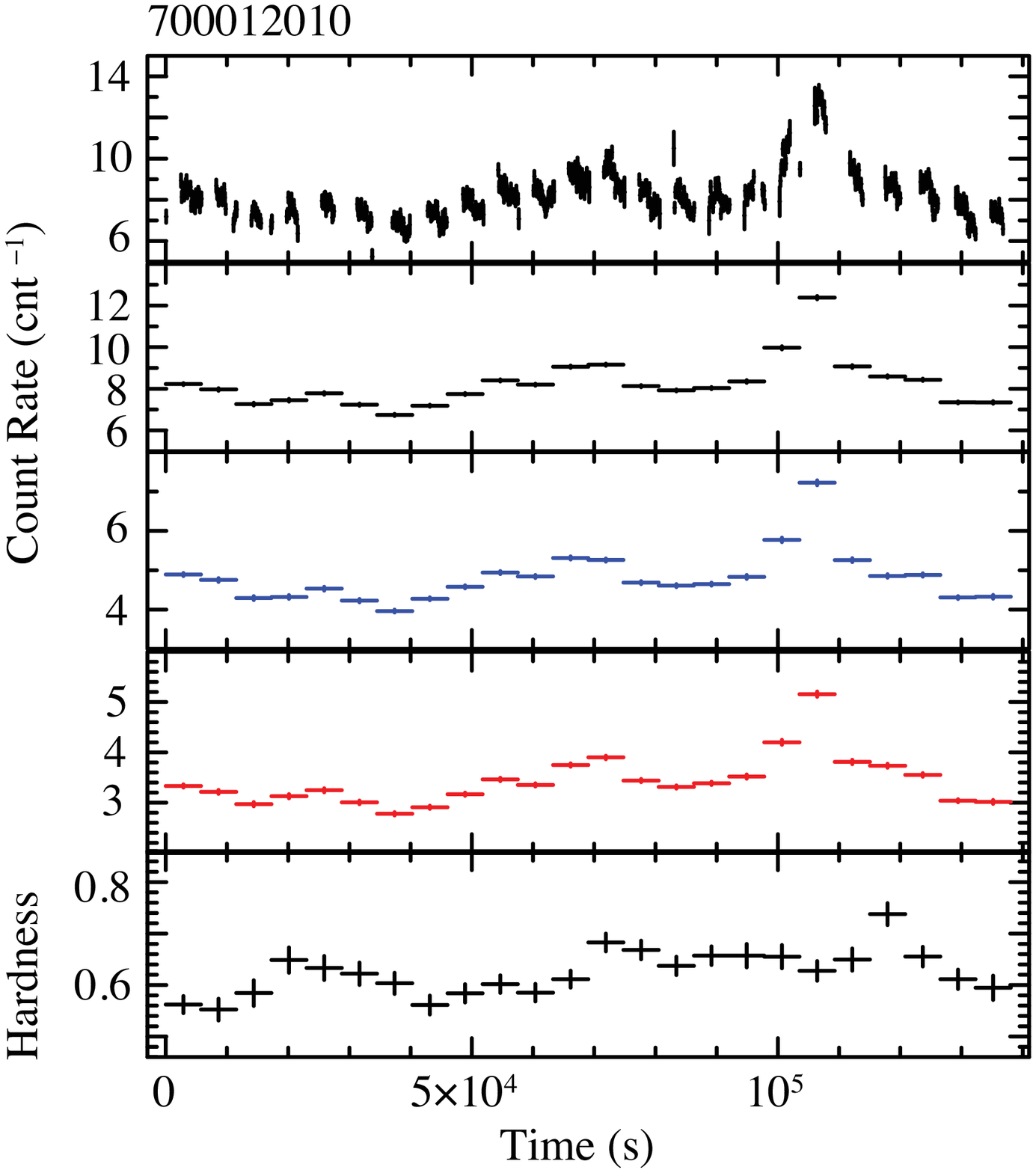}  &
\includegraphics[width=5.7cm,angle=0]{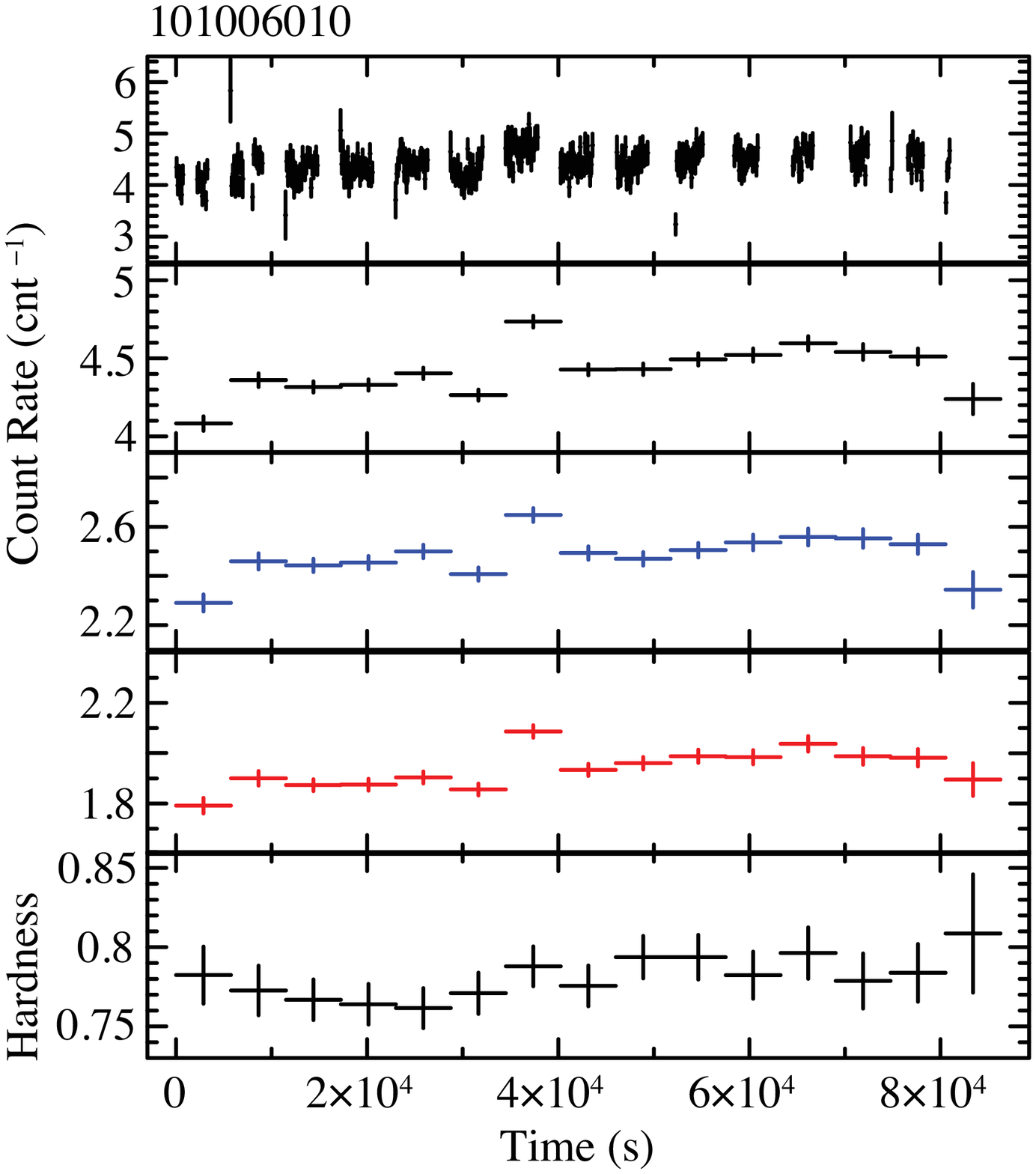}  &
\includegraphics[width=5.7cm,angle=0]{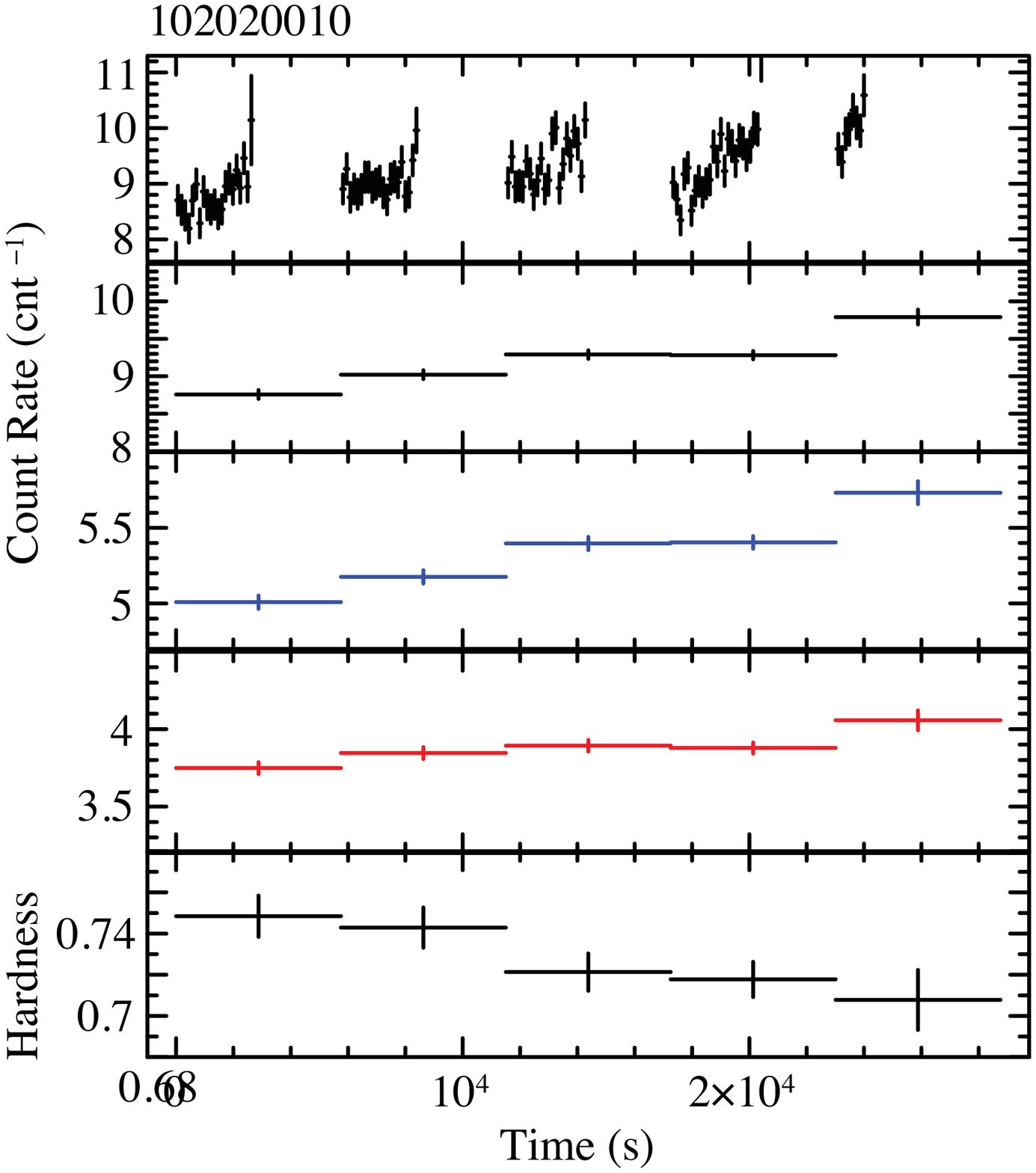}  \\
\includegraphics[width=5.7cm,angle=0]{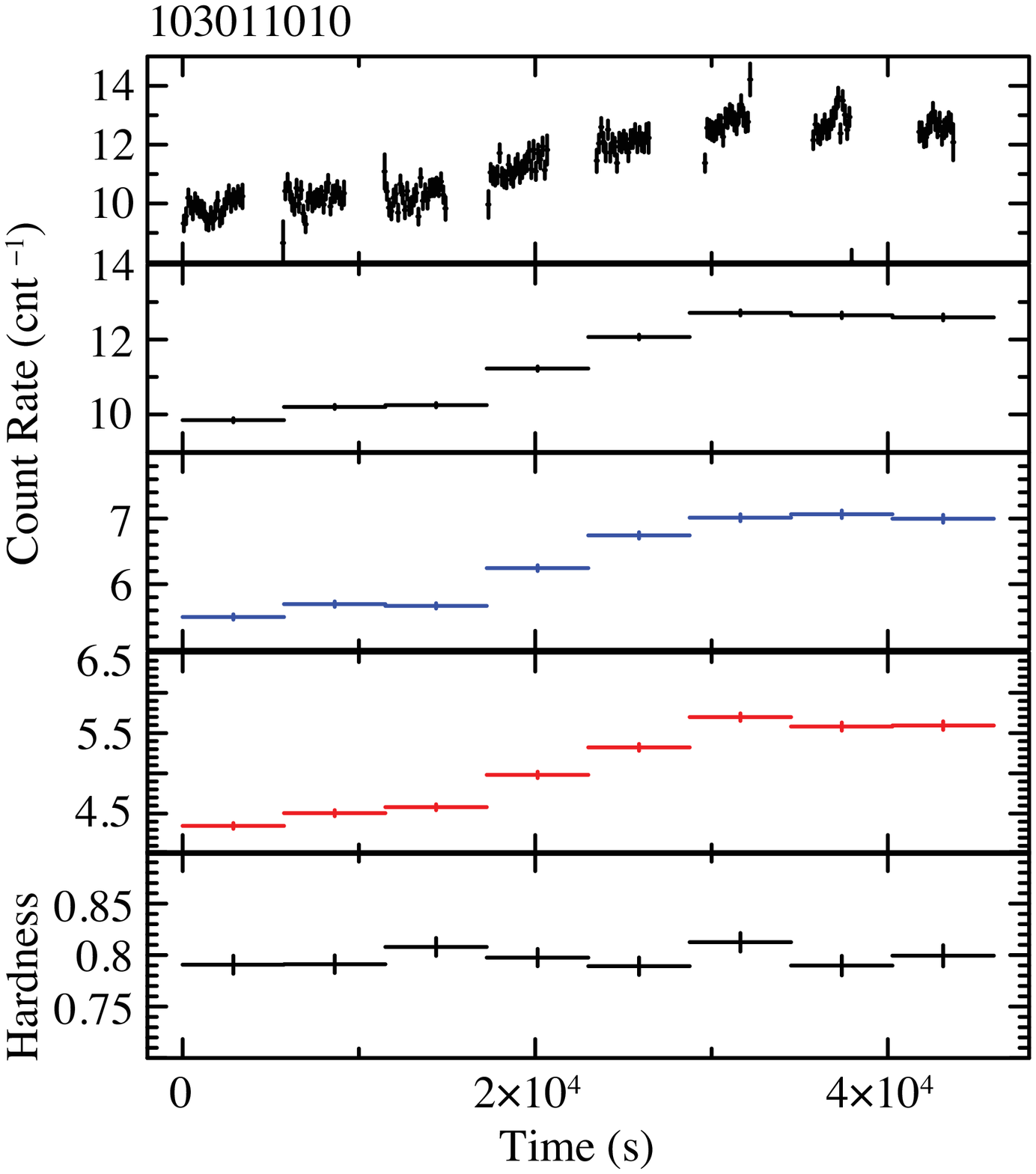}  &
\includegraphics[width=5.7cm,angle=0]{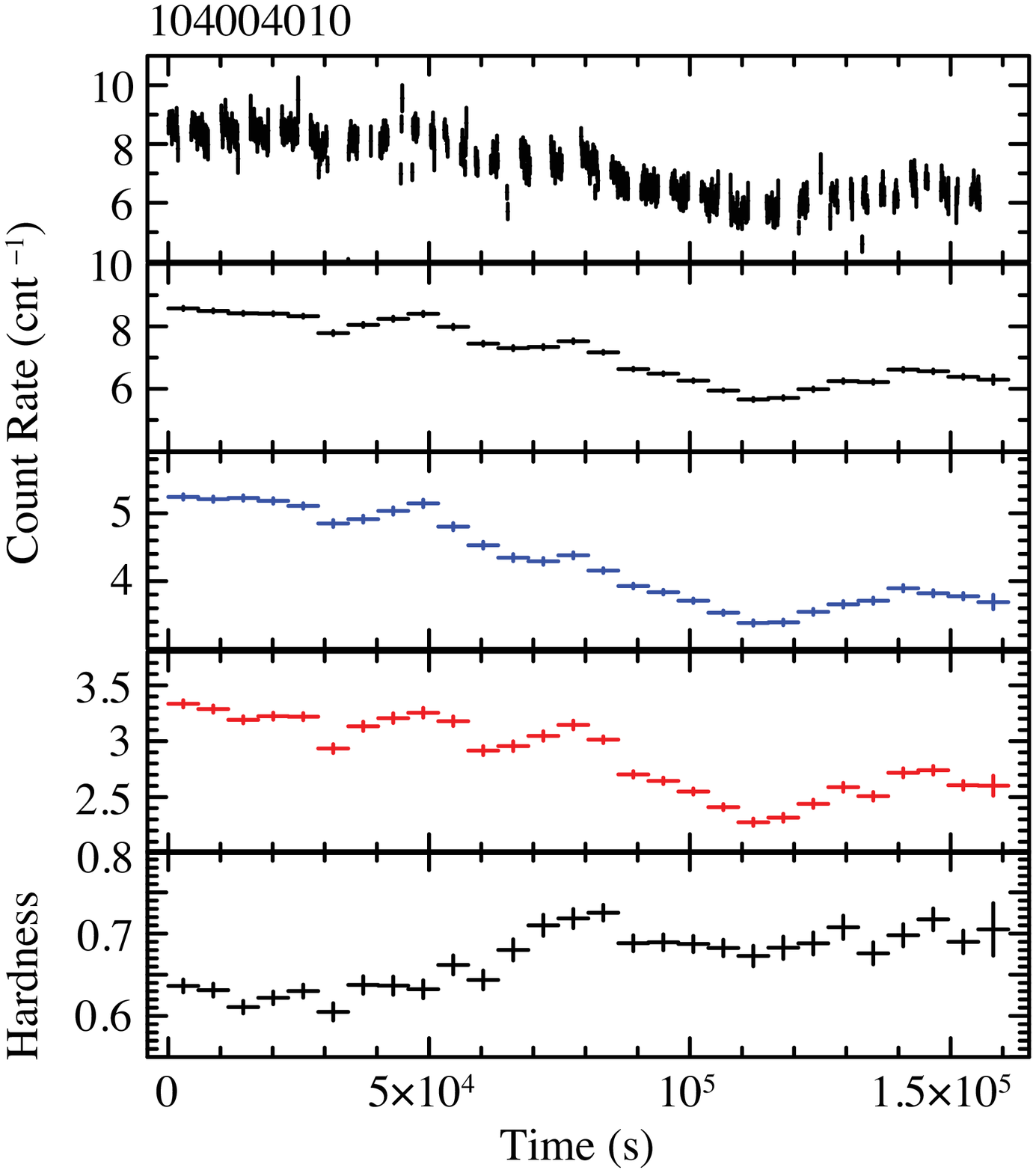}  &
\includegraphics[width=5.7cm,angle=0]{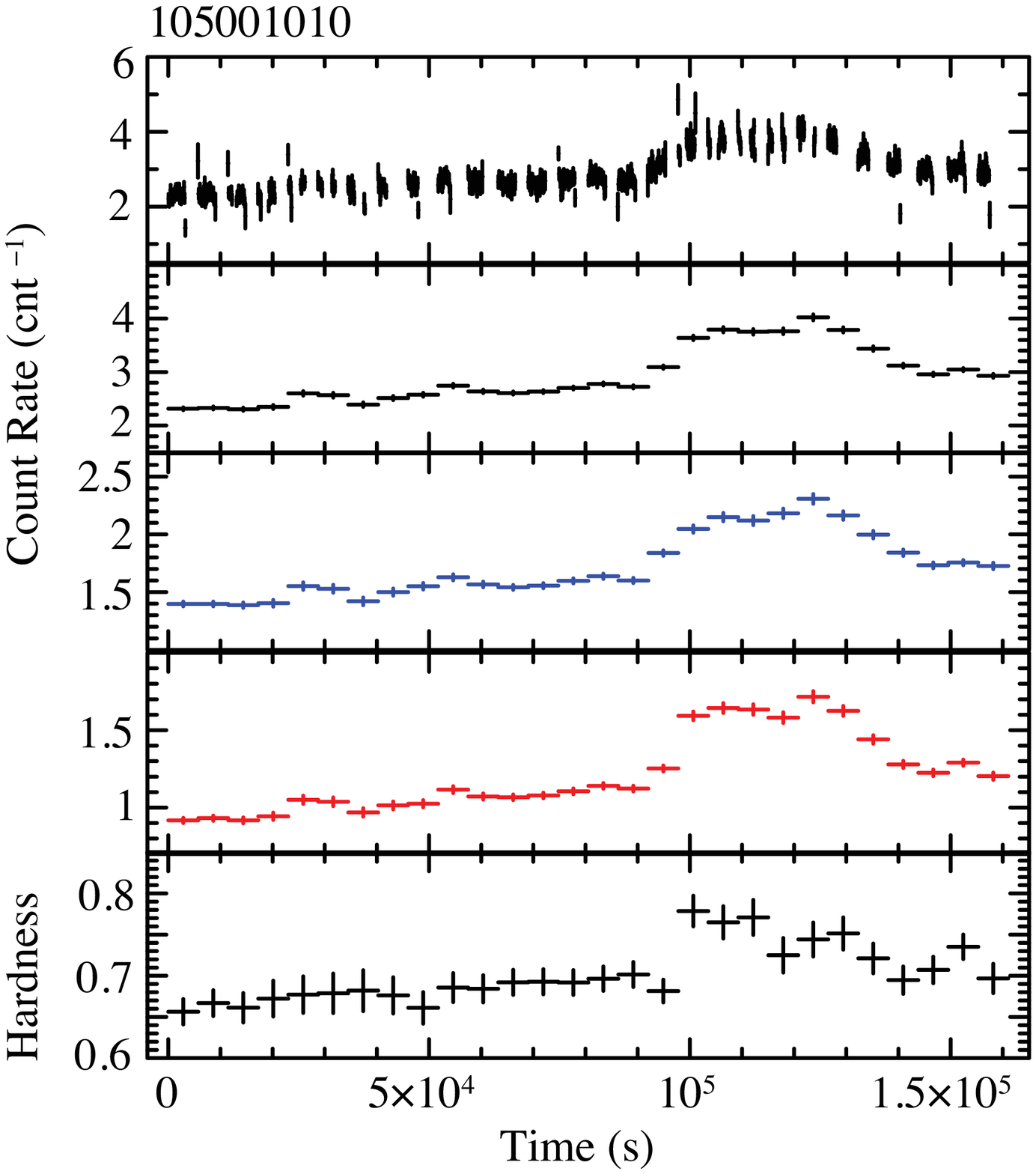}  \\
\includegraphics[width=5.7cm,angle=0]{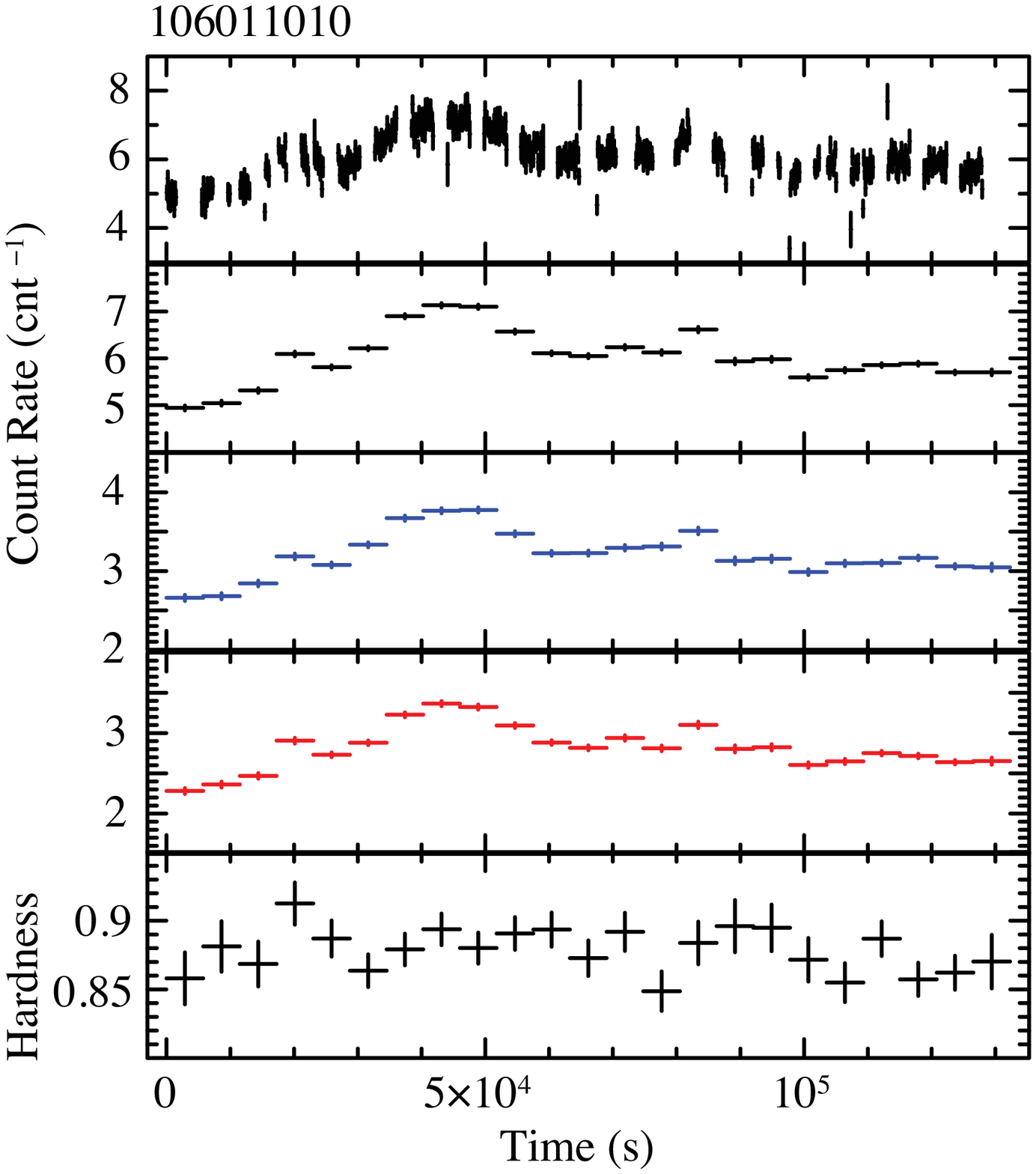}  &
\includegraphics[width=5.7cm,angle=0]{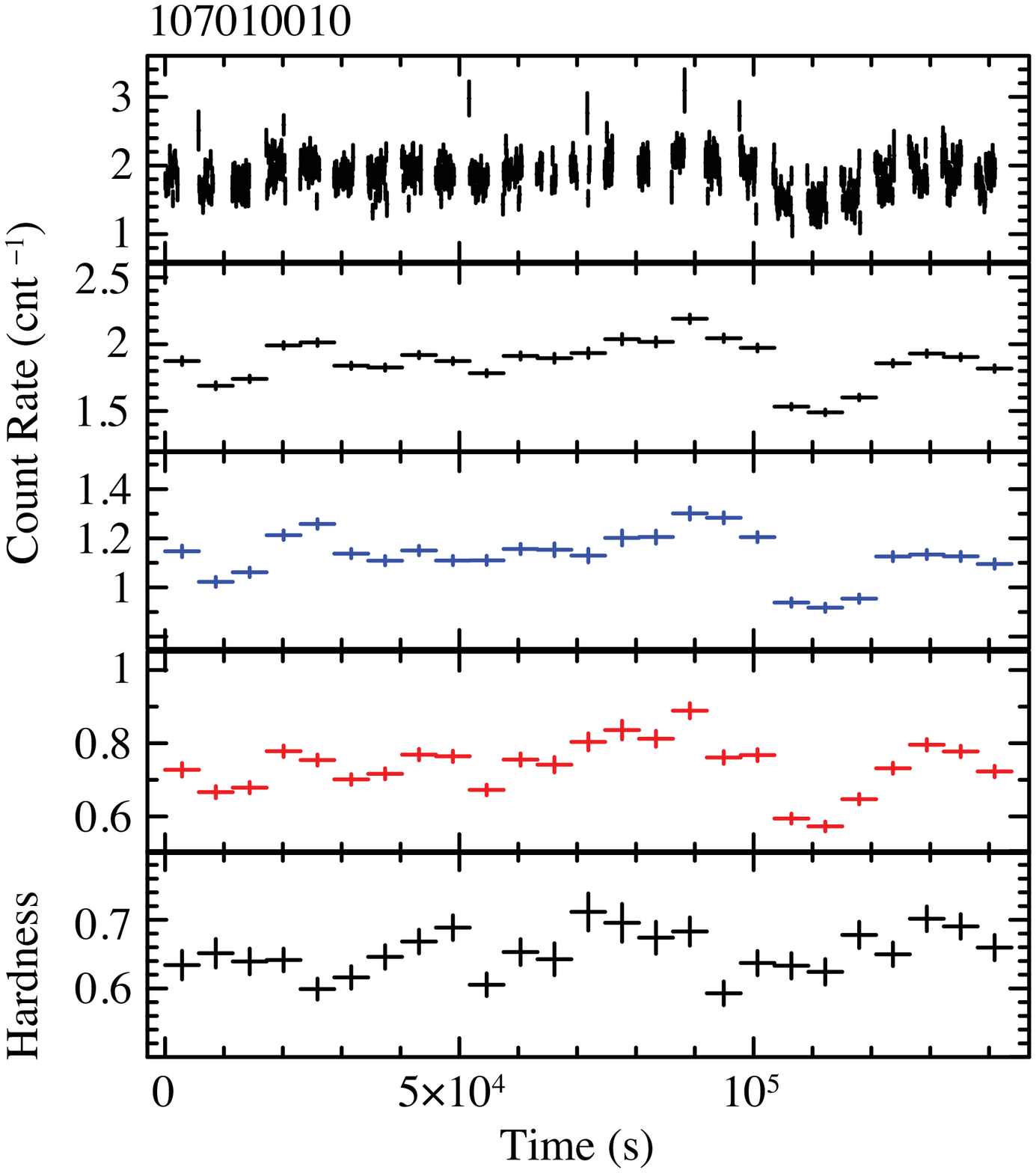}  &
\includegraphics[width=5.7cm,angle=0]{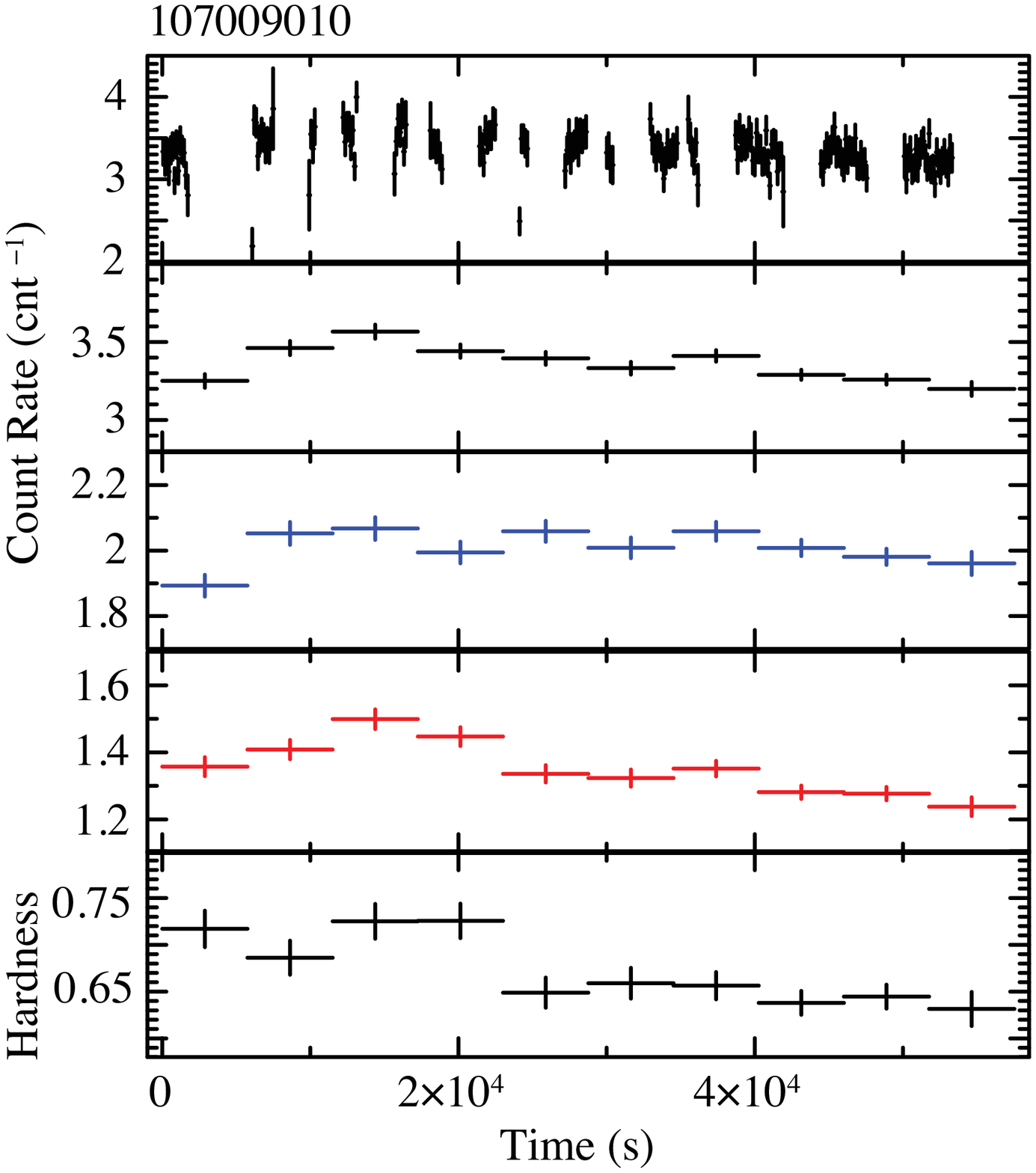}  \\
 \end{tabular}
\caption{Background-subtracted light curves and hardness ratios of the 13 observations, using the count rates extracted from the source region on XIS $0+3$.  From the top panel to the bottom of each subfigure: the full XIS (0.8$-$8 keV) LC in 128-second bins; the full XIS (0.8$-$8 keV) LC in 5752-second bins; the XIS soft (0.8--1.5 keV) LC; the XIS hard (1.5--8 keV) LC; the hardness ratios.}
\label{fig:curve}
\end{figure*}

\addtocounter{figure}{-1}
\begin{figure*}
    \centering
    \begin{tabular}{lcr}
\includegraphics[width=5.7cm,angle=0]{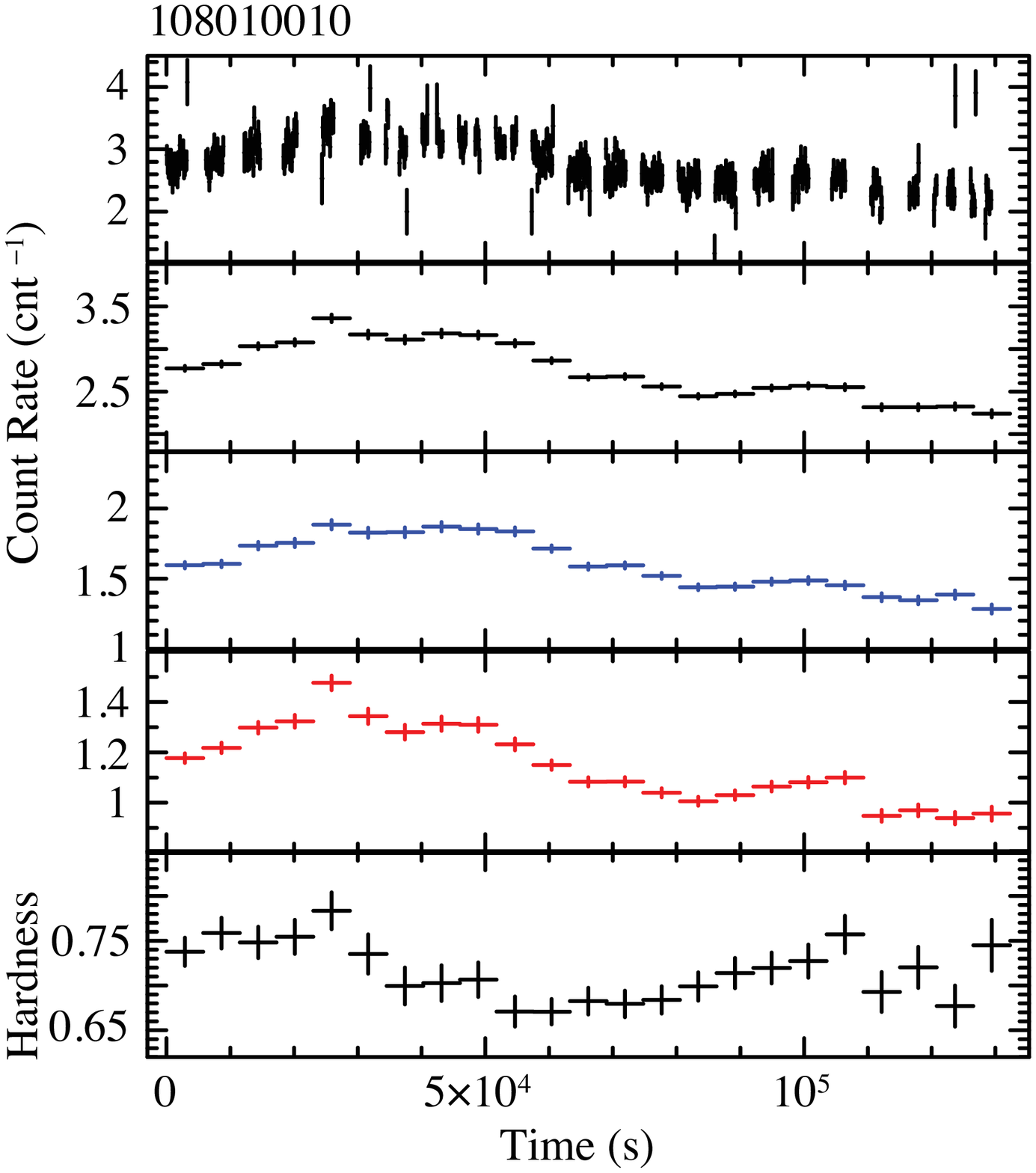}  &
\includegraphics[width=5.7cm,angle=0]{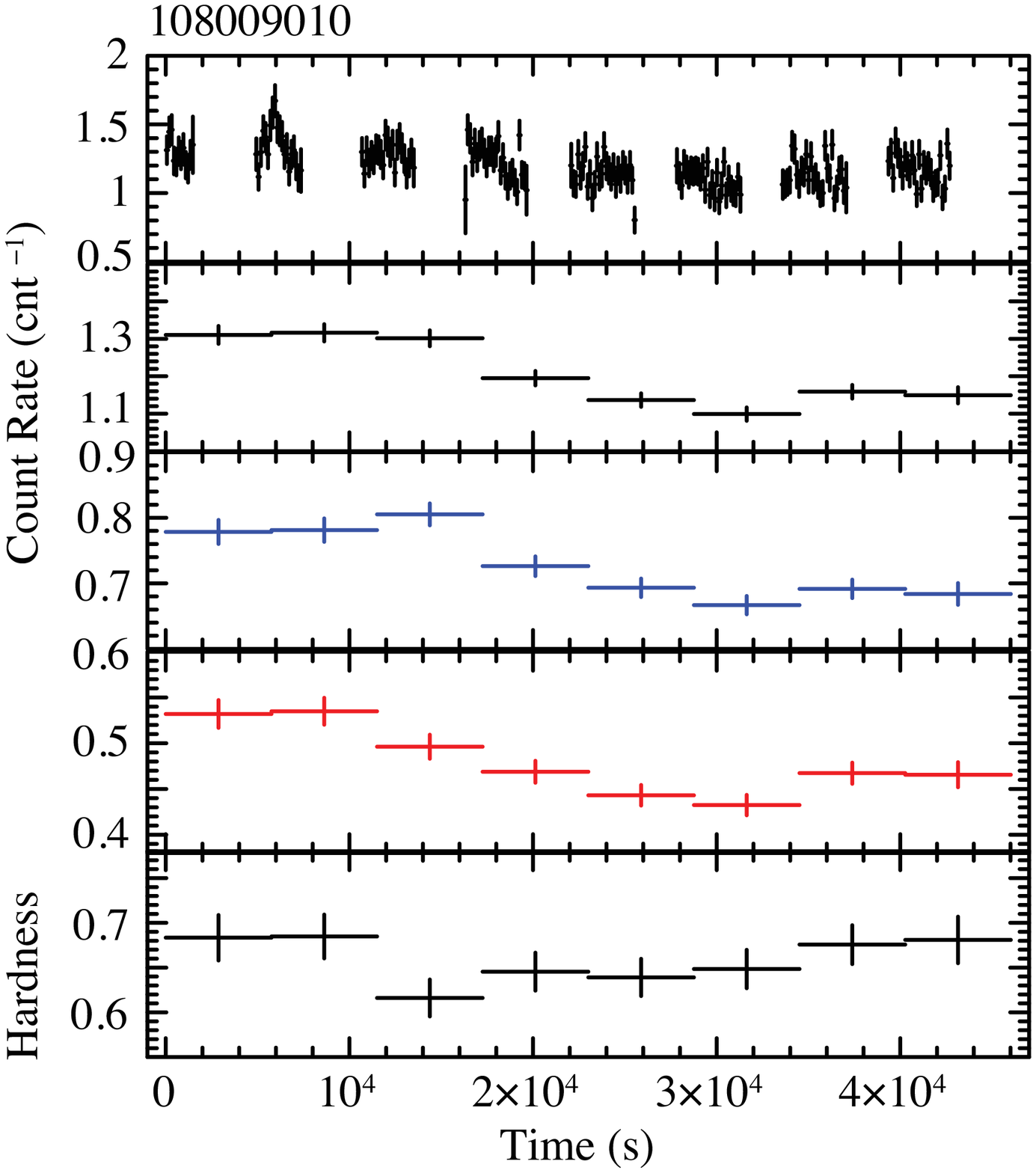}  &
\includegraphics[width=5.7cm,angle=0]{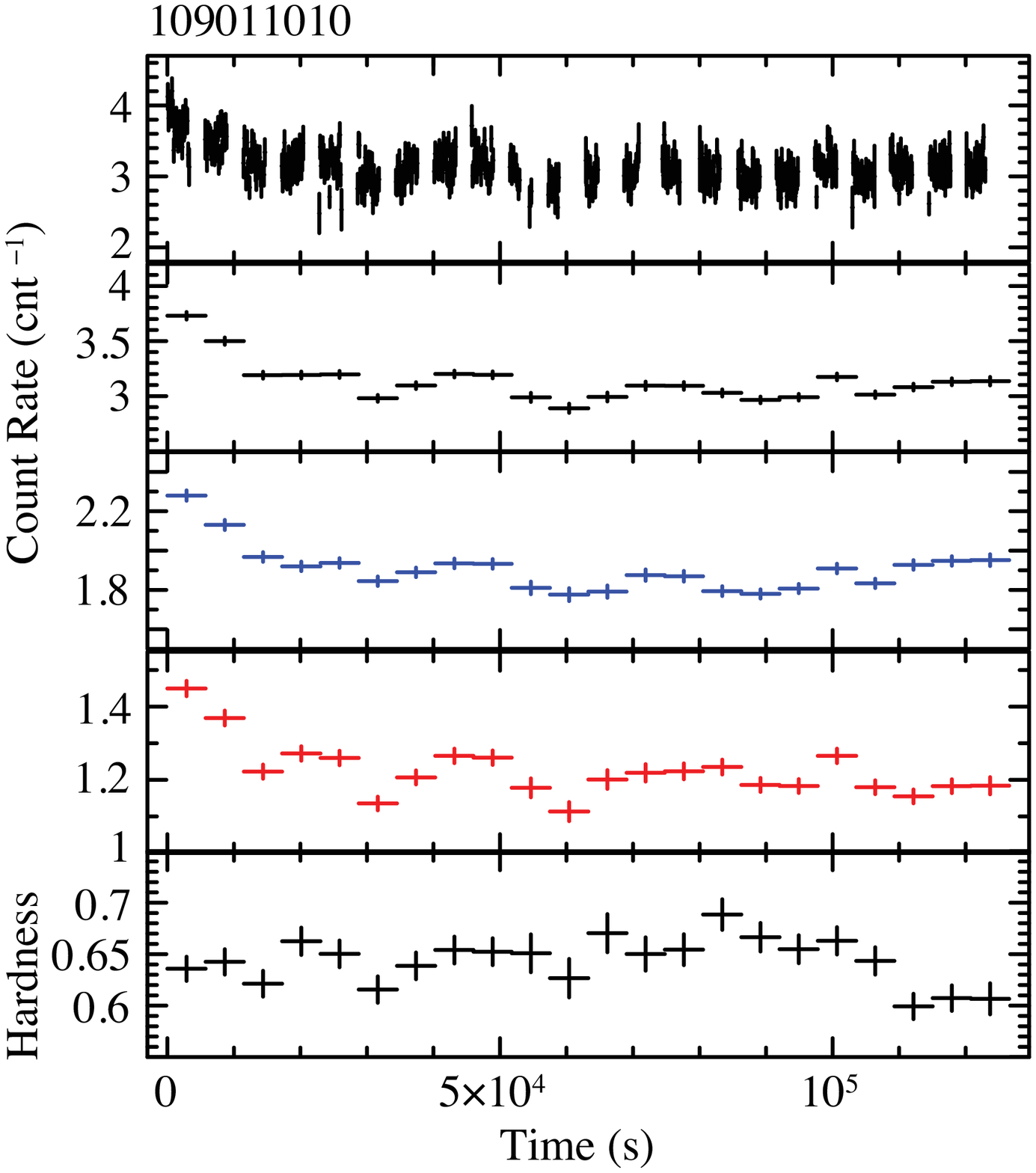}  \\
& \includegraphics[width=5.7cm,angle=0]{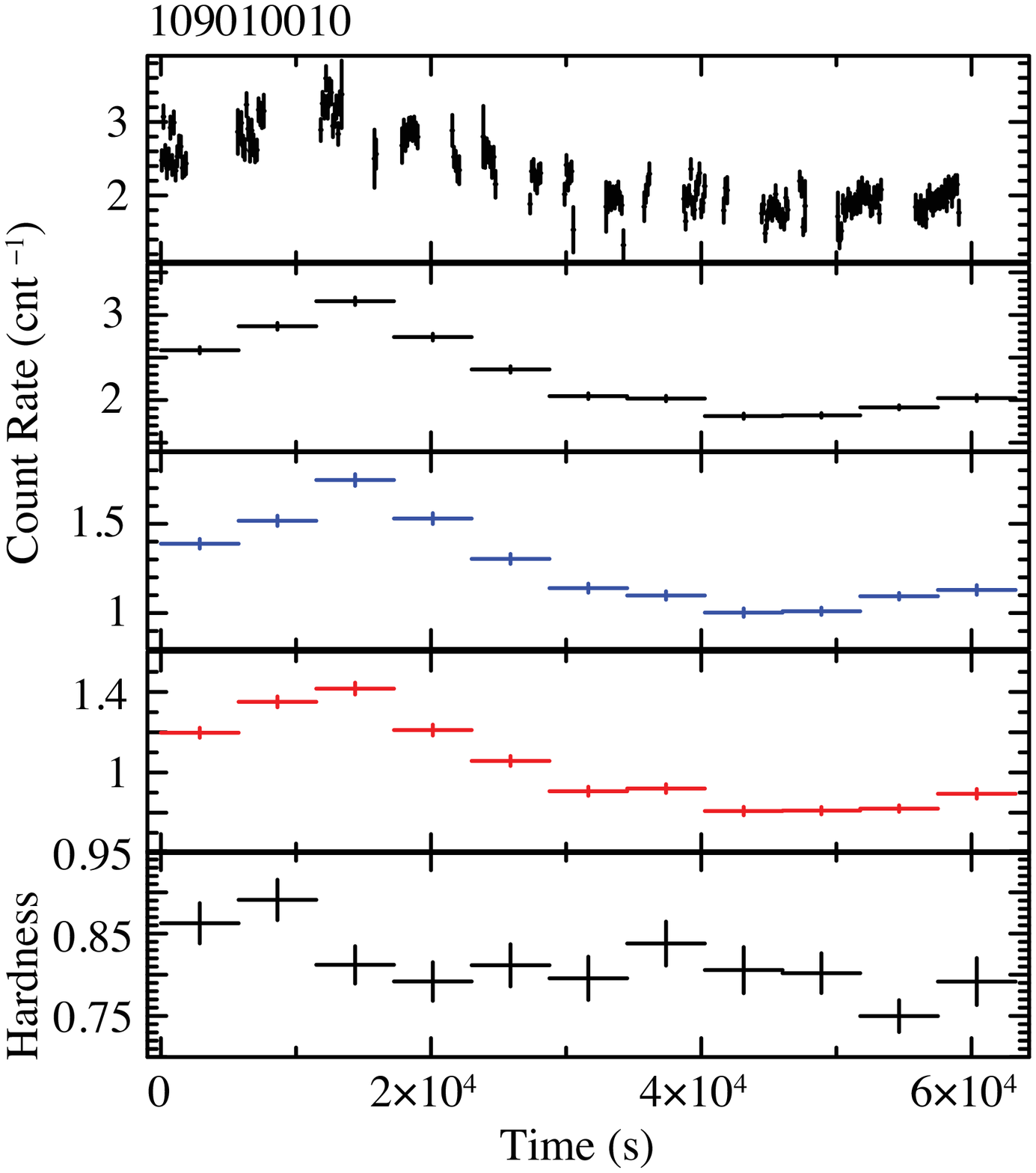}  &    \\
\end{tabular}
\caption{-- Continued.}
\end{figure*}

\begin{table*}
\centering
\caption{X-ray variability parameters.}
 \label{tab:var_par}
 \begin{tabular}{ccccc} \hline \hline
                      & \multicolumn{3}{c}{ $F_{var}(percent)$}    &  \multicolumn{1}{c} {$\tau_{var}$ (ks)} \\\cline{2-4} 
Observation           & Soft                & Hard              & Total                 &  Total  \\
ID                    & (0.8 $-$ 1.5 keV)   & (1.5 $-$ 8 keV)   & (0.8 $-$ 8 keV)       & (0.8 $-$ 8 keV)   \\\hline         
700012010             &  13.7 $\pm$ 0.2   &  14.6 $\pm$ 0.2 &  14.0 $\pm$ 0.1     & ~18.5 $\pm$ ~0.6      \\
101006010             &  ~3.3 $\pm$ 0.4   &  ~3.6 $\pm$ 0.4 &  ~3.5 $\pm$ 0.3     & ~58.9 $\pm$ ~6.1      \\
102020010             &  ~5.0 $\pm$ 0.4   &  ~2.7 $\pm$ 0.5 &  ~4.1 $\pm$ 0.3     & 108   $\pm$  24       \\
103011010             &  10.5 $\pm$ 0.3   &  10.7 $\pm$ 0.3 &  10.6 $\pm$ 0.2     & ~63.2 $\pm$ ~5.2      \\
104004010             &  15.2 $\pm$ 0.2   &  11.4 $\pm$ 0.3 &  13.6 $\pm$ 0.2     & ~73.7 $\pm$ ~9.0      \\
105001010             &  16.1 $\pm$ 0.3   &  20.8 $\pm$ 0.4 &  18.1 $\pm$ 0.3     & ~35.3 $\pm$ ~3.4      \\                  
106011010             &  ~9.1 $\pm$ 0.2   &  ~9.8 $\pm$ 0.3 &  ~9.4 $\pm$ 0.2     & ~42.0 $\pm$ ~3.9      \\                  
107010010             &  ~8.4 $\pm$ 0.4   &  ~9.5 $\pm$ 0.5 &  ~8.6 $\pm$ 0.3     & ~22.8 $\pm$ ~1.8      \\                  
107009010             &  ~2.2 $\pm$ 0.6   &  ~5.7 $\pm$ 0.6 &  ~3.2 $\pm$ 0.4     & ~91.7 $\pm$ 27      \\                 
108010010             &  11.8 $\pm$ 0.4   &  13.0 $\pm$ 0.4 &  12.2 $\pm$ 0.3     & ~59.2 $\pm$ 13      \\  
108009010             &  ~6.9 $\pm$ 0.8   &  ~8.6 $\pm$ 0.3 &  ~7.1 $\pm$ 0.6     & ~67.4 $\pm$ 18      \\                  
109011010             &  19.3 $\pm$ 0.6   &  21.5 $\pm$ 0.7 &  20.2 $\pm$ 0.5     & ~62.2 $\pm$ ~9.1      \\                  
109010010             &  ~6.0 $\pm$ 0.3   &  ~5.8 $\pm$ 0.4 &  ~5.8 $\pm$ 0.2     & ~38.6 $\pm$ ~5.6      \\\hline
\end{tabular} \\
\noindent
$F_{var}$= the fractional rms variability amplitude. \\
$\tau_{var}$= the flux variability timescale.\\
\end{table*}

\section{Analysis Techniques}
\label{sec:analysis}

\noindent
Blazars display strong and rapid flux variations on diverse timescales and the strength of flux variability is often quantified by the excess variance, $\sigma_{XS}$, and the fractional rms variability amplitude, $F_{var}$ \citep[e.g.,][]{2002ApJ...568..610E}.  The excess variance is a measure of a blazar's intrinsic variance, estimated by removing the variance arising from measurement errors from the total variance of the observed LC; $F_{var}$ is the square root of that excess variance normalized by the square of the mean value of the flux. 
We define and compute $F_{var}$ and its uncertainty following  \citet{2003MNRAS.345.1271V} as we did previously for Mrk 421 \citep{2019ApJ...884..125Z}.\\
\\
For estimating variability timescales, we adopted the method described in \citet{2018A&A...619A..93B} and \citet{2019ApJ...884..125Z} which we also briefly describe here. As explained in \citet[][]{1974ApJ...193...43B}, a weighted, or flux normalized,  variability timescale can be estimated by the following equation

\begin{equation}
   \tau_{var}=  \left | \frac{\Delta t}{ \Delta lnF} \right |,
  \end{equation}

\noindent
where $\Delta t$ is the time interval between variable flux, $F$, measurements  \citep[see also][]{2008ApJ...672...40H}. 
To compute the uncertainties in $\tau_{var}$, we adopted the standard error propagation method \citep[similar to Equation 3.14 given in][]{2003drea.book.....B}
to estimate the
uncertainties in $\tau_{var}$ as
\begin{equation}
\Delta \tau _{var}
\simeq \sqrt{\frac{F_{1}^{2} \Delta F_{2}^{2} +F_{2}^{2} \Delta F_{1}^{2}}{F_{1}^{2}F_{2}^{2}\left (  ln \left [ F_{1}/F_{2} \right ]\right )^{4}}}\ \Delta t ,
\end{equation}
\noindent
where $F_{1}$ and $F_{2}$ are the fluxes (in count sec$^{-1}$) used to estimate the shortest variability timescales, and $\Delta F_{1}$ and $ \Delta F_{2}$ are their corresponding uncertainties. \\
\\
The hardness ratio (HR) is the simplest way to characterize spectral variations of X-ray emission and is defined, as usual \citep[e.g.][]{2019ApJ...884..125Z} as HR = $H/S$, where $H$ and $S$ are the net count rates in the hard and soft energy bands, respectively.  The error, $\sigma_{\rm HR}$, is calculated from the individual errors in the bands as 
\begin{equation}
\sigma_{HR} = {\frac {2} {(H + S)^2}} \sqrt{(H^{2}\sigma^{2}{_S} + S^{2}\sigma^{2}{_H})} .
\end{equation}
\noindent
To study the spectral variability of the TeV blazar PKS 2155$-$304 with {\it Suzaku}, we divided the XIS instrument energy into 0.8 -- 1.5 keV (soft) and 1.5 -- 8.0 keV (hard) bands to use for our HR analysis. \\
\\
To search for correlations between LCs in two energy bands, we used a discrete correlation function (DCF) analysis, introduced by \citet{1988ApJ...333..646E}. A  description of the calculation of the DCF and the way in which we use it are given in detail in \citet{2019ApJ...884..125Z}. \\
\\
Most of the DCFs between soft and hard X-ray bands (shown in Figure \ref{fig:dcf}) are broad, so we fit them with a Gaussian function:
\begin{equation}
DCF(\tau)=a \times {\rm exp}\Bigl[\frac{-(\tau - m)^{2}}{2 \sigma^{2}}\Bigr] .
\end{equation}

\noindent
Here, $m$ is the time lag at which the DCF peaks, $a$ is that peak value of the DCF, and $\sigma$ is the width of the Gaussian function.
These calculated parameters are given in Table \ref{tab:dcffit}.\\
\\
The single most important technique used in searching for the nature of temporal flux variations, especially for any possible periodicities or QPOs, is the periodogram analysis producing a power spectral density (PSD). This method involves calculating the Fourier transform of the LC and then fitting the red noise variability of the PSD to the power-law. If the significance of any peak rising above the red-noise is $3\sigma$ (99.73\%) or more, one normally considers it to provide a significant QPO detection. We used the approach of \citet{2005A&A...431..391V}  to test for an QPOs in the PSDs.
Each PSD is calculated and its normalization $N$ is defined so that the units of the periodogram are (rms/mean)$^{2}$ Hz$^{-1}$ see Eqn.\ (2) of \citet{2003MNRAS.345.1271V}. We assume a power-law in the form of $P(f) = N~f^{\alpha}$ to fit the red-noise part of the spectrum $P(f)$ as a function of the frequency $f$,   where  $\alpha \leq 0$ is the power spectral index  \citep{1989ARA&A..27..517V}. The best fit line to the PSD is the red-noise level which is calculated as described by Eqns.\ (4--6) of \citet{2005A&A...431..391V}. The significance levels are obtained by adding an appropriate term to the power spectrum.

\section{Results}

\noindent
The 13 publicly archived {\it Suzaku} observations of the TeV blazar PKS 2155$-$304 we analyzed  have individual elapsed times between 24 ks and 157.5 ks and were taken over a $\sim$ 9 year time span: the earliest pointed observation of this blazar  was taken on 30 November 2005 and the last one on 30 October 2014. These observations provided us an excellent opportunity to study flux and spectral variability of the blazar PKS 2155--304 on intraday and long timescales.    

\subsection{Intraday Flux Variability} 

\noindent
We generated LCs of individual observation IDs with these 13 observations using three XIS energy data sets (soft,  hard, and total) and we have plotted them in different subfigures of Figure \ref{fig:curve}. From top to bottom, the panels of the subfigures represent: the full 0.8 -- 8 keV LC in 128-second bins, the entire 0.8 -- 8 keV LC in 5752-second bins, the soft 0.8 -- 1.5 keV LC in 5752-second bins, the hard 1.5 -- 8 keV LC in 5752-second bins, and the hardness ratios defined as hard over soft, respectively. On visual inspection of Figure \ref{fig:curve}, it is clear that the LCs of all 13 observation IDs in all the energy bands represented in the top four panels show evidence of IDVs. To quantify the IDV variability results we found the fractional rms variability amplitude and its error 
for all the LCs of these, and the results are reported in Table \ref{tab:var_par}. We also calculated the weighted, or flux normalized, variability timescales and their errors for the XIS total 0.8 -- 8 keV LCs 
and the results are also given in Table \ref{tab:var_par}.

\subsection{Intraday Spectral Variability}

\noindent
As noted above, X-ray spectral variations can be characterized by the hardness ratio (HR), even when spectroscopic data are unavailable. We have plotted the HR with respect to time in the bottom panel of all subfigures of Figure \ref{fig:curve}, for all these observations of PKS 2155$-$304. It is seen from Table \ref{tab:var_par} that in the hard band the variability amplitude is usually (10 of 13 cases) higher than in the soft band and from Figure \ref{fig:curve} that the HR often follows nearly the same pattern as the overall LC.  So these HR versus time plots in Figure \ref{fig:curve} show that in general this blazar becomes harder when brighter and softer when dimmer.  However, it does occasionally show a stable HR and/or a slightly softer when brighter trend. 

\subsection{Intraday Cross Correlated Variability}
\begin{figure*}
    \centering
    \begin{tabular}{lcr}
\includegraphics[width=4.5cm,angle=0]{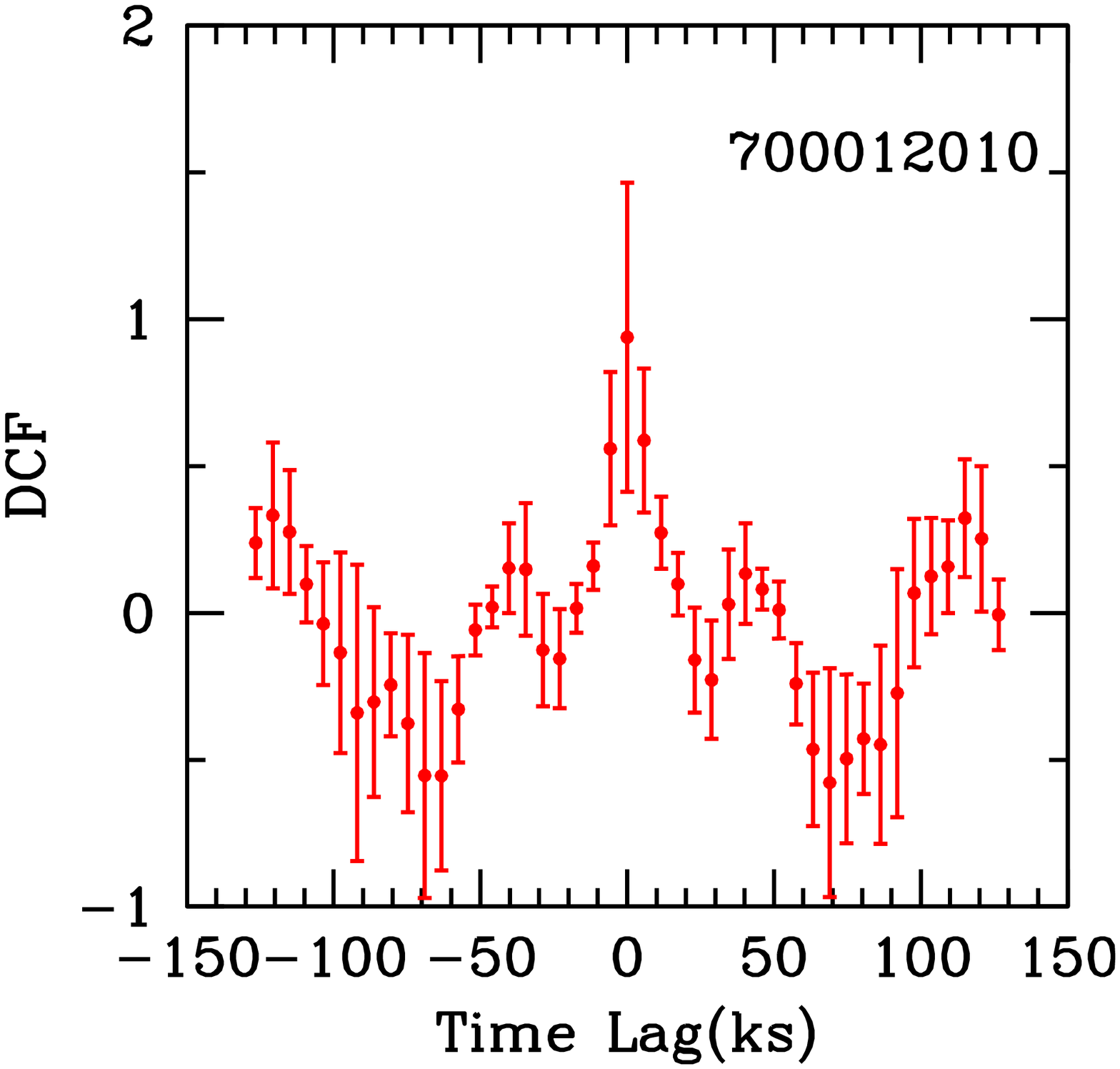} &
\hspace*{0.5cm} \includegraphics[width=4.5cm,angle=0]{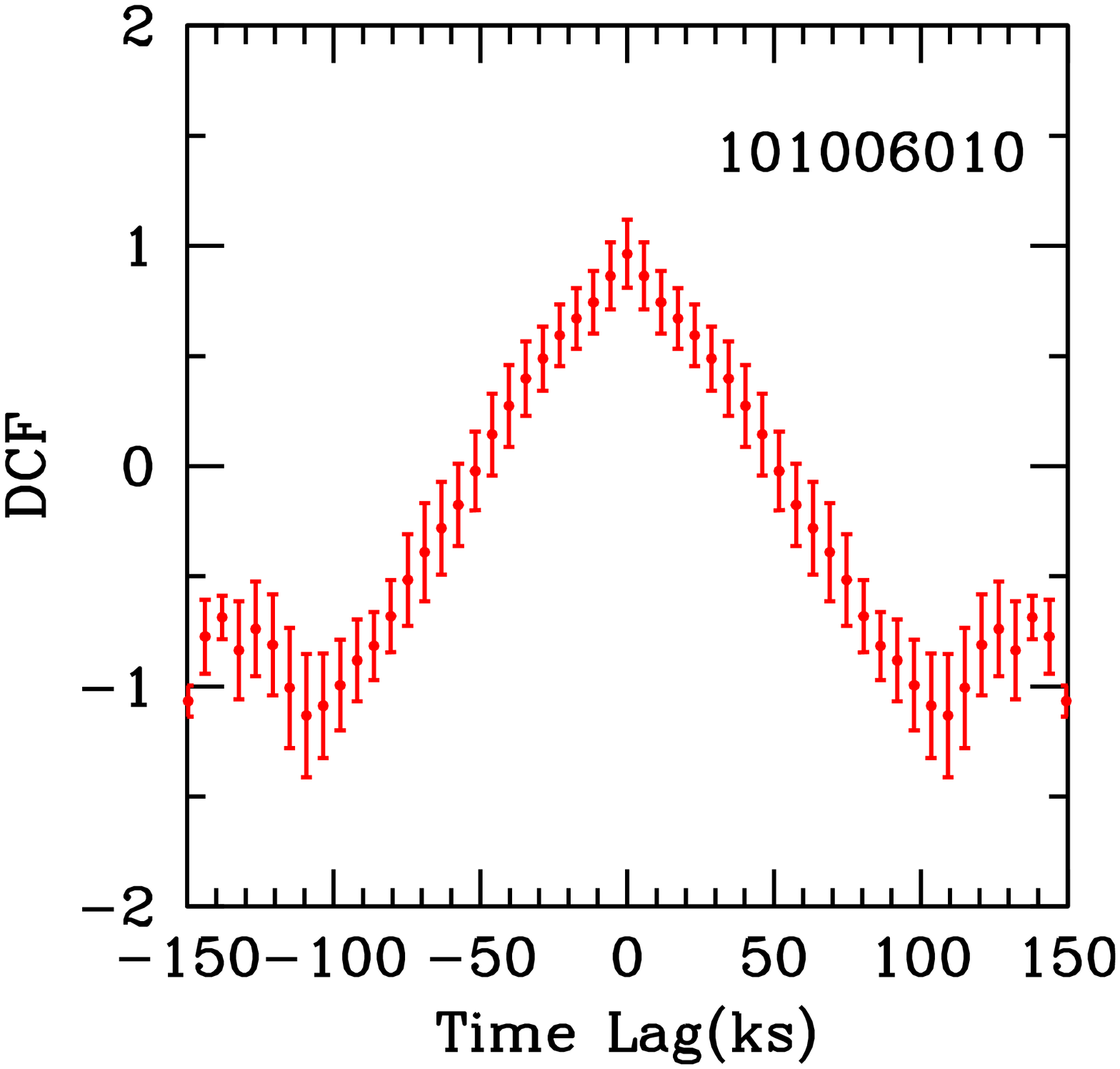} &
\hspace*{0.5cm} \includegraphics[width=4.5cm,angle=0]{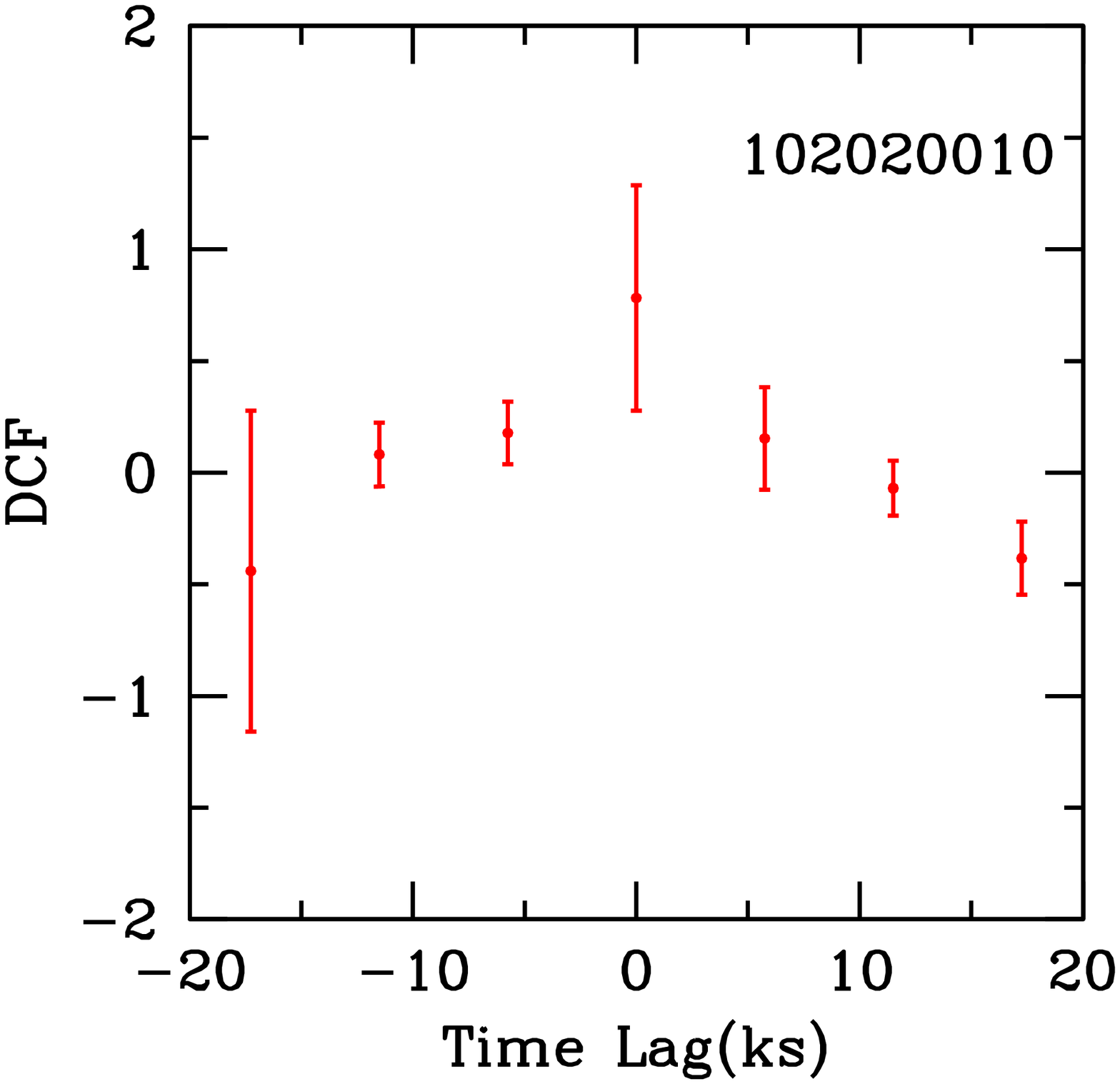} \\ 
\includegraphics[width=4.5cm,angle=0]{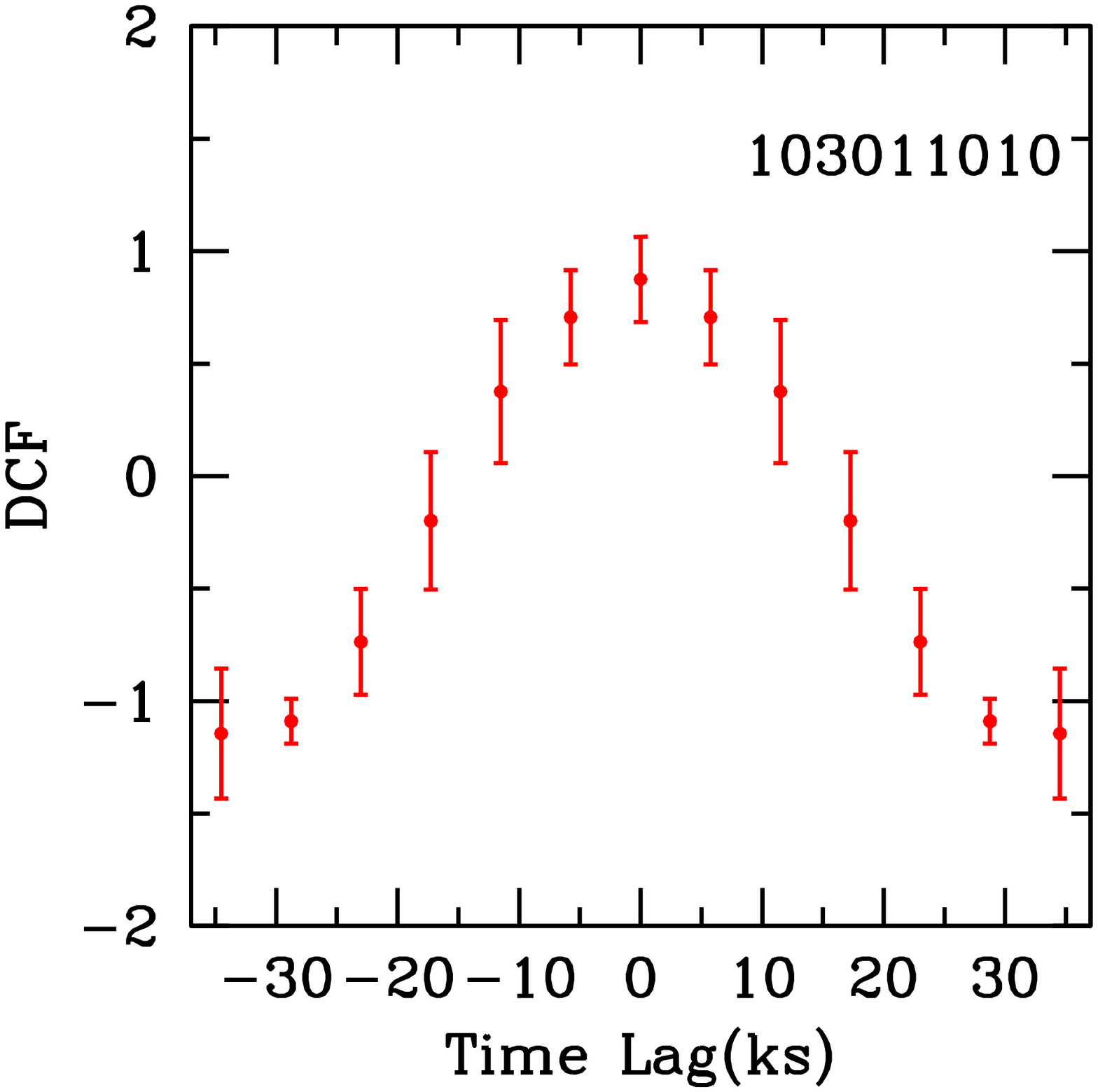} &
\hspace*{0.5cm} \includegraphics[width=4.5cm,angle=0]{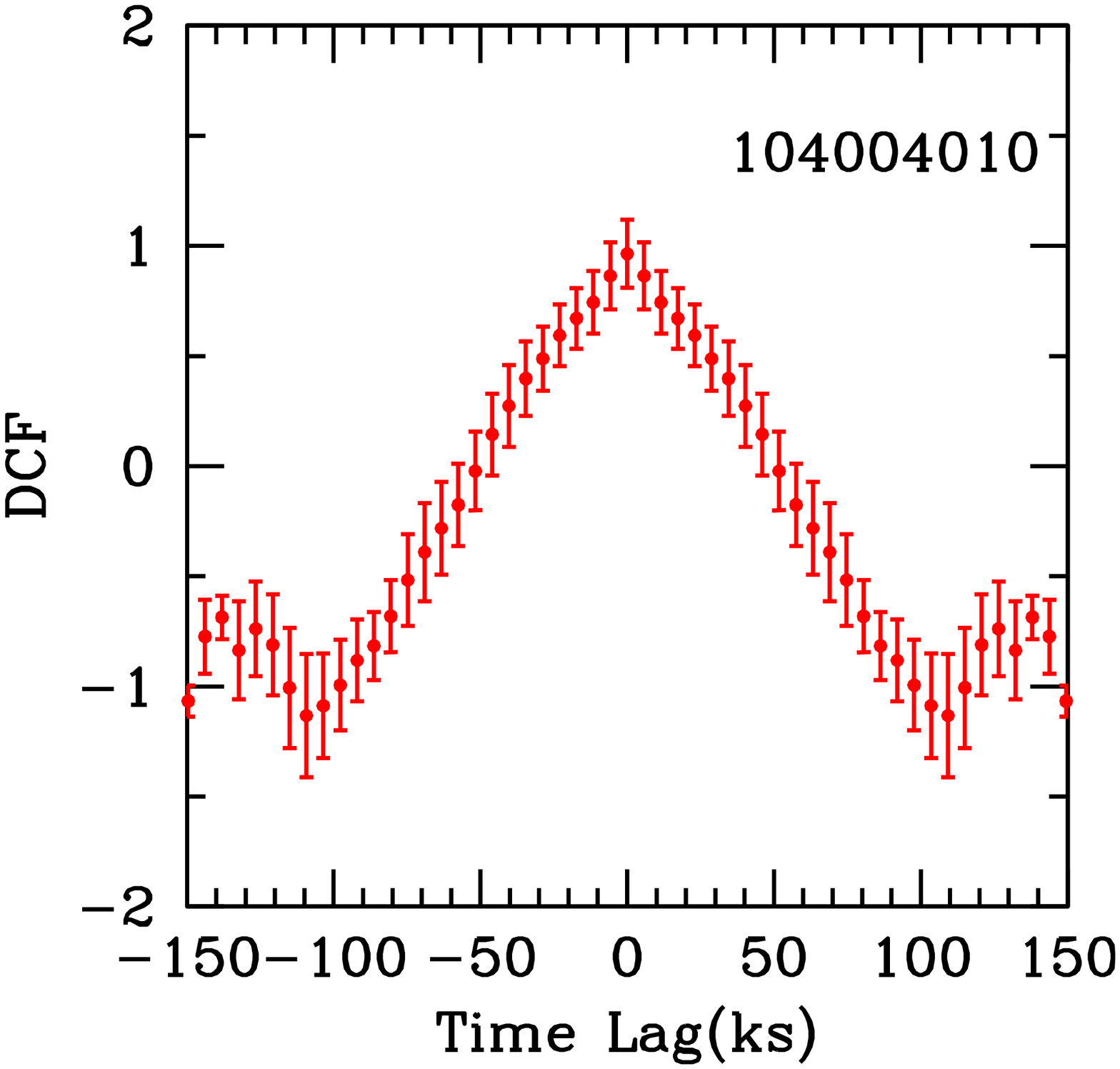} &
\hspace*{0.5cm} \includegraphics[width=4.5cm,angle=0]{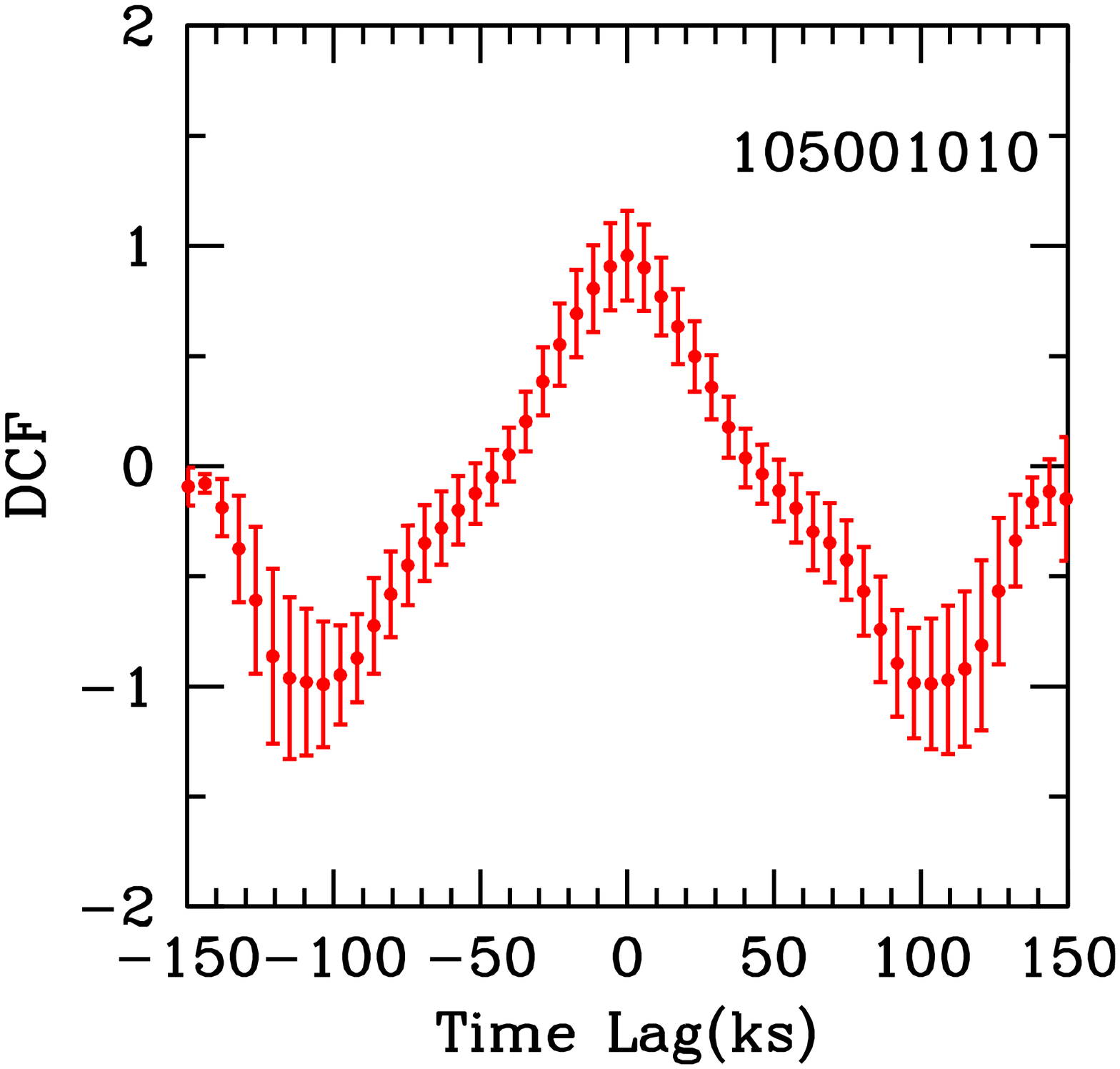} \\
\includegraphics[width=4.5cm,angle=0]{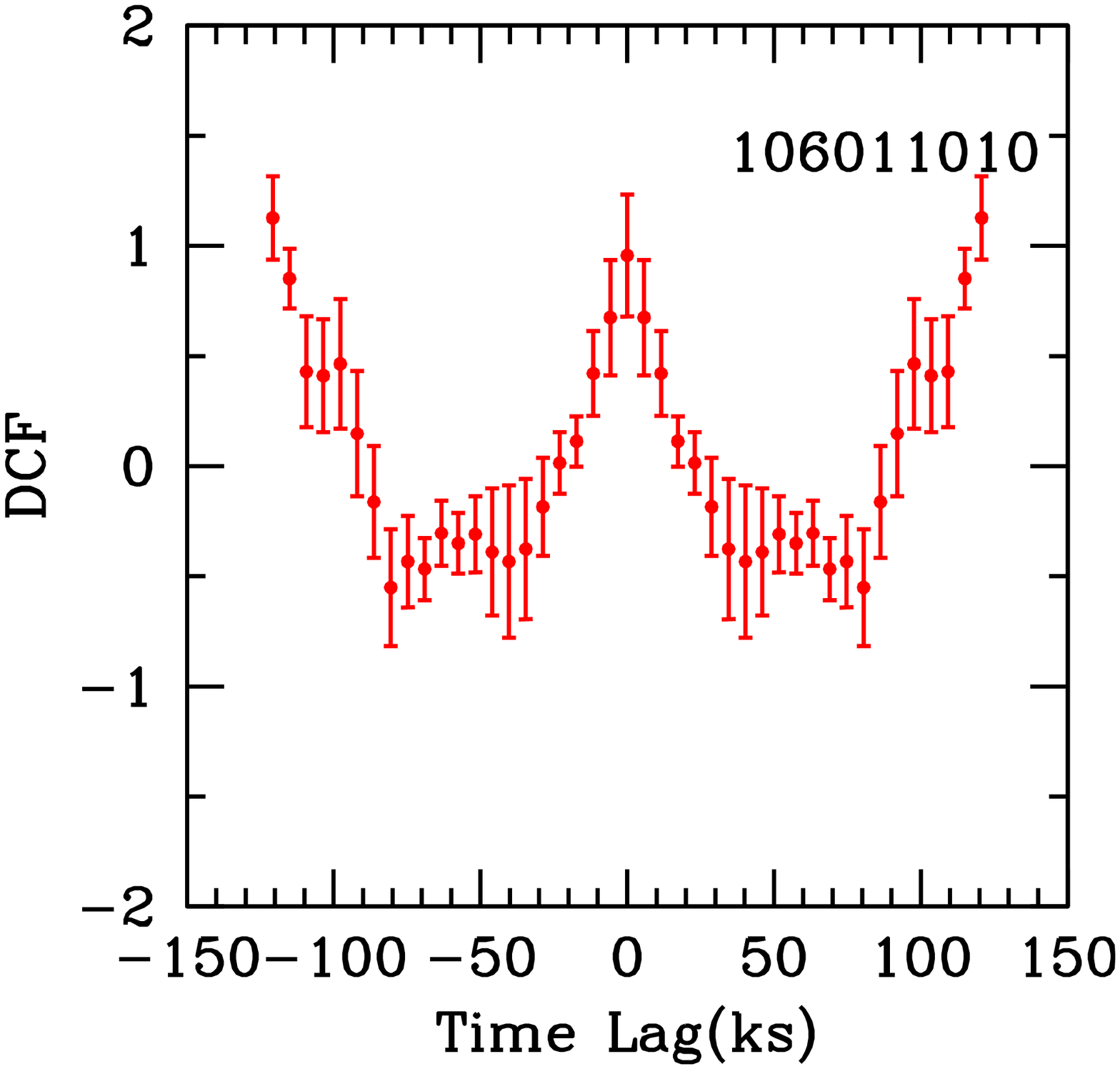} &
\hspace*{0.5cm} \includegraphics[width=4.5cm,angle=0]{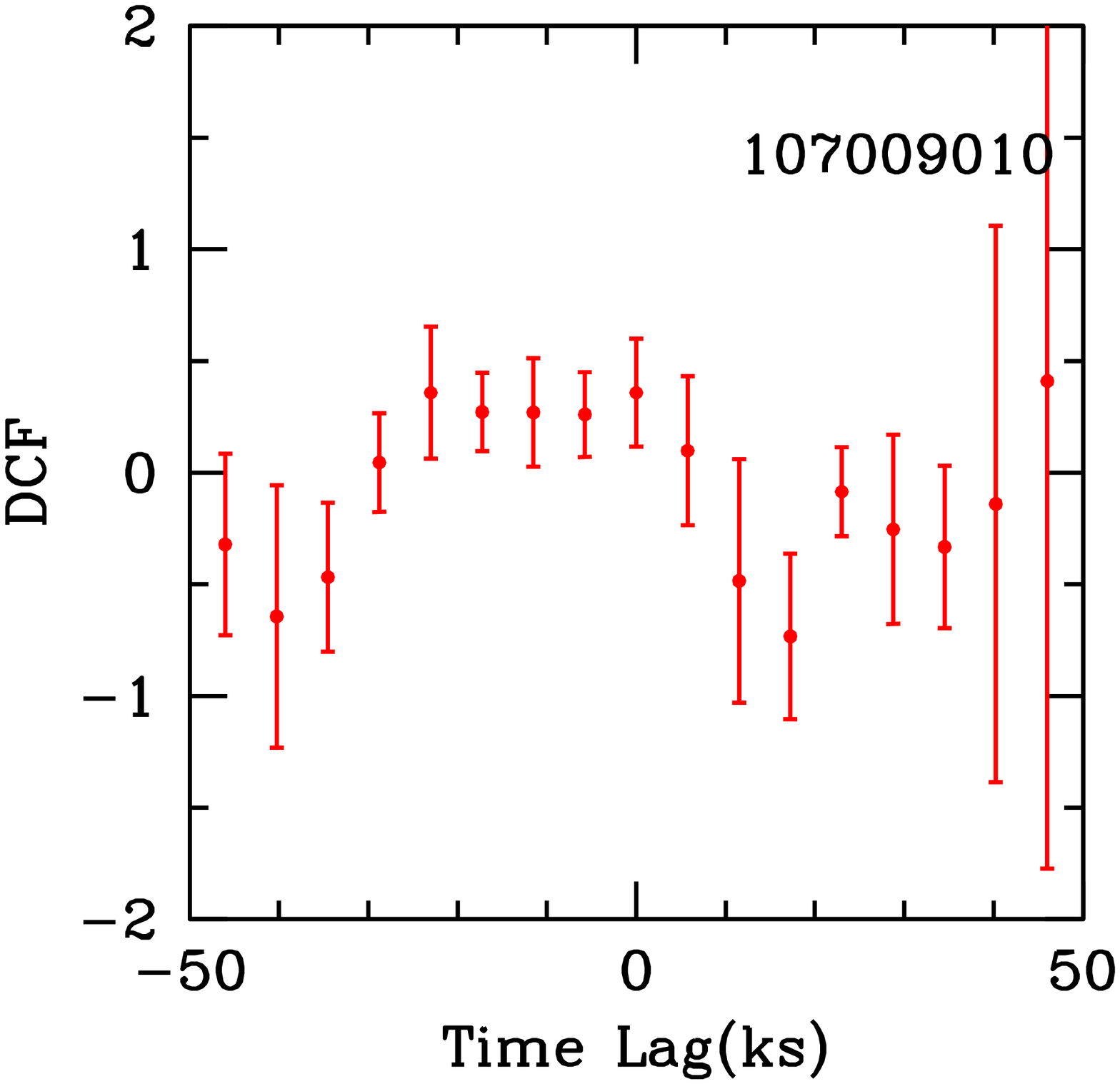} &
\hspace*{0.5cm} \includegraphics[width=4.5cm,angle=0]{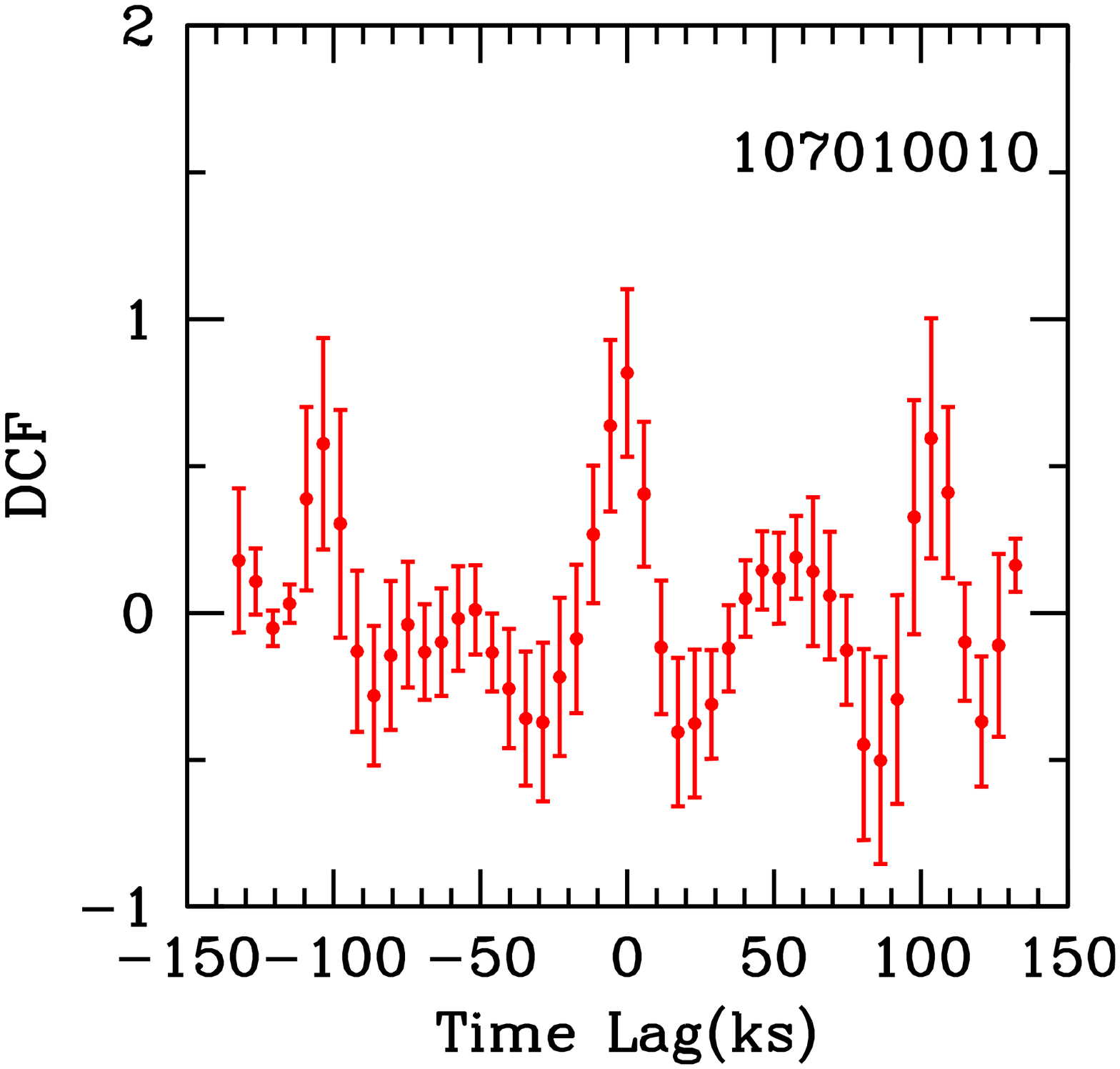} \\
\includegraphics[width=4.5cm,angle=0]{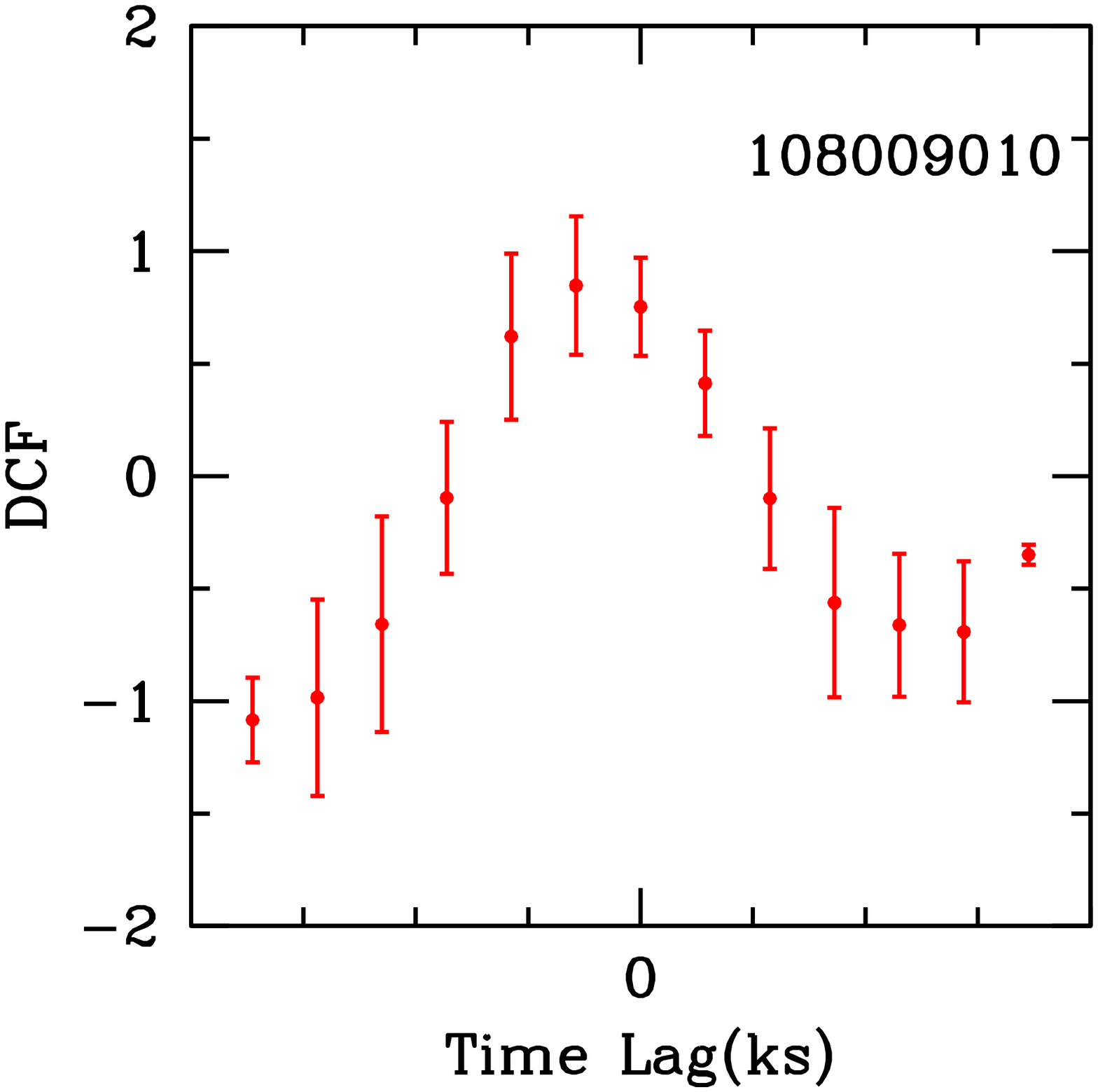} &
\hspace*{0.5cm} \includegraphics[width=4.5cm,angle=0]{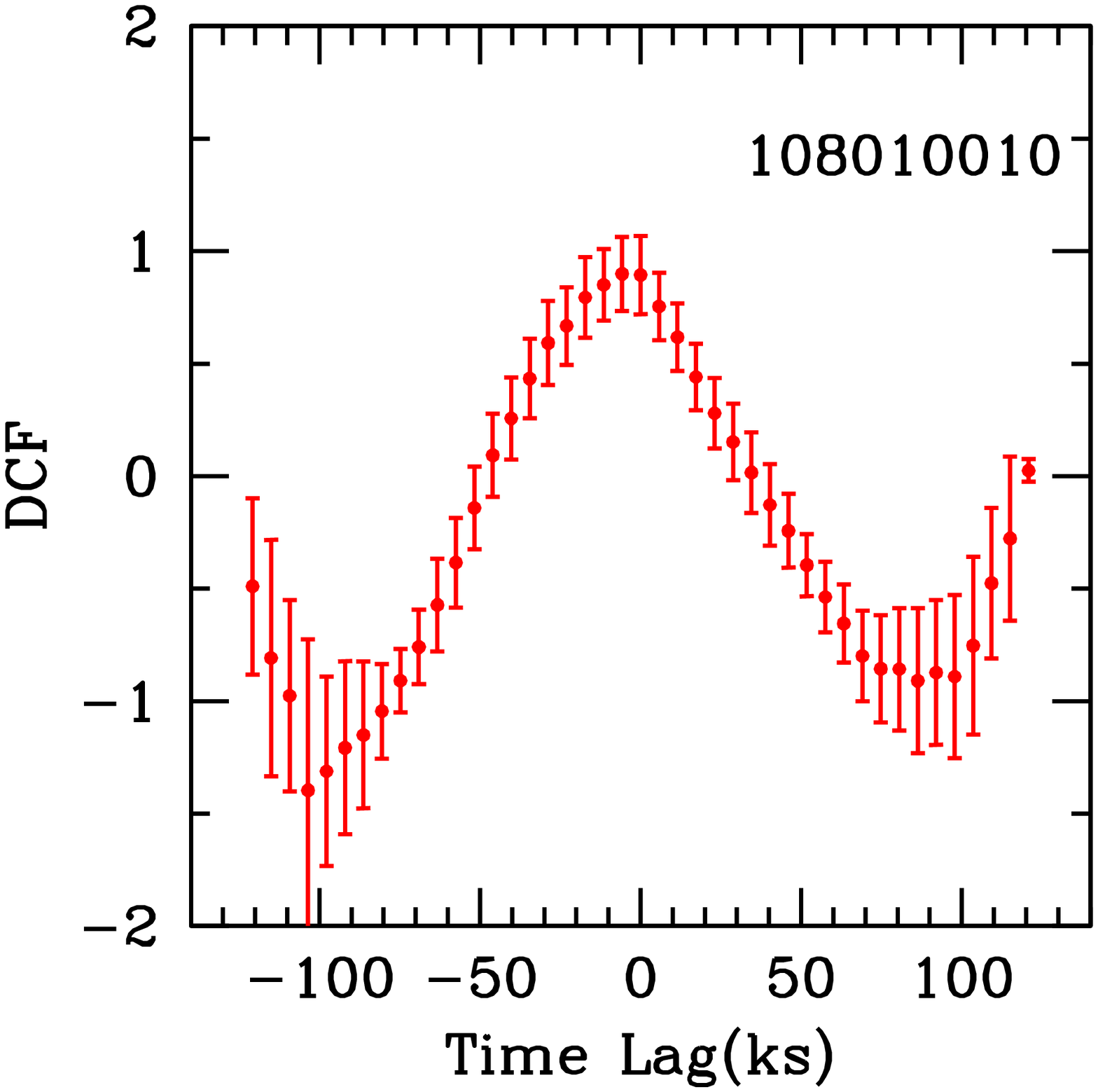} &
\hspace*{0.5cm} \includegraphics[width=4.5cm,angle=0]{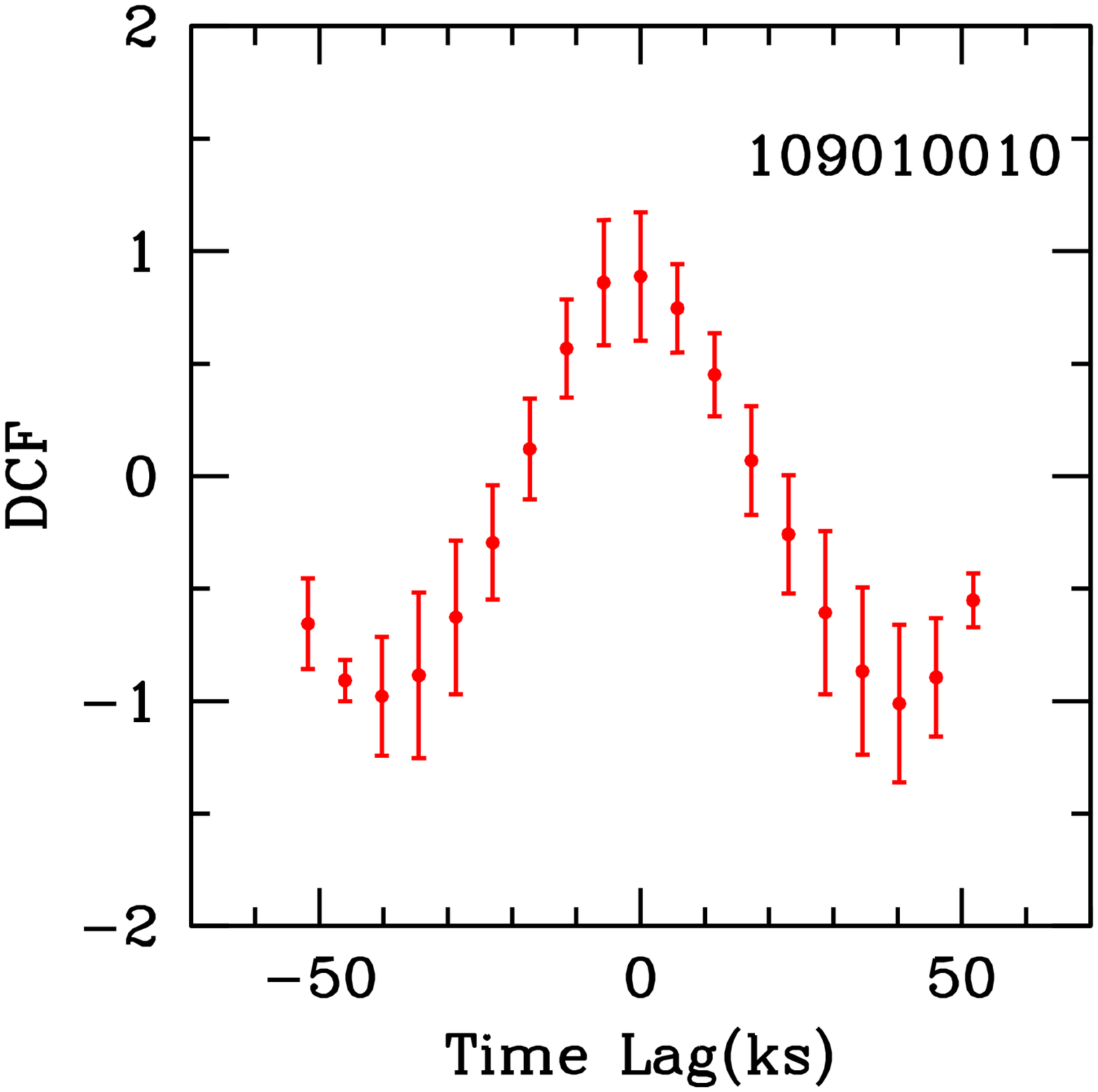} \\
& \hspace*{0.5cm} \includegraphics[width=4.5cm,angle=0]{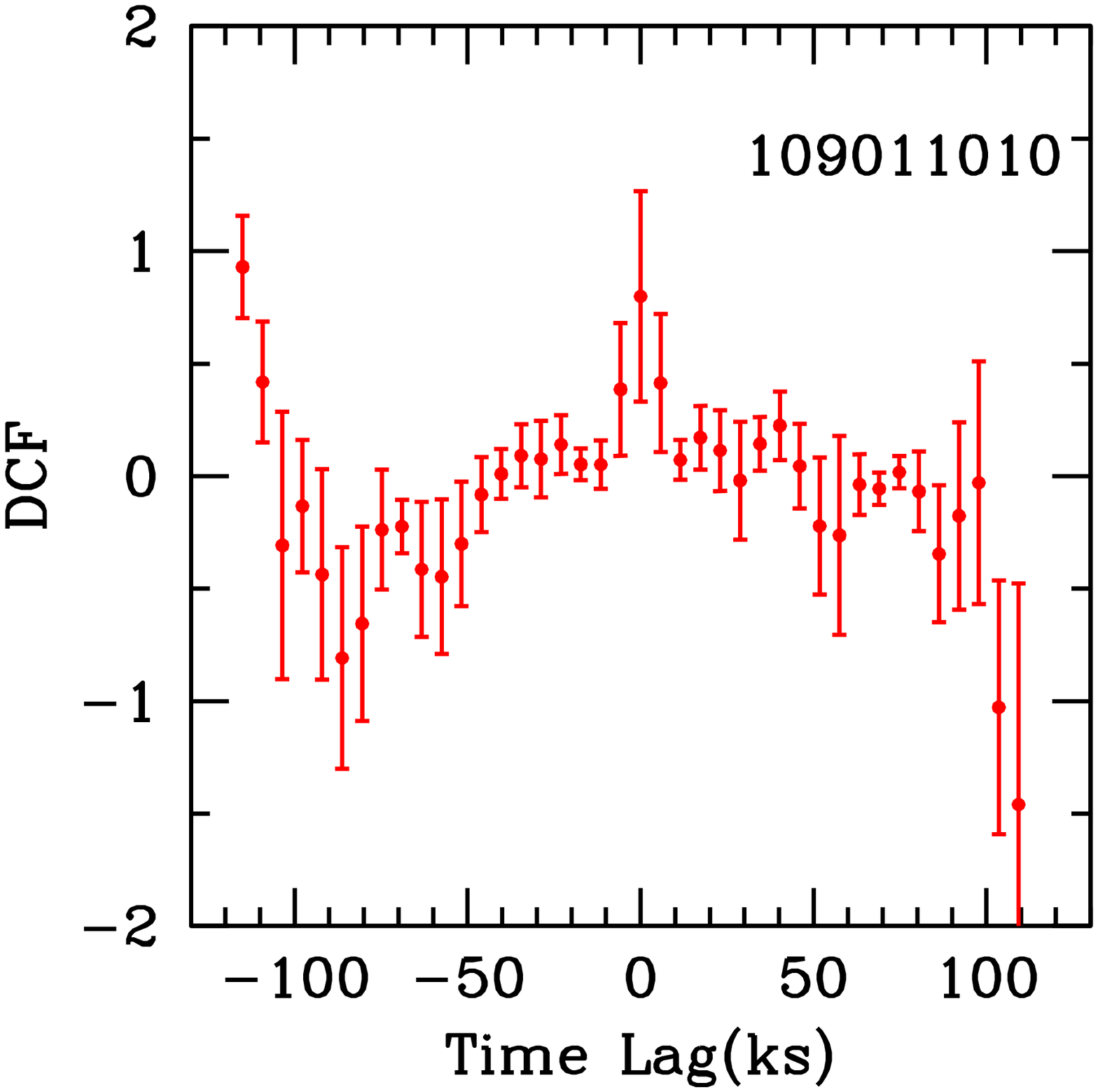}  &  \\
\end{tabular}
\figcaption{Cross-correlation analysis by DCF for soft (0.8 $-$ 1.5) keV and hard (1.5 $-$ 8.0) keV for all 13 XIS observations. Observation IDs are given in the DCF panels.}  
\label{fig:dcf}
\end{figure*}

\noindent
To estimate the time lags between soft and hard X-ray energies, we performed DCF analyses between {\it XIS} soft (0.8 -- 1.5 keV) 
versus {\it XIS} hard (1.5 -- 8.0 keV) bands for all 13 observations. All the DCF plots are displayed in Figure \ref{fig:dcf}. These DCF
plots are fitted with a Gaussian function given by Eqn.\ (4), and the fitting parameters are provided for all 13 observations 
in Table \ref{tab:dcffit}. We find that all soft versus hard DCF plots are well correlated with small time lags, all of which are consistent with zero time lag, though most of the peaks are quite broad. These well correlated DCF plots for the soft and the hard X-ray bands give an
 indication that the soft and the hard band X-ray emissions are nearly cospatial and probably arise from the same population of leptons. However,
 as noted in the Introduction, there may be an IC component to the harder X-ray emission at some times, and given the breadth of the DCF peaks, we cannot tightly constrain the possibility that more than one populations of leptons are involved.

\subsection{Intraday Power Spectral Density Analysis}
\begin{figure*}
    \centering
    \begin{tabular}{lcr}
\includegraphics[width=5.2cm,angle=0]{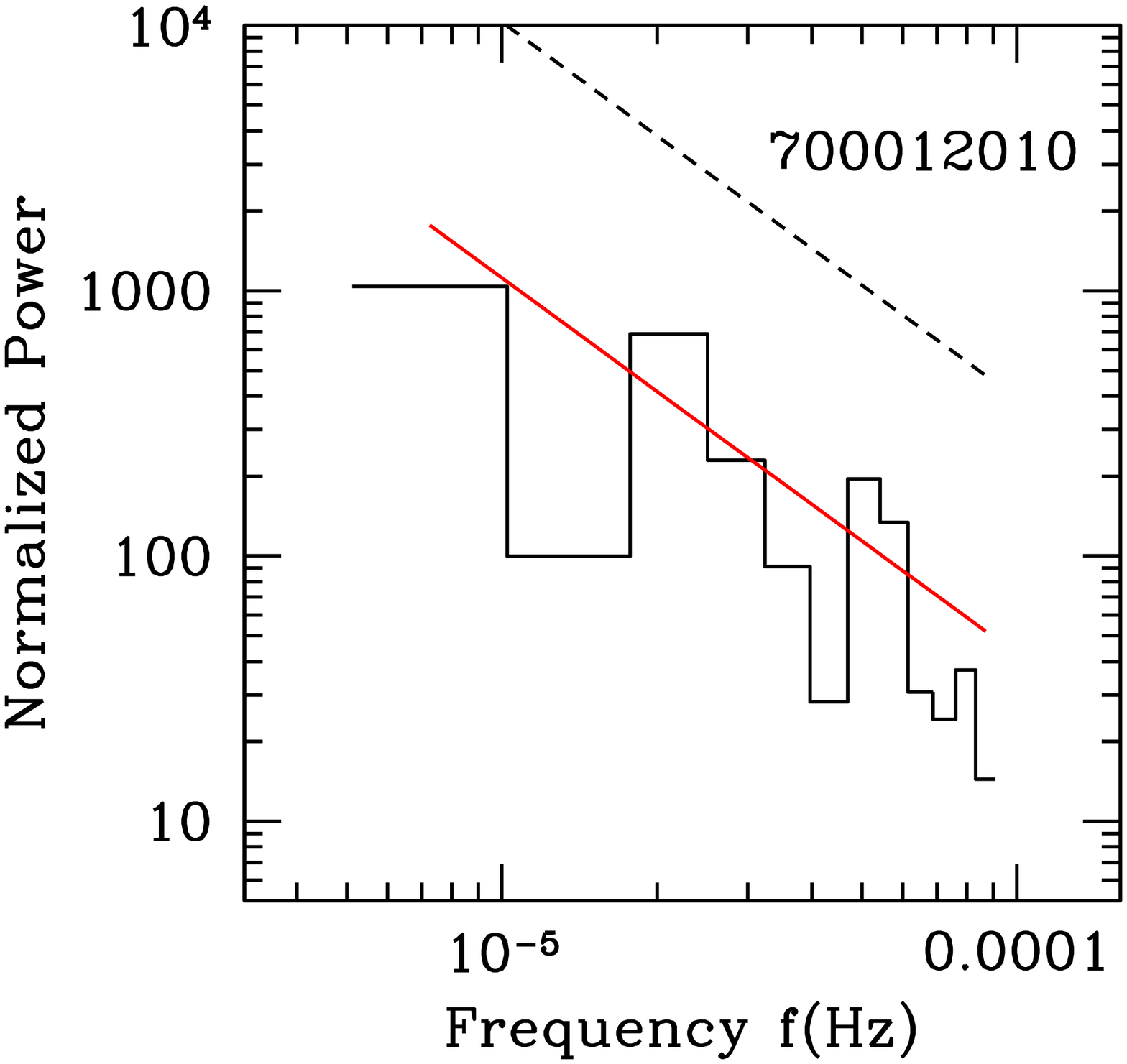} &
\includegraphics[width=5.2cm,angle=0]{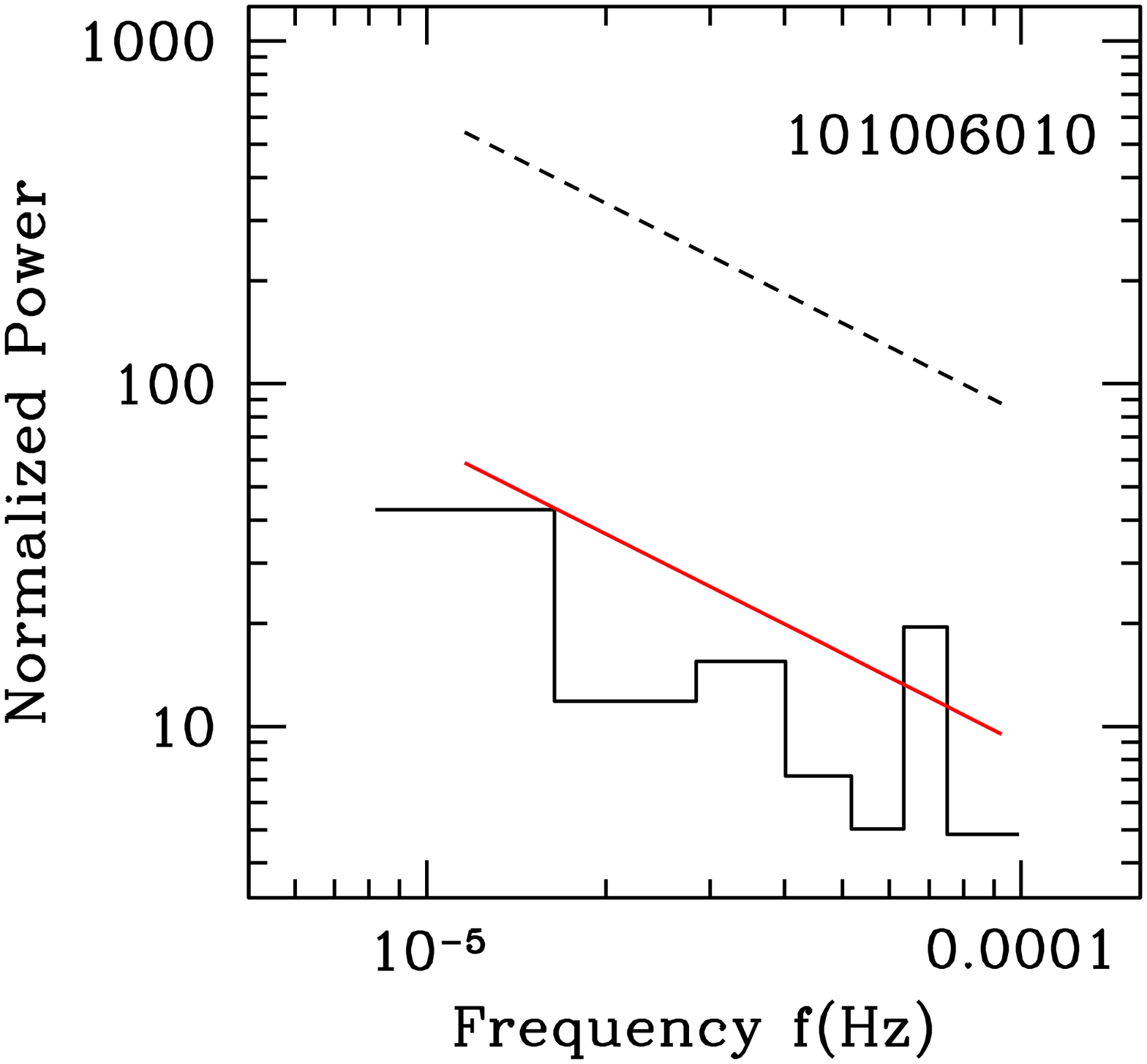} &
\includegraphics[width=5.2cm,angle=0]{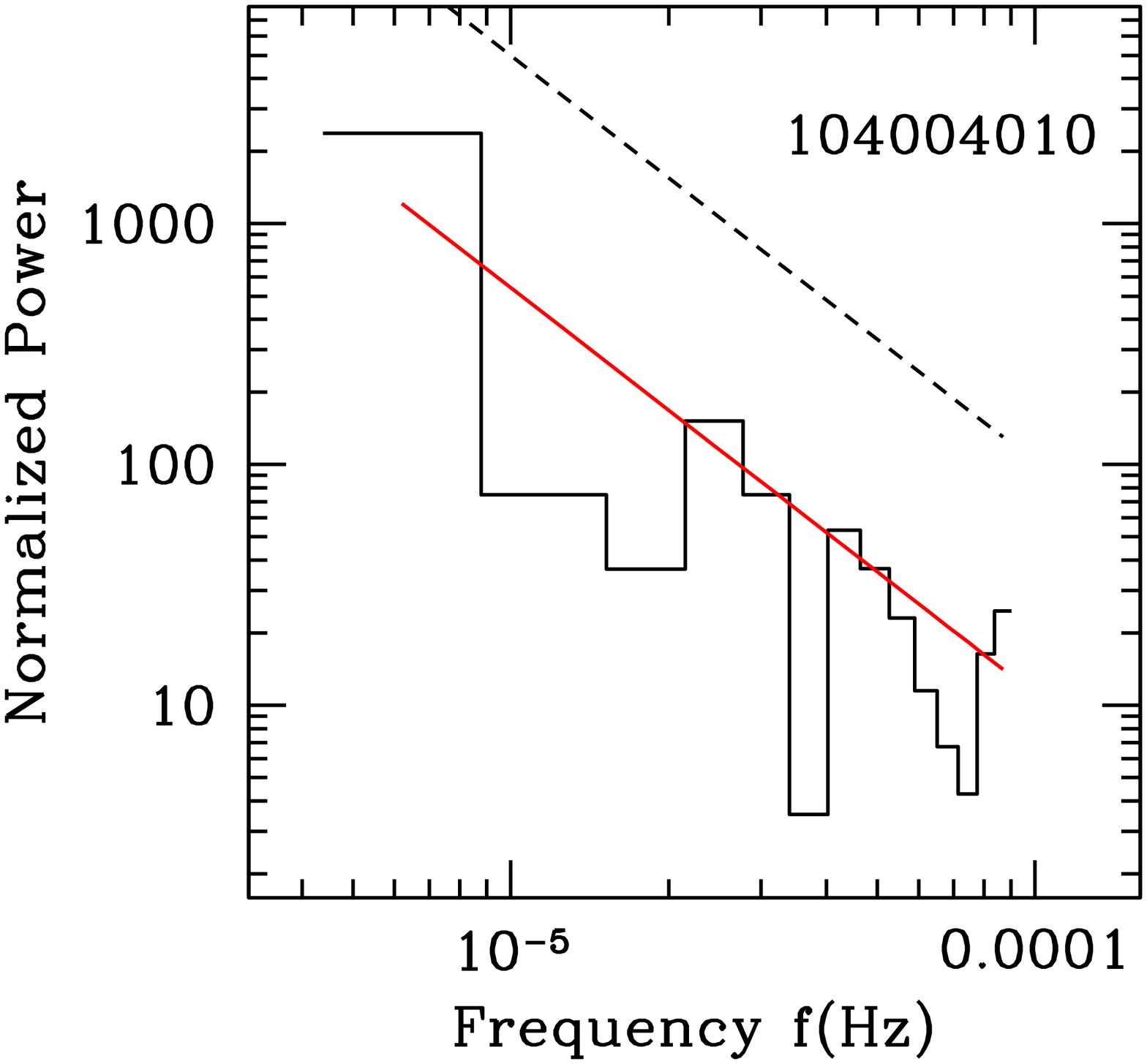} \\
\includegraphics[width=5.2cm,angle=0]{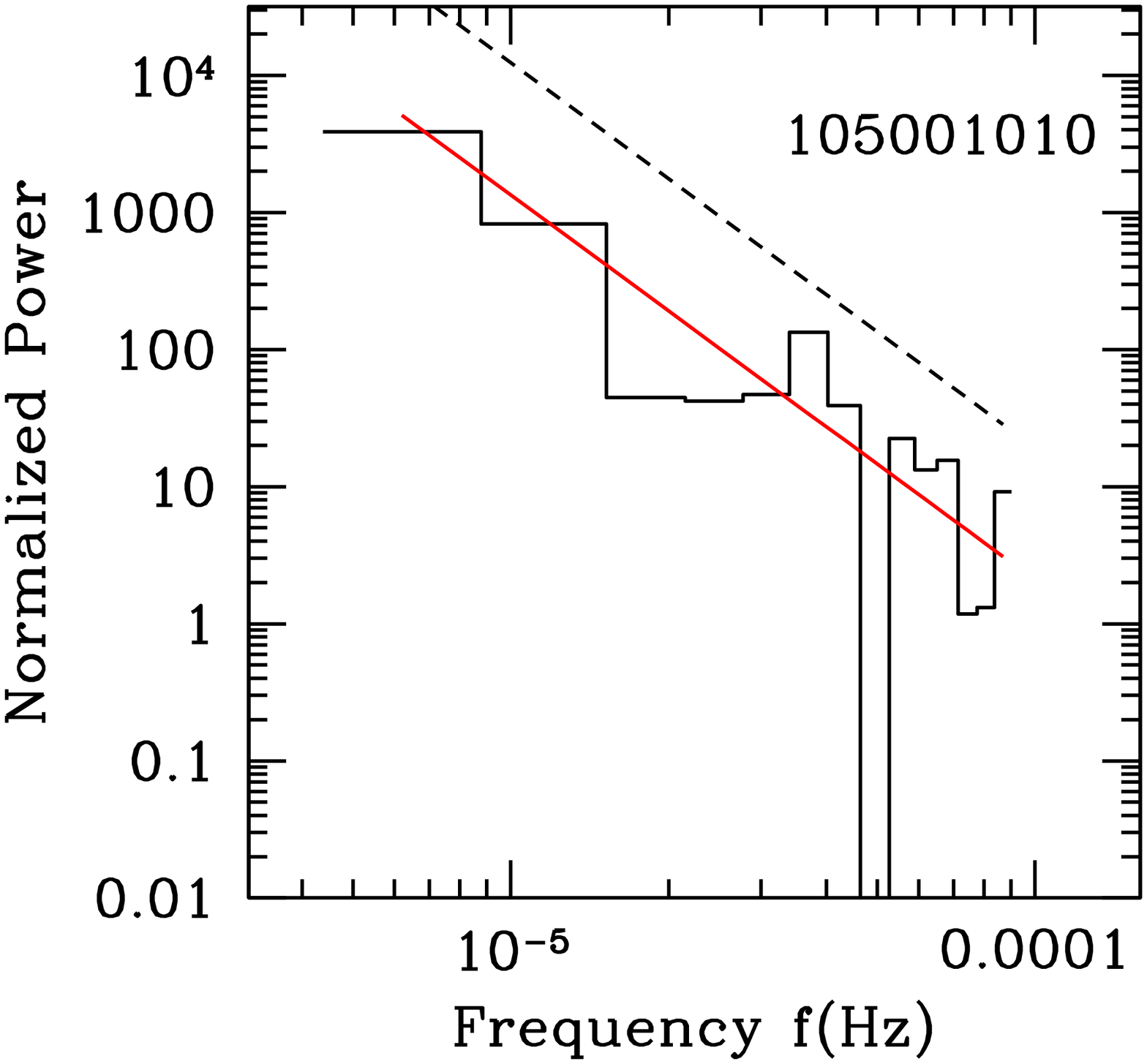} &
\includegraphics[width=5.2cm,angle=0]{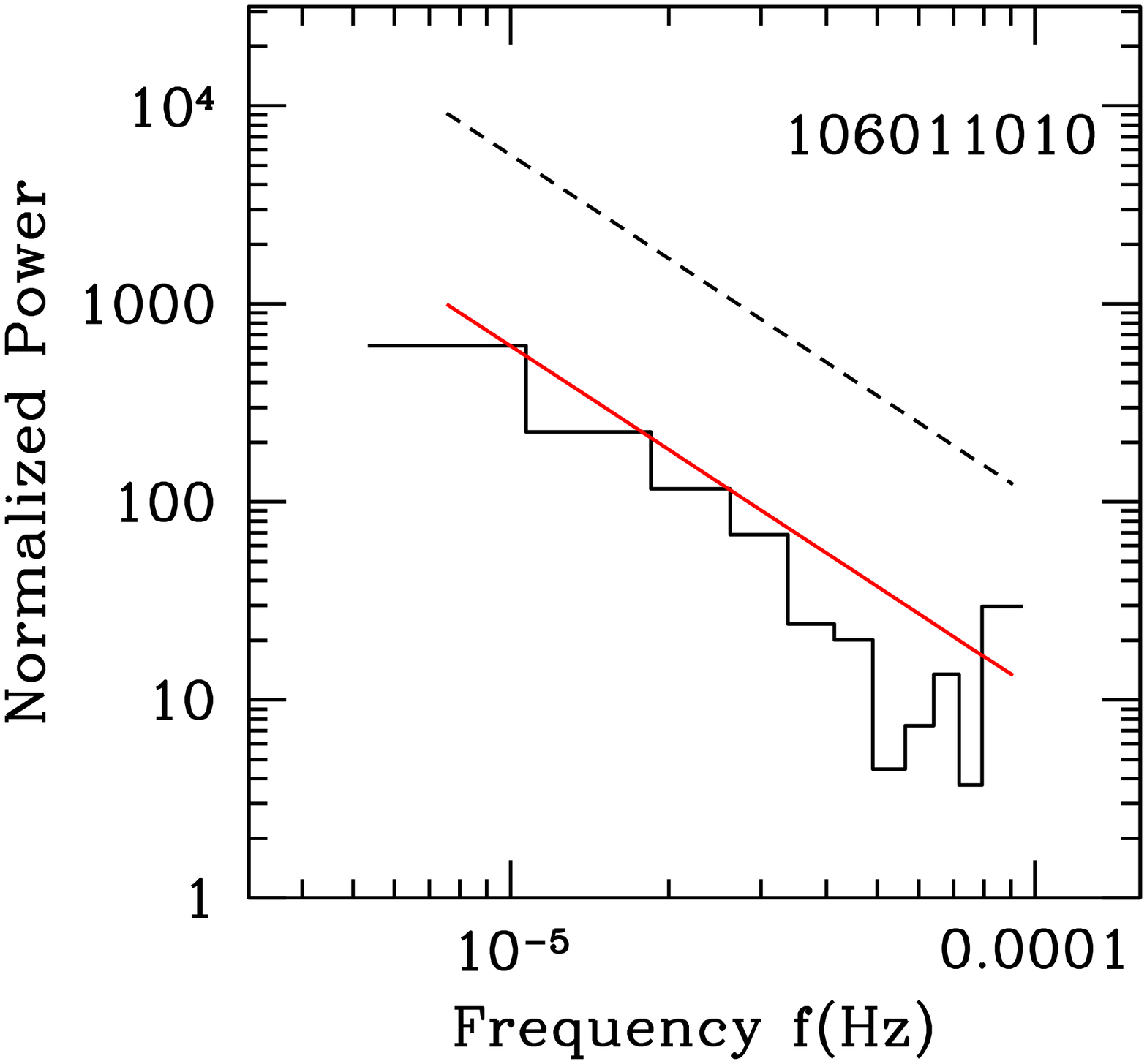} & 
\includegraphics[width=5.2cm,angle=0]{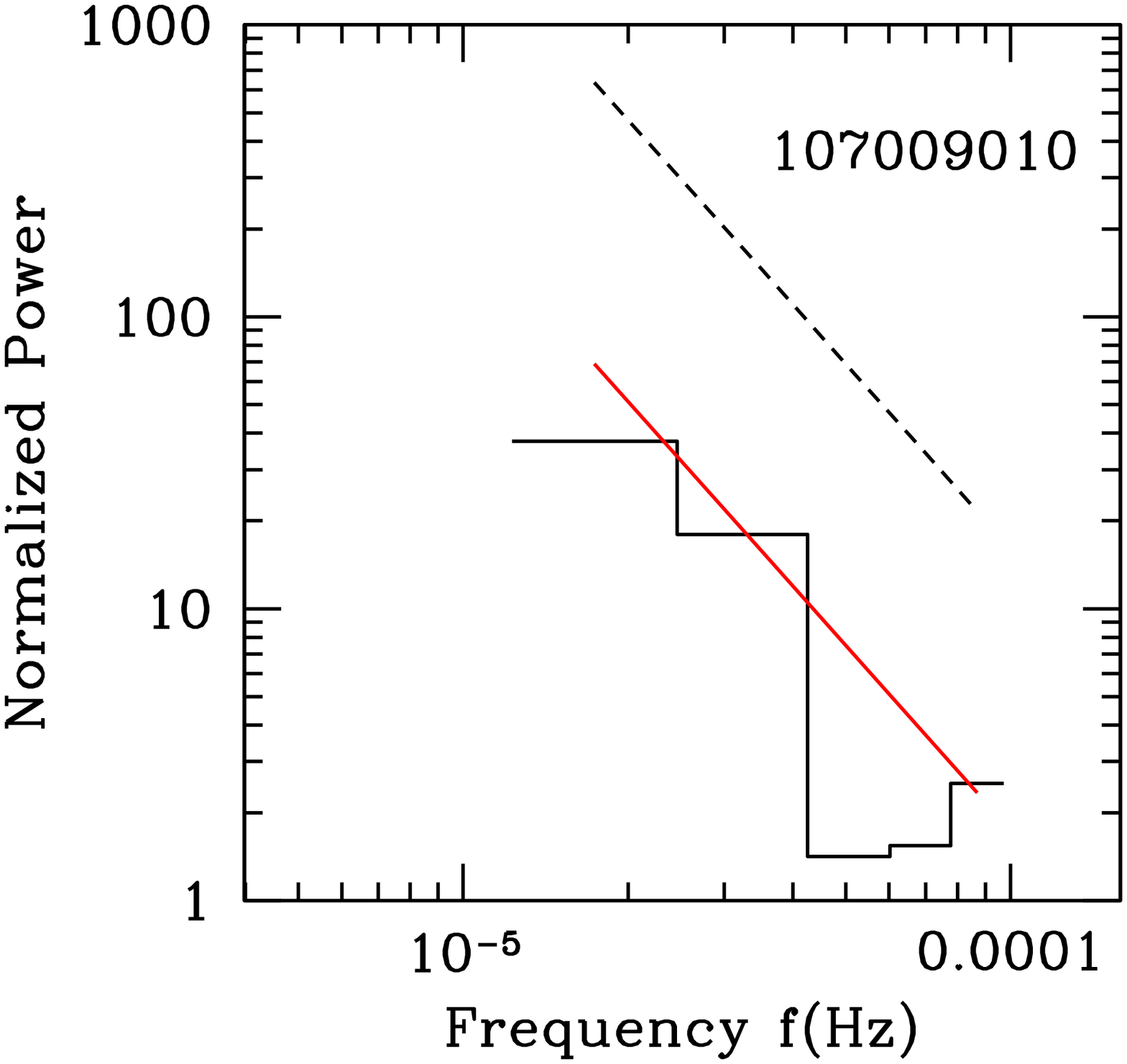} \\
\includegraphics[width=5.2cm,angle=0]{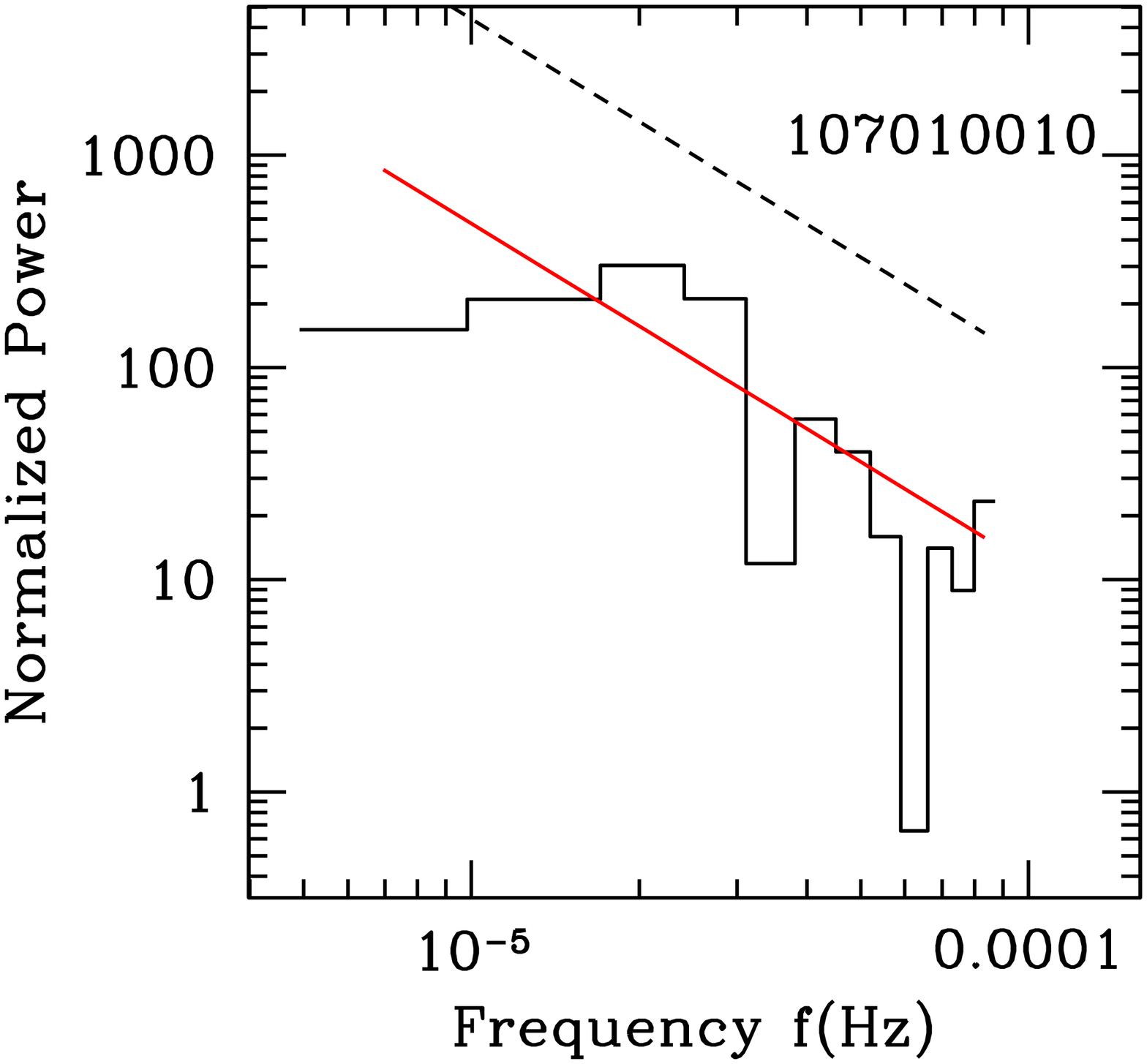} &
\includegraphics[width=5.2cm,angle=0]{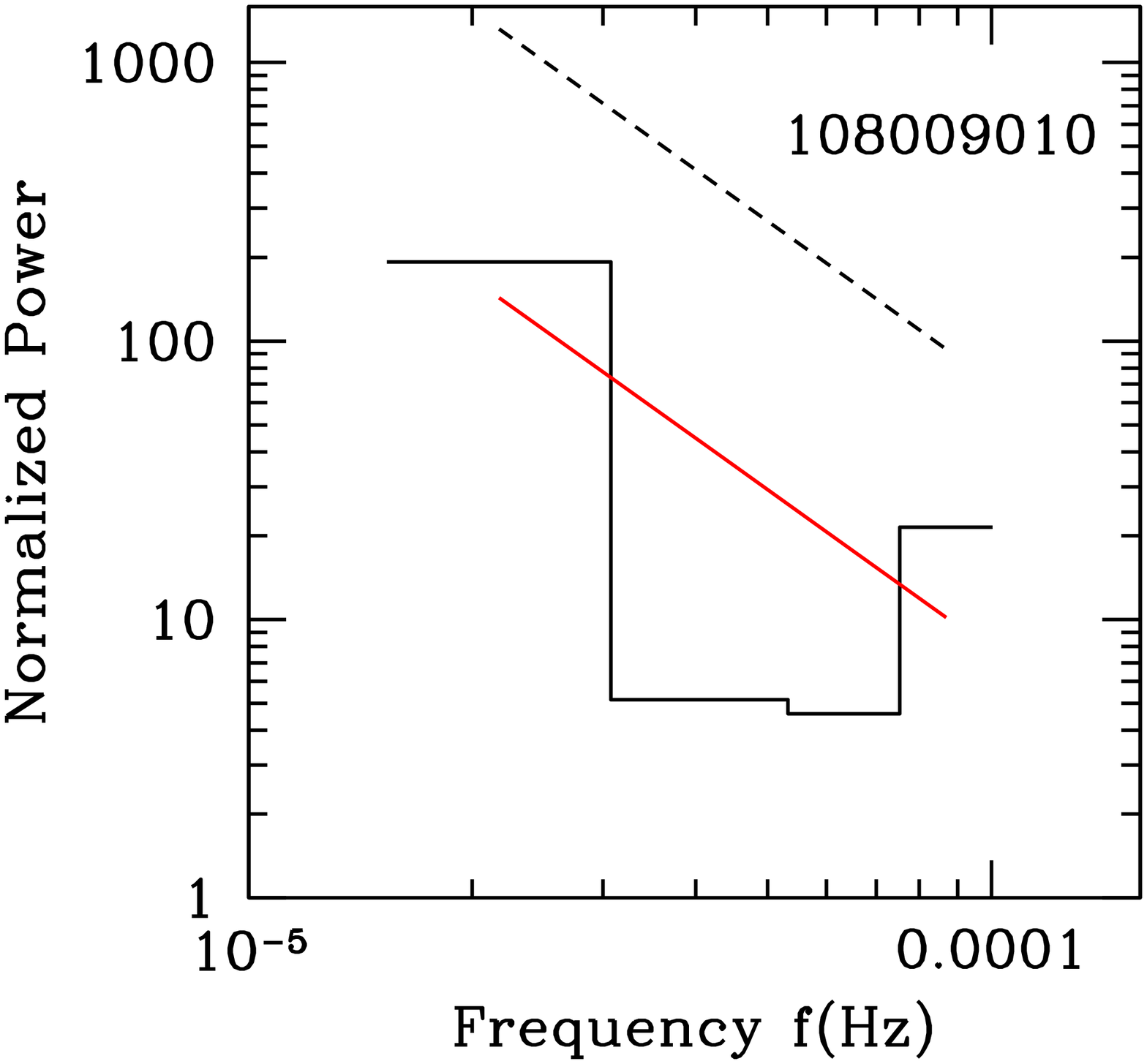} & 
\includegraphics[width=5.2cm,angle=0]{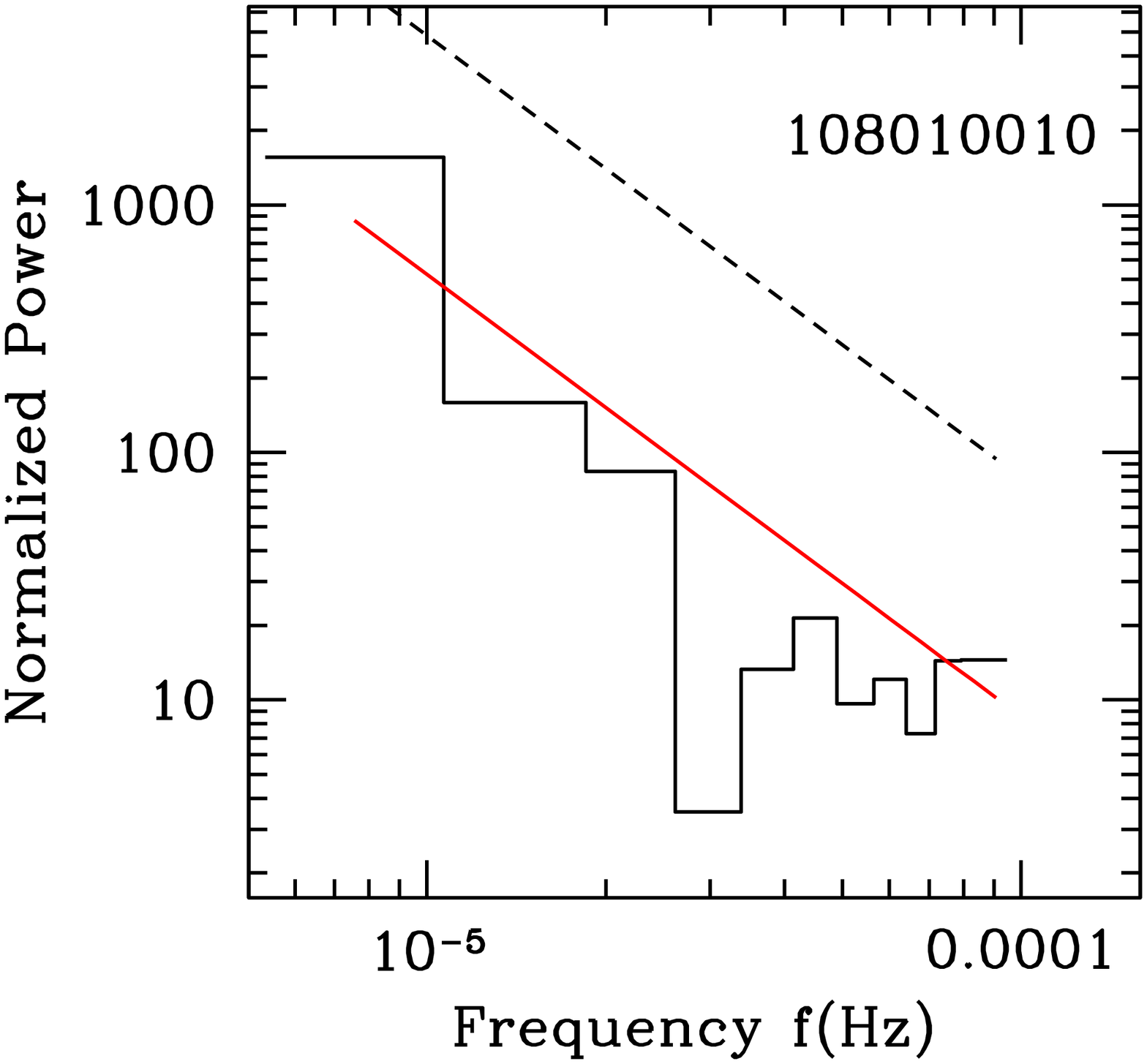} \\ 
\includegraphics[width=5.2cm,angle=0]{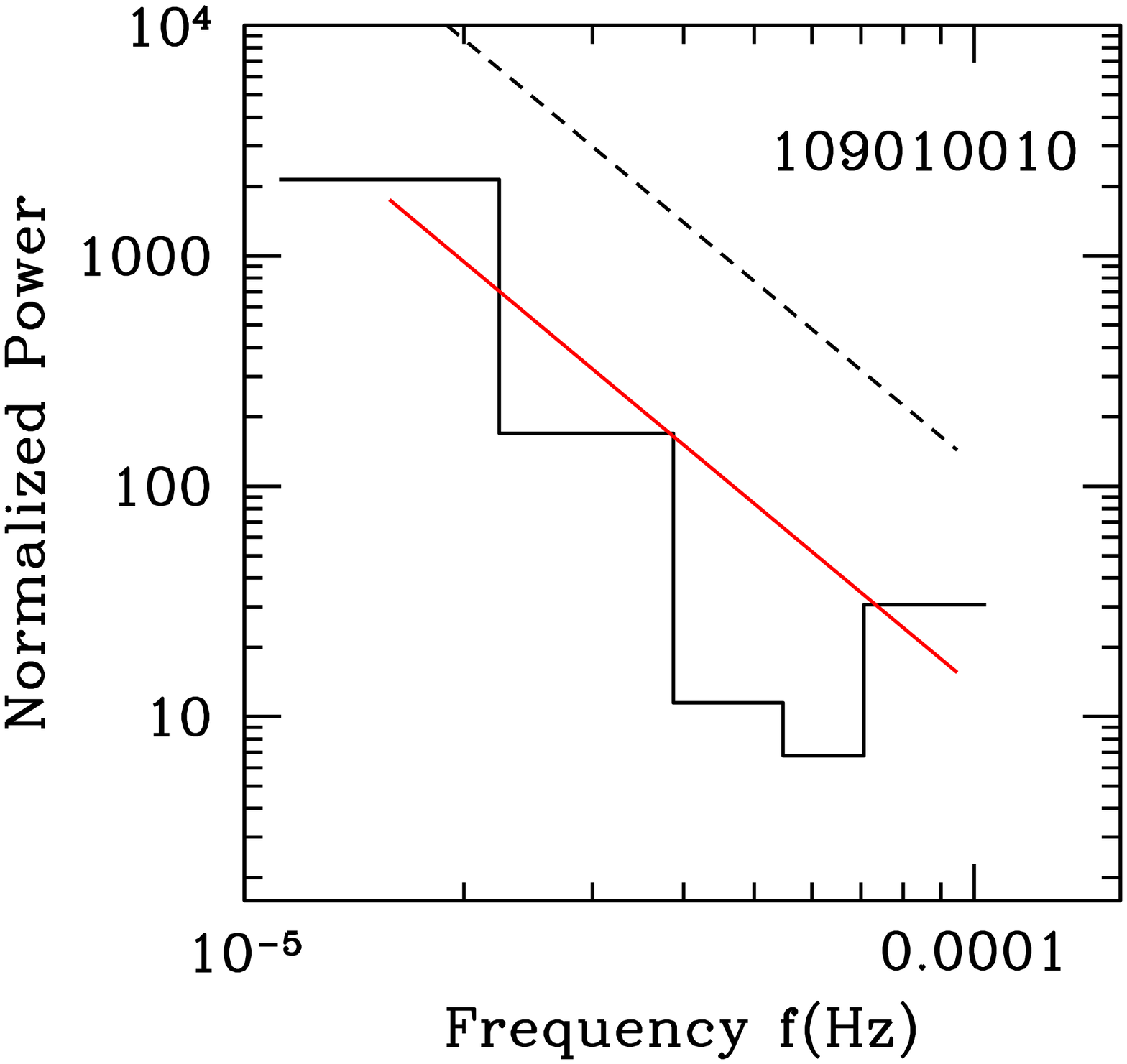} &
\includegraphics[width=5.2cm,angle=0]{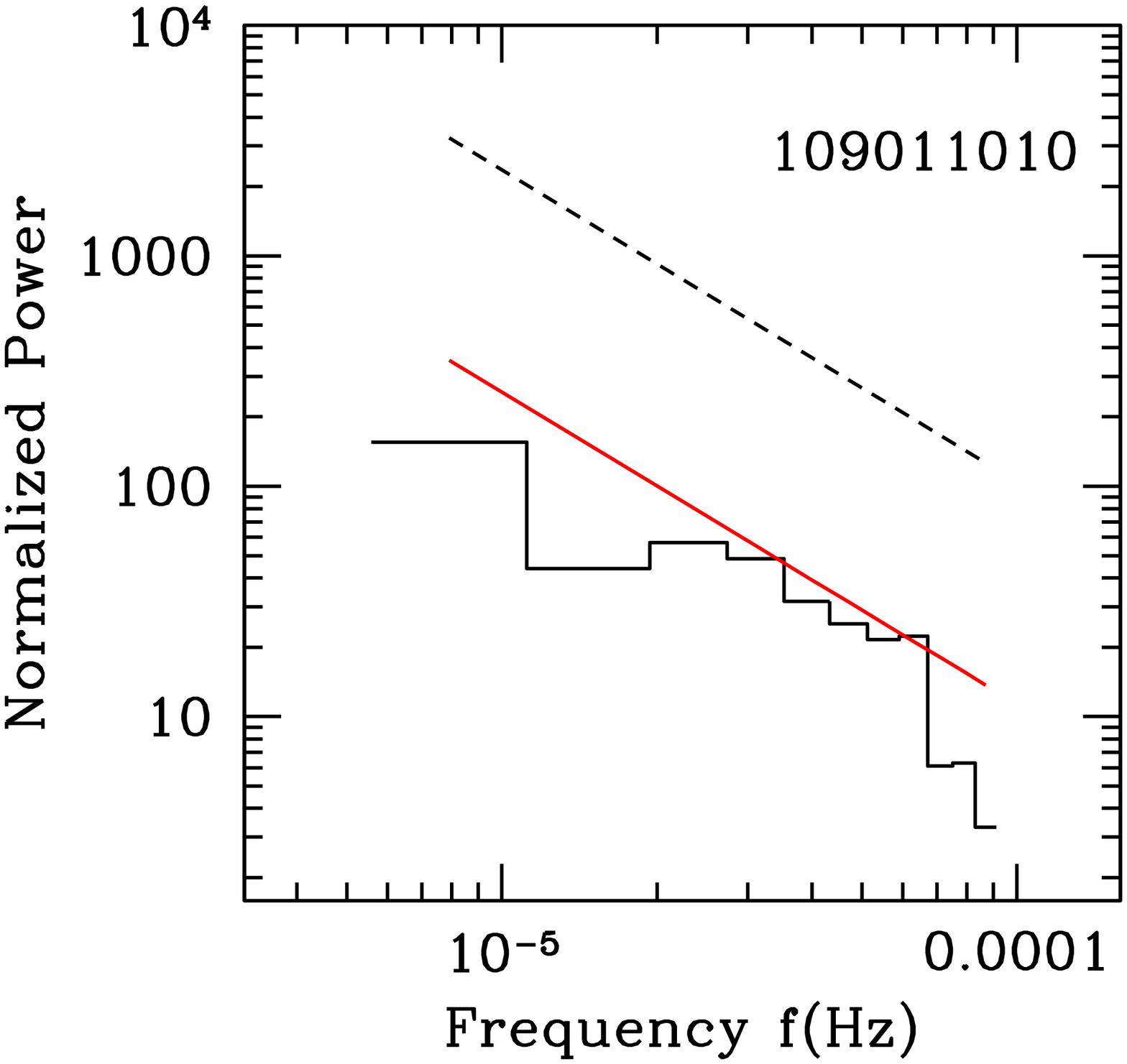} &
\end{tabular}
\figcaption{Power spectral densities (PSDs) of all XIS total (0.8 $-$ 8.0 keV) LCs of 11 out of 13 observations. Observation IDs are given in the PSD panels; the continuous red lines are the fitted red-noise and the dotted black line shows 
the 99.73\% (3$\sigma$) confidence level for any QPO within the red-noise model.}  
\label{fig:psd}
\end{figure*}

\noindent
To characterize the temporal intraday flux variations, and to search for short term periodicity  or quasi-periodicity in the X-ray emission of PKS 2155$-$304, we have performed PSD analysis on the {\it Suzaku XIS} IDV LCs. Out of the 13 LCs, 2 Observation IDs 102020010 and 103011010, have only a few data points, so the PSD analysis could not be performed for them. We performed PSD analyses on the other 11 LCs of their {\it XIS} total energy (0.8 -- 8.0 keV) LCs and the PSD plots are presented in Figure \ref{fig:psd}. These plots indicate that these observations appear to show red-noise dominated PSDs over the limited (approximately one decade) span of frequency for which most of them could be obtained. Nonetheless, the results are quite uncertain because they are based on low signal-to-noise light curves of relatively short length. The key result is that there is no detection of any  quasi-periodicity. The values of the power-law slopes and normalization constants for the 11 viable observations are reported in Table \ref{tab:psdfit}. The power-law slopes $\alpha$ of the red noise of PSDs span a surprisingly wide range, from $-$2.81 $\pm$ 0.87 to $-$0.88 $\pm$ 0.29 (with an average of $-$1.81 $\pm$ 0.55) with logarithmic normalization constants $N$ in the range of $-$11.16 $\pm$ 3.87 to $-$2.81 $\pm$ 1.27.  The results presented here for the PSDs are, overall, consistent with those computed for a large number of AGNs studied with {\it XMM-Newton} \citep{2012A&A...544A..80G}. We recall that our results for the slopes of the PSDs of the three {\it Sukaku} LCs of Mrk 421 were $-1.51 \pm 0.27$, $-3.12 \pm 0.44$, and $-1.40 \pm 0.11$  \citep{2019ApJ...884..125Z}, so also spanning a significant range, but perhaps a bit steeper than PKS 2155$-$304. We looked for a correlation between normalization constant log($N$) and variability amplitude $F_{var}$ but found none; however, this might be because of the low  signal-to-noise ratio of the light curves. 

\subsection{Correlation between Flux and IDV Timescale}

\noindent
Our sample of all 13 {\it Suzaku} X-ray LCs of the blazar PKS 2155$-$304 is spread over almost nine years (2005 November 30 to 2014 October 30). 
In the 0.8 -- 8.0 keV energy band of this instrument, all the LCs are found to be clearly variable on IDV timescales. We estimated variability 
timescales for all these LCs. In Figure 5, we plot the average flux versus IDV timescales. We find that there 
is no correlation between the average flux and the IDV timescales of this blazar. \\
\\
The most accepted emission model for blazars is that the vast majority of the radiation is emitted by the plasma in the relativistic jet oriented at a small viewing 
angle, and that is strongly affected by relativistic beaming. This mechanism leads to a shortening of the observed timescales by a factor of $\delta^{-1}$, where $\delta =[\Gamma \left (1-\beta \cos \theta \right )]^{-1}$, the bulk Lorentz factor 
$\Gamma=1/\sqrt{1-\beta^{2}}$, $\theta$ is the angle between the jet and the line-of-sight (LOS), and $\beta=V/c$ in terms of the bulk velocity $V$ of the emitting region.  
A negative  
correlation between average flux and variability timescales is expected if the main source of the variability is changes in the viewing 
angle and/or the velocity of the relativistic charged particles. But we did not find any such correlation, which implies that the X-ray 
emission variability from the blazar PKS 2155$-$304 is not primarily from bulk velocity changes. 
     
\subsection{Long Term Flux and Spectral Variability}

\noindent
The blazar PKS 2155$-$304 was observed by {\it Suzaku} for almost a decade giving us an opportunity to study its flux and spectral 
variability on LTV timescales in X-rays with a single instrument. The overall change in the total X-ray flux with time is shown 
in the top panel of Figure 6. On visual inspection, there appears to be a
trend of a roughly linearly decreasing flux with time from 2005 to 2014. A least-squares fit to these long term flux versus time data yields a slope of $-$0.0017 counts s$^{-1}$ yr$^{-1}$, with a correlation coefficient for the fit of $-$0.7876 and its corresponding null hypothesis of 0.0013. However, as only 13 data points are present in the roughly 9 years of the span of these observations it is difficult to make a strong claim that this nominal decline is real.  The overall HR (spectral change) with time during this extended observing period is shown in the bottom panel of Figure 6. On visual inspection, there is no clear trend of increasing or decreasing HR with respect to time is seen. We found no correlation between HR with time on this long time scale. \\
\\
In Figure 7, we present plots of the HR versus flux 
 of the blazar PKS 2155--304 for six temporal intervals. It is widely accepted that the lower energy hump of emission in blazars involve processes that are 
accelerating particles in the relativistic jet to extremely high energies, and when these relativistic particles 
encounter the inhomogeneous magnetic field in the jet significant synchrotron radiation, which dominates 
the cooling process, extends into the X-ray band. In Figure 7, we found indications of anti-clockwise loops (hard lags) most of the time (Epochs 1, 2, 4, 5, 6). The anti-clockwise direction indicates that during 
these epochs, particles were accelerated to their highest speeds indicating some dominance of the particle acceleration 
mechanism \citep[e.g.][and references therein]{2002ApJ...572..762Z}. 
Only Epoch 3 shows a part of a clockwise loop (soft lag) in  Figure 7 that is consistent with the synchrotron cooling 
mechanism being dominant. Similar results were found for the blazars 3C 273 and Mrk 421 using observations from {\it XMM-Newton} and {\it Chandra}, 
respectively \citep{2015MNRAS.451.1356K,2018MNRAS.480.4873A}. 

\noindent
\begin{table}
\caption{Correlation analysis between XIS soft and XIS hard bands for all 13 observations.}
\label{tab:dcffit}
\begin{center}
\begin{tabular}{crc} \hline \hline
ObsID           & $m$    (ks)        &  $\sigma$ (ks)  \\\hline
700012010       &   ~0.5 $\pm$ ~2.0 &  ~~6.6 $\pm$ ~2.0    \\
101006010       &   ~3.1 $\pm$ ~4.7 &  ~~7.1 $\pm$ ~4.7    \\
102020010       &$-$~0.2 $\pm$ ~8.5 &  ~~3.3 $\pm$ ~5.1    \\
103011010       &$-$~0.0 $\pm$ ~6.2 &  ~~7.0 $\pm$ ~6.2    \\
104004010       &   ~0.0 $\pm$ ~8.9 &  ~22.3 $\pm$ ~8.9    \\
105001010       &$-$~0.7 $\pm$ ~5.9 &$-$18.6 $\pm$ ~5.9    \\
106011010       &   ~0.0 $\pm$ ~4.2 &  ~~8.6 $\pm$ ~4.2    \\
107009010       &$-$11.4 $\pm$ 17.0 &  ~~8.8 $\pm$ 17.0    \\
107010010       &$-$~1.8 $\pm$ ~1.8 &  ~~5.6 $\pm$ ~1.8    \\
108009010       &$-$~3.9 $\pm$ ~4.4 &  ~~6.0 $\pm$ ~4.4    \\
108010010       &$-$~7.3 $\pm$ ~7.3 &  ~17.8 $\pm$ ~7.3    \\
109010010       &$-$~0.9 $\pm$ ~4.6 &  ~~8.3 $\pm$ ~4.6    \\ 
109011010       &   ~0.2 $\pm$ ~4.2 &  ~~5.0 $\pm$ ~3.9    \\\hline
\end{tabular}
\end{center}
\noindent
$m$= time lag at which DCF peaks \\
$\sigma$= width of the Gaussian function\\
\end{table}

\begin{table}
\caption{Power-law fits to the PSDs$^{a}$ of 11 observations.}
\label{tab:psdfit}
\begin{center}
\begin{tabular}{ccc} \hline \hline
ObsID              & $\alpha$          & log (N)      \\\hline
   700012010       &$-$1.4 $\pm$ 0.4 &$-$~4.3 $\pm$ 1.5  \\
   101006010       &$-$0.9 $\pm$ 0.3 &$-$~2.8 $\pm$ 1.3  \\
   104004010       &$-$1.7 $\pm$ 0.4 &$-$~6.0 $\pm$ 1.7  \\
   105001010       &$-$2.8 $\pm$ 0.9 &$-$11.2 $\pm$ 3.9  \\
   106011010       &$-$1.7 $\pm$ 0.4 &$-$~6.2 $\pm$ 1.5  \\
   107009010       &$-$2.1 $\pm$ 0.6 &$-$~8.4 $\pm$ 2.8  \\
   107010010       &$-$1.6 $\pm$ 0.5 &$-$~5.6 $\pm$ 2.4  \\
   108009010       &$-$1.9 $\pm$ 1.5 &$-$~7.0 $\pm$ 6.7  \\
   108010010       &$-$1.8 $\pm$ 0.4 &$-$~6.5 $\pm$ 1.8  \\
   109010010       &$-$2.6 $\pm$ 0.9 &$-$~9.7 $\pm$ 3.9  \\
   109011010       &$-$1.4 $\pm$ 0.2 &$-$~4.6 $\pm$ 1.0  \\\hline
\end{tabular}
\end{center}
\noindent
$^{a}$ A power-law model is assumed with: $P(f) \propto f^{\alpha}$ for $\alpha <$ 0.
\end{table}

\section{Discussion}

\noindent
We have studied 13 archival pointed observations of the blazar PKS 2155$-$304 in a period spanning 2005 
to 2014. These observations were taken by XIS instrument onboard {\it Suzaku}. The observations were carried out 
with total elapsed times in the range of 24.0 ks to 157.5 ks with GTIs (good time intervals) in the range of 12.0 ks to 64.0 ks. 
The blazars 3C 273 and PKS 2155$-$304 have been used as X-ray calibrators for the International Astronomical Consortium for High 
Energy Calibration (IACHEC) in the northern and southern hemispheres, respectively. Out of 13 pointed observations of 
PKS 2155$-$304, four observation IDs (700012010, 101006010, 102020010, and 103011010) were used for cross spectral calibration 
of {\it Suzaku, XMM-Newton}, and {\it Chandra} \citep{2011PASJ...63S.657I}, and one (108010010) was used 
for cross spectral calibration of {\it Chandra, NuSTAR, SWIFT, Suzaku}, and {\it XMM-Newton} \citep{2017AJ....153....2M}. 
The total set of {\it Suzaku} observations of PKS 2155$-$304 presented here were taken over the course of the entire operational period of the satellite and 
have not been used before to study the X-ray variability properties of this blazar. These observations provided us with an 
excellent opportunity to study some of the most puzzling properties of blazars, e.g., flux and spectral variability, and here we have performed
cross-correlations between the soft and the hard X-ray bands, evaluated the PSDs of the blazar in X-ray energies on IDV timescales for 
the first time with {\it Suzaku} observations. Although the general results are not surprising, the large number of different LCs for a single source obtained and analyzed in a common manner provides unique results. \\

\begin{figure}
%\begin{center}
\epsscale{1.0}
\includegraphics[width=8.5cm,angle=0]{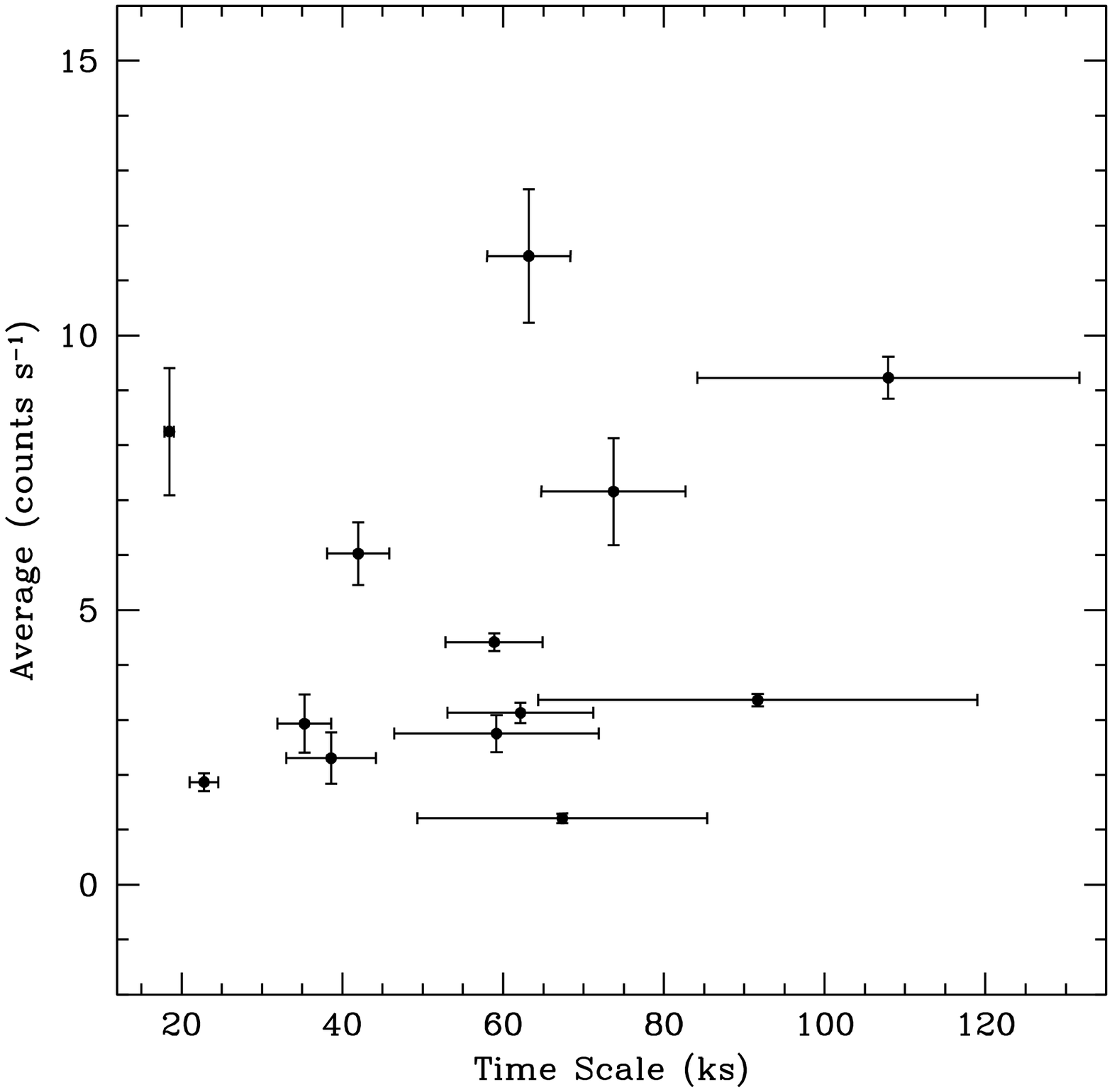} 
\figcaption{Average flux (in counts s$^{-1}$) versus IDV timescales for all LCs of the 13 observations in 
total energy 0.8 -- 8.0 keV.}
\label{fig:flux}
%\end{center}
\end{figure}

\begin{figure}
%\begin{center}
\epsscale{1.0}
\includegraphics[width=8.5cm,angle=0]{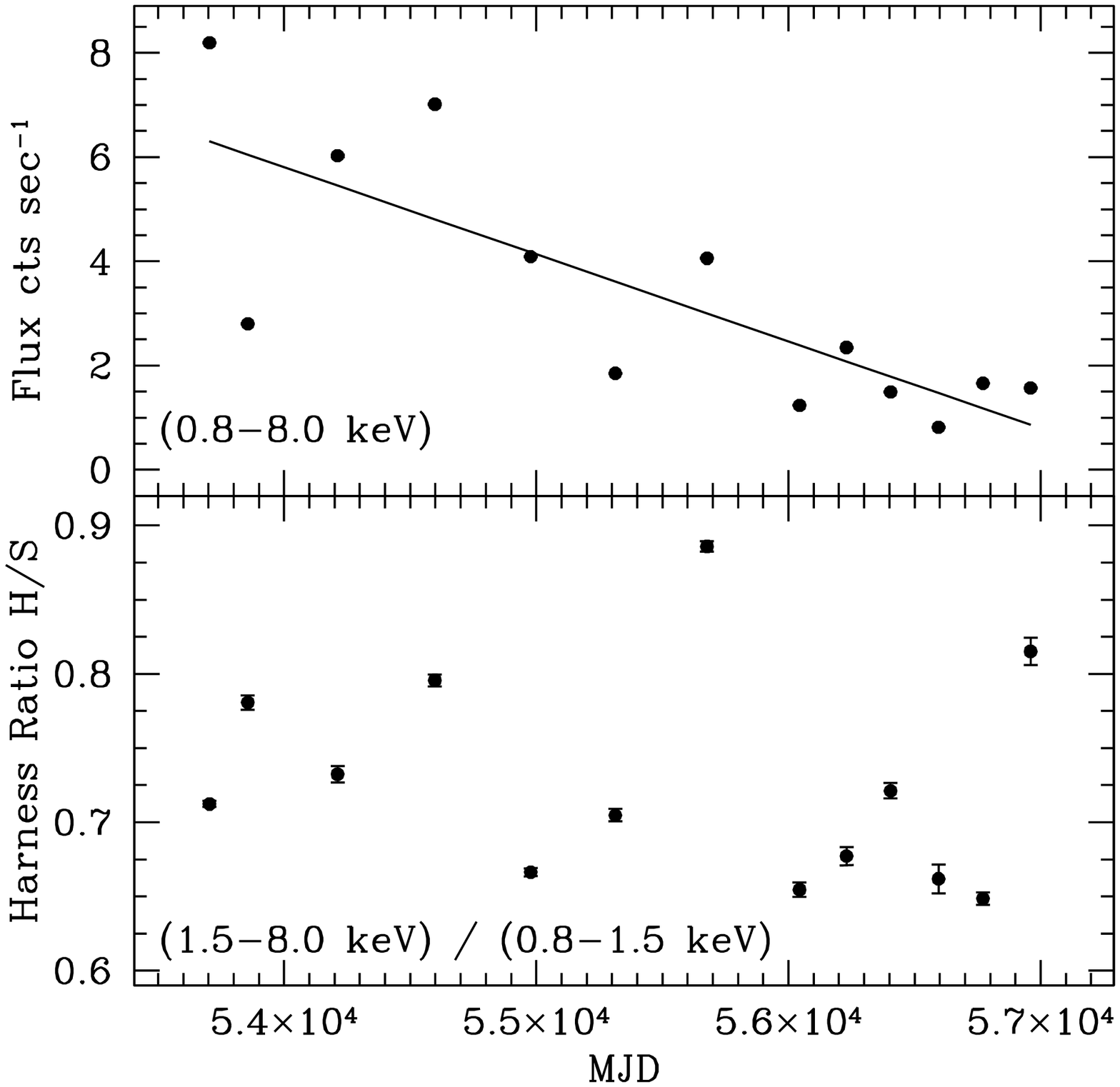} 
\figcaption{Long-term X-ray flux variability (top panel) and spectral variability (bottom panel) of PKS 2155--304.}
\label{fig:ltv}
%\end{center}
\end{figure}

\begin{figure}
%\begin{center}
\epsscale{1.0}
\includegraphics[width=8.5cm,angle=0]{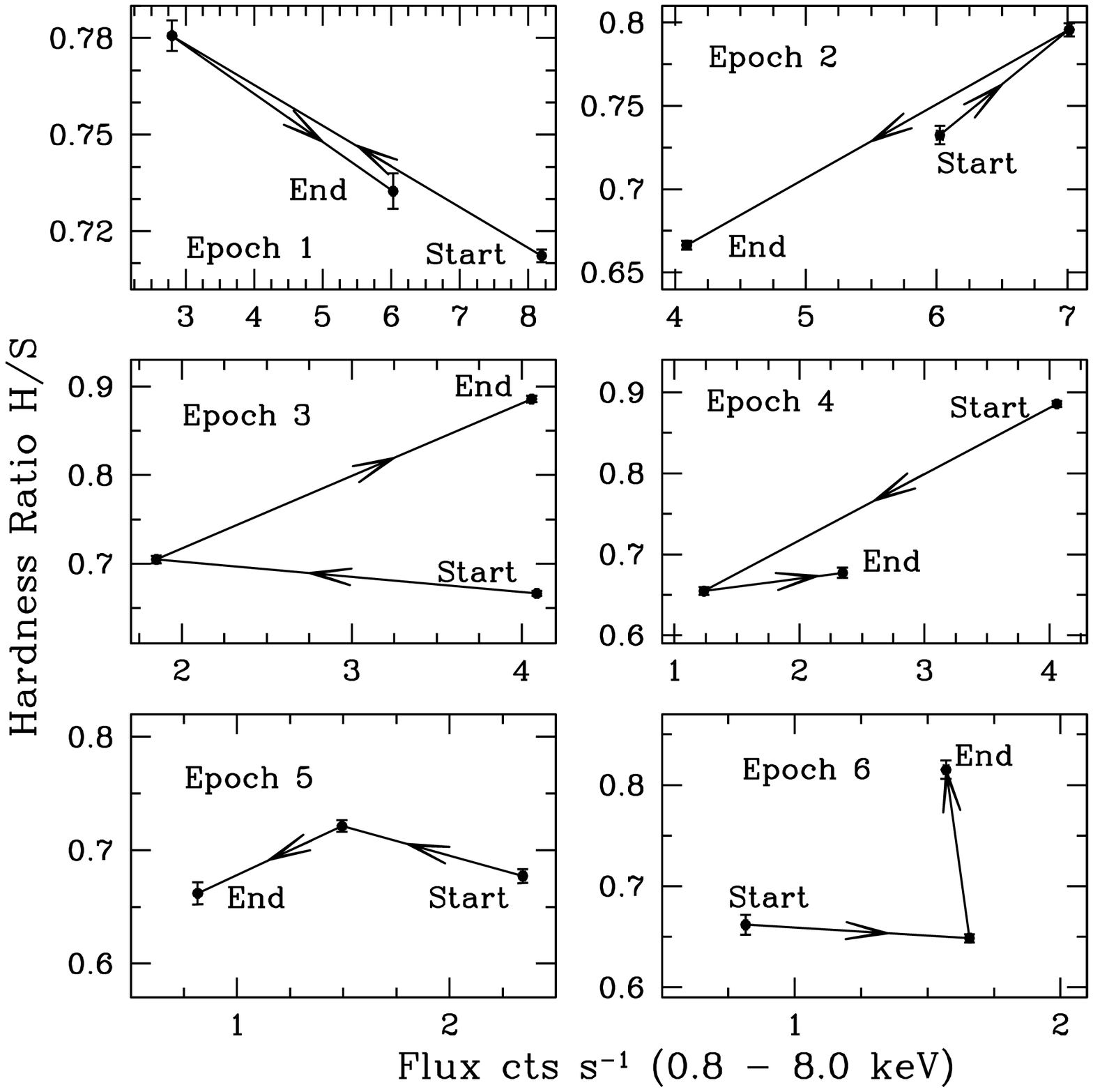} 
\figcaption{Spectral variations of PKS 2155--304 in various epochs with start and end points marking the loop directions. Each epoch corresponds to the 
time interval during which the data were acquired for each loop, considered from Epoch 1 to Epoch 6: Epoch 1: 30-11-2005 to 22-04-2007; Epoch 2: 22-04-2007
to 27-05-2009; Epoch 3: 27-05-2009 to 26-04-2011; Epoch 4: 26-04-2011 to 30-10-2012; Epoch 5: 30-10-2012 to 30-10-2013; and Epoch 6: 30-10-2013 to
30-10-2014.}
\label{fig:hr}
%\end{center}
\end{figure}

\noindent
In the case of blazars, the IDV across the complete EM spectrum is understood to be intrinsic in nature except for low-frequency radio observations 
where it may be extrinsic and arise from interstellar scintillation \citep{Wagner1995}. Low-frequency radio observations may also 
have IDV arising from a mixture of intrinsic and extrinsic natures and separating their different contributions  has been a long-lasting problem in blazar astrophysics \citep[e.g.,][]{1991ApJ...372L..71Q,2012MNRAS.425.1357G}. 
In a unified model of AGN, blazars are seen nearly face-on, so, any fluctuations arising from the accretion disk as well as the emission due to the jet should 
be directly visible \citep{Urry1995}. The IDV detected when a blazar is in a low-state can be explained by some types of disk related instabilities, because when the source is fainter the  jet emission is less dominant over the thermal emission from the disk.  Such fast accretion disk fluctuations could arise from
hot spots on or above the accretion disk, a tilted disk, or a dynamo 
\citep[e.g.,][]{Chakrabarti1993,Mangalam1993,Henisey2012,Sadowski2016}.    \\ 
\\
For any AGN, IDV studies can provide an important tool to learn about properties of its central SMBH, particularly if the emission
 mechanism functions in its close vicinity.   The observed flux variability timescales, given in Table 2 
 can be used to estimate the size, and possibly even the location, of the emitting region \citep[e.g.,][]{2003A&A...400..487C}. Using 
these 13 observations, we estimated that the shortest weighted IDV timescale was 18.5 ks,  from the 
observation ID 700012010.  Employing the simplest causality argument, for emission from a relativistic jet we can use the shortest variability timescale 
$\tau_{var}$ for estimating the upper limit for the size of the emitting region, $R$, as  

\begin{equation}
R\leq\frac{\delta }{\left ( 1+z \right )}c\tau_{var} .
\end{equation}

\noindent
The value of $\delta$ for PKS 2155$-$304
has been estimated in a broad range of $\sim$ 6--58 in different flux states and in different EM bands 
\citep[e.g.,][]{1997ApJ...486..799U,2008MNRAS.386L..28G,Gaur2010,2017ApJ...850..209G,2020MNRAS.495.2162P}. 
Using the shortest variability timescale 18.47 ks and this full range of $\delta$,
we find that the size of the emitting region lies in the range of $\sim$ 3.0 $\times$ 10$^{15}$ -- 2.9 $\times$ 10$^{16}$ cm. \\
\\
The mass of the central SMBH in any AGN is probably the most fundamental parameters needed to understand the object. The primary methods of estimation of masses of SMBHs in AGNs are reverberation mapping, and stellar or gas kinematics techniques \citep[e.g.,][]{Vestergaard2004}. These techniques are based 
on spectroscopic methods and require that emission lines be detected in the AGN spectrum, so they work for most quasars and Seyfert galaxies and even weaker AGN.  However, PKS 2155$-$304 is a BL Lac object whose spectrum is essentially 
a featureless continuum, so these primary methods cannot be directly used to determine the mass of its SMBH. 
In the case of a BL Lac object one can more crudely estimate the mass of its SMBH 
using the absolute magnitude of the elliptical host galaxy. In this fashion a SMBH mass for PKS 2155$-$304 of
$\sim$ 10$^{9}$ M$_{\odot}$ was proposed \citep{1991ApJ...380L..67F,1998A&A...336..479K}. \\ 
\\
There are alternative methods which can be used to estimate the SMBH mass in case neither the primary nor such indirect methods 
can be used. One such method is to estimate the period from the detection of periodic or quasi-periodic oscillations
in the time series data of a blazar.  Only occasionally in a few blazars have detections of  QPOs been claimed 
\citep[see for reviews,][and references therein]{2014JApA...35..307G,2018Galax...6....1G}. PKS 2155$-$304 is one of 
these blazars which have shown possible QPOs in the time series data in different EM bands on various occasions, but on diverse timescales
\citep[e.g.,][]{1993ApJ...411..614U,2000A&A...355..880F,2009A&A...506L..17L,Gaur2010,2014RAA....14..933Z,2014ApJ...793L...1S,2016MNRAS.462L..80P}. The central SMBH mass of PKS 2155$-$304 estimated using these putative QPO periods is in the huge range of
1.8 $\times$ 10$^{7}$ M$_{\odot}$ -- 2.2 $\times$ 10$^{9}$ M$_{\odot}$ \citep[e.g.,][]{1993ApJ...411..614U,2009A&A...506L..17L,Gaur2010}. \\
\\
Another alternative method of estimating the SMBH mass of a blazar is based on the shortest detected viability timescale.
Using that method and fastest detected very high energy (VHE) variability in PKS 2155$-$304, and by assuming the emission region 
has a size comparable to the Schwarzschild radius $R_{S}$ where $R_{S} = 2GM_{BH}/c^{2}$, the SMBH mass was estimated to be 
$\sim$ 10$^{9}$ M$_{\odot}$ \citep{2007ApJ...664L..71A}. If we assume that the detected X-ray emissions by XIS {\it Suzaku} were 
emitted in the close vicinity to the SMBH and not from the jet, and at around $R = 5R_{S}$, the mass of SMBH can be expressed, 
as in \citet{2012NewA...17....8G} as
\begin{equation}
M_{BH} \approx {\frac {c^{3}t}{10~G~(1+z)}}.
\end{equation}    

\noindent
Using the shortest weighted variability timescale seen by {\it Suzaku} of 18.47 ks, we would arrive at an estimate of the mass of SMBH of PKS 2155$-$304
to be $\approx$ 3.4 $\times$ 10$^{8}$ M$_{\odot}$. This SMBH mass estimation lies within the wide range of previous mass estimates of 
1.8 $\times$ 10$^{7}$ M$_{\odot}$ -- 2.2 $\times$ 10$^{9}$ M$_{\odot}$ 
\citep[e.g.,][]{1991ApJ...380L..67F,1993ApJ...411..614U,1998A&A...336..479K,2005ApJ...629..686Z,2007ApJ...664L..71A,2009A&A...506L..17L,Gaur2010}. 
However, if the observed flux perturbations 
in the blazar on IDV timescales arise from fluctuations in the accretion disk but are advected into the jet, then an additional Doppler boosting factor $\delta$ should be introduced in the SMBH mass estimation \citep{2007AJ....133.2187D}, or
$M_{BH} (\delta) = \delta \times M_{BH}$.
By using the above assumptions, and taking the previously mentioned extensive range of $6 \le \delta \le 58$, we would estimate the mass of SMBH of the blazar PKS 2155$-$304
to be in the range of 2.0 $\times$ 10$^{9}$ to 2.0 $\times$ 10$^{10}$ M$_{\odot}$. The SMBH mass estimate obtained this way  overlaps with the highest  previously estimated mass of this SMBH but extends an order of magnitude higher. 
For PKS 2155$-$304, the most recent independent estimation is $\delta = 22.3$ by using optical polarization \citep{2020MNRAS.495.2162P}. 
Also recently, $\delta = 35$ was adopted to model its multi-wavelength SED  \citep{2019MNRAS.484..749C}. Another recent 
multi-epoch SED model was best fit by fixing the final Lorentz factor of the jet at $\Gamma  = 15$ and the LOS angle $\theta = 2.5^{\circ}$, leading to
a value of $\delta \approx 22$ \citep{2019MNRAS.482.4798L}. \\
\\
To estimate some important quantities for this blazar 
we now focus on the more likely situation that the emission is jet dominated and does not fundamentally arise from the accretion disk. To do that, we adopt an intermediate 
value of $\delta = 30$, based on the above recent estimates. For electron acceleration in blazar jets, the acceleration timescale of the diffusive 
shock acceleration mechanism is often assumed to be responsible \citep[e.g.,][]{1987PhR...154....1B}. \citep{2000ApJ...536..299K,2002ApJ...572..762Z} have estimated 
the acceleration timescale (in the observer's frame) of an electron with energy ${\it E = \gamma m_{e}c^{2}}$ for the diffusive shock acceleration mechanism as
\begin{equation}
t_{acc}(\gamma) \simeq 3.79\times10^{-7} \frac{(1+z)}{\delta} \xi B^{-1}\gamma ~{\rm s},
\end{equation}
here $\xi$, $B$, and $\gamma$ are the acceleration parameter, magnetic field in Gauss, and the Lorentz factor of the ultrarelativistic 
electrons in the jet frame, respectively. PKS 2155$-$304 is a TeV emitting object and belongs to the HBL/HSP subclass of the 
blazars\footnote{http://tevcat.uchicago.edu}. In such blazars, the X-ray emission is almost certainly dominated by synchrotron radiation in the first SED hump, since the inverse Compton emission extends to such extreme $\gamma$-rays.   \\
\\
We can then proceed with an analysis exactly along the lines of our discussion of Mrk 421 in \citet{2019ApJ...884..125Z}, and specifically refer the reader to equations (17)--(21) in that paper and to the discussion around Eq.\ (5) of \citet{2015ApJ...811..143P}.  In summary, we use the synchrotron cooling timescale for an electron of Lorentz factor $\gamma$ 
\citep[see, e.g.,][]{Rybicki1979} 
and combine it with the critical synchrotron emission frequency in the {\it Sukaku} energy range we consider here (0.8 -- 8.0 KeV) and the shortest weighted variability timescale for PKS 2155$-$304, or 18.5 ks, for which $F_{var} \simeq 14$\%.  By requiring the cooling timescale to not be shorter than the minimum variability timescale one obtains a constraint on the minimum value of the magnetic field of
\begin{equation}
B \geq 0.19 (1+z)^{1/3} \delta^{-1/3} \nu^{-1/3}_{18} ~{\rm G},
\end{equation}
\noindent where 0.193 $< \nu_{18} \equiv \nu/10^{18}{\rm Hz} <$ 1.93.  Using the moderate value of $\delta = 30$ and recalling that  $z = 0.116$ for 
 PKS 2155$-$304, this becomes 
$B \geq 0.06~ \nu^{-1/3}_{18} {\rm G}$.
In turn, this produces  a constraint on the electron Lorentz factor of
\begin{equation}
\gamma \leq 0.38 \times 10^{6} ~\nu^{2/3}_{18}.
\end{equation}
\noindent
If we simply adopt $\nu_{18} \approx 1$ because it lies between the limits of 0.193 and 1.93, we get $B \geq$ 0.06 G
and $\gamma \leq$ 0.38 $\times$ 10$^{6}$. There are several previous estimations of the magnetic field $B$ for PKS 2155$-$304 found by various methods; they range from 0.018 G to
1.2 G, though they were made at different epochs and at different flux states \citep{2000ApJ...528..243K,2007ApJ...657L..81F,2009ApJ...696L.150A,2012A&A...539A.149H,2017ApJ...850..209G}. We note that our estimate of the magnetic field  is consistent with those arising from the earlier approaches.
  
\section{Conclusions}
\noindent
We examined the light curves of 13 X-ray pointed observations taken by {\it Suzaku} satellite for the TeV emitting HSP blazar
PKS 2155$-$304. These observations were taken over the entire operational period of the {\it Suzaku} satellite and the data were 
available in the public archive. We searched for IDV and its timescales, HRs,  time lags between soft and hard X-ray energies, and did
PSD analyses of the LCs for the total X-ray energy range  to search for the power-laws of the red-noise and any possible 
QPOs, and we also searched for correlation between the average fluxes and IDV timescales for each  individual pointed 
observation. Our conclusions are  as follows: 
\begin{enumerate}
\item[{\bf 1.}] The fractional variability $F_{var}$ measurements clearly show that this blazar shows large-amplitude IDV for all 13 pointed 
observations in both soft and hard X-ray bands of {\it XIS} onboard the {\it Suzaku} satellite. In general $F_{var}$ is lower in 
soft band than hard band. But in three observations, the opposite trend is found. The weighted IDV timescale for the total energy
(0.8 -- 8.0 keV) are estimated to range between 18.47 ks to 107.92 ks. We use the shortest IDV timescale of 18.47 ks  to calculate  various 
parameters of the blazar emission.    
\item[{\bf 2.}] For all 13 pointed observation IDs, our HR analyses in the soft (0.8 -- 1.5 keV) versus the hard (1.5 -- 8.0 keV) bands show
in general similar patterns as the LCs in the total (0.8 -- 8.0 keV) X-ray energy band. PKS 2155$-$304 exhibits in general a
harder-when-brighter trend often seen in HSP blazars. But occasionally, the HR analysis exhibits either a stable hardness ratio or a slightly 
softer-when-brighter trend. 
\item[{\bf 3.}] Cross-correlations between the soft (0.8 -- 1.5 keV) and the hard (1.5 -- 8.0 keV) X-ray were performed 
by the DCF analysis method for all 13 pointed observations. All DCF plots are well correlated with lags consistent with zero, indicating that the emissions in the soft and the hard bands are likely cospatial and  emitted from the same population of leptons.
\item[{\bf 4.}] PSD analyses in the total (0.8 -- 8.0 keV) X-ray energy band could be performed for 11 out of the 13 pointed observations.
The PSDs are red-noise dominated, and there is no evidence of QPO detection on any of the PSD plots.  The PSD power-law slopes are found to vary
in a large range from $-2.8$ to $-0.88$, which is one of the more surprising results of our study.
\item[{\bf 5.}] We find no correlation between average flux and IDV timescale.
\item[{\bf 6.}] On the long timescale spanning these 9 years of data the flux a shows decreasing trend while the HR does not show any trend with time. On intermediate long timescales, the HR versus flux 
plots frequently show portions of anti-clockwise loops (hard lags) which may imply that the timescales for particle acceleration mechanisms dominate the variable emission of this blazar.    
\item[{\bf 7.}] If we make the rather unlikely assumption that the X-ray emission in the blazar comes from the close vicinity of the SMBH, we estimate its mass to be $\approx$ 3.4 $\times$ 10$^{8}$ M$_{\odot}$. But if the flux variability arises due to such perturbations carried into the jet, then the emission
is Doppler boosted, which yields much larger SMBH mass estimates.        
\item[{\bf 8.}] If we assume the more likely situation where the emission independently arises in the jet then constraints are obtained on the magnetic field strength ($B \gtrsim 0.06$G) and the Lorentz factor of the ultrarelativistic electrons emitting the X-rays ($\gamma \lesssim 3.5 \times 10^5$).  
\end{enumerate}     

\section*{ACKNOWLEDGMENTS}
\noindent
We thank the referee for comments that led to significant improvements in the presentation of these results. This research has made use of data obtained from the {\it Suzaku} satellite, a collaborative mission between 
the space agencies of Japan (JAXA) and the USA (NASA). \\  
\\
This work is funded by the National Key R\&D Programme of China (under grants No. 2018YFA0404602 and 2018YFA0404603). ZZL is thankful for support from the Chinese Academy of Sciences (CAS) for the Talented Program. ACG is partially supported by Chinese Academy of Sciences (CAS) President's International Fellowship Initiative (PIFI) (grant no. 2016VMB073). HG acknowledges  financial support from the Department of Science \& Technology (DST), Government of India through the INSPIRE faculty award IFA17-PH197 at ARIES, Nainital, India. H.G.X. is supported by the Ministry of Science and Technology of China (grant No. 2018YFA0404601), and the National Science Foundation of China (grants No. 11621303, 11835009, and 11973033).

\software{HEADAS \citep[v6.18;][]{Blackburn1995}}

\clearpage

\end{document}